\documentclass[longauth]{aa}

\usepackage{txfonts}			
\usepackage[utf8]{inputenc}
\usepackage{amsmath}
\usepackage{nccmath}		
\usepackage[english]{babel}	
\usepackage{xcolor}			
\usepackage{graphicx}			
\usepackage{booktabs}		
\usepackage{gensymb}		
\usepackage[
colorlinks = true,
anchorcolor = blue,
linkcolor = blue,
urlcolor  = blue,
citecolor = blue,
]{hyperref}
\usepackage[colorinlistoftodos]{todonotes}
\usepackage{textcomp}
\usepackage{multirow}
\usepackage{orcidlink}
\usepackage{comment}
\usepackage{makecell}
\usepackage{subcaption}
\usepackage{longtable,booktabs,array}
\usepackage{needspace}
\usepackage{alphalph}
\usepackage{etex}

\let\oldsim\sim
\renewcommand{\sim}{{\oldsim}}

\makeatletter
\renewcommand\thesubfigure{\alphalph{\value{subfigure}}}
\renewcommand\p@subfigure{\thefigure}
\makeatother

\makeatletter
\providecommand{\lcpanel}{}
\renewcommand{\lcpanel}[3][]{
  \begin{minipage}{0.24\textwidth}\centering
    \includegraphics[width=\linewidth]{#2}
    \refstepcounter{subfigure}
    \if\relax\detokenize{#1}\relax\else\label{#1}\fi
    \\[0.2em]\textbf{(\thesubfigure)}\ #3
  \end{minipage}%
}
\makeatother

\begin{document}
\title{Constraining the size, shape, and albedo of the large Trans-Neptunian Object (28978) Ixion with multi-chord stellar occultations}
\titlerunning{Physical properties of the large TNO Ixion}

\author{
Y.~Kilic\inst{1,2}\orcidlink{0000-0001-8641-0796}
\and
F.~Braga-Ribas\inst{3,4,2}\orcidlink{0000-0003-2311-2438}
\and
C.~L.~Pereira\inst{5,4}\orcidlink{0000-0003-1000-8113}
\and
J.~L.~Ortiz\inst{1}\orcidlink{0000-0002-8690-2413}
\and
B.~Sicardy\inst{6}\orcidlink{0000-0003-1995-0842}
\and
P.~Santos-Sanz\inst{1}\orcidlink{0000-0002-1123-983X}
\and
O.~Erece\inst{7,8}\orcidlink{0000-0002-9723-6823}
\and
J.~L.~Rizos\inst{1}\orcidlink{0000-0002-9789-1203}
\and
J.~M.~Gómez-Limón\inst{1}\orcidlink{0009-0006-8584-1416}
\and
G.~Margoti\inst{5,4}\orcidlink{0000-0002-2103-4408}
\and
D.~Souami\inst{2,9}\orcidlink{0000-0003-4058-0815}
\and
B.~Morgado\inst{10}\orcidlink{0000-0003-0088-1808}
\and
A.~Gomes-Junior\inst{11,4,12}\orcidlink{0000-0002-3362-2127}
\and
L.~M.~Catani\inst{10,1}\orcidlink{0000-0002-1960-888X}
\and
J.~Desmars\inst{13,6}\orcidlink{0000-0002-2193-8204}
\and
M.~Kretlow\inst{1,14}\orcidlink{0000-0001-8858-3420}
\and
F.~Rommel\inst{15}\orcidlink{0000-0002-6085-3182}
\and
R.~Duffard\inst{1}\orcidlink{0000-0001-5963-5850}
\and
A.~Alvarez-Candal\inst{1}\orcidlink{0000-0002-5045-9675}
\and
J.~I.~B.~Camargo\inst{5,4}\orcidlink{0000-0002-1642-4065}
\and
M.~Kaplan\inst{16}\orcidlink{0000-0002-9723-6823}
\and
N.~Morales\inst{1}\orcidlink{0000-0003-0419-1599}
\and
D.~Herald\inst{17}\orcidlink{0000-0003-2026-1630}
\and
M.~Assafin\inst{10,4}\orcidlink{0000-0002-8211-0777}
\and
G.~Benedetti-Rossi\inst{2,5,4}\orcidlink{0000-0002-4106-476X}
\and
R.~Sfair\inst{12,18,2}\orcidlink{0000-0002-4939-013X}
\and
R.~Savalle\inst{19}\orcidlink{0000-0002-4826-246X}
\and
J.~Arcas-Silva\inst{5}\orcidlink{0009-0007-1444-9502}
\and
L.~Bernasconi\inst{20}
\and
T.~Blank\inst{21}
\and
M.~Bonavita\inst{22}\orcidlink{0000-0002-7520-8389}
\and
N.~Carlson\inst{21}
\and
B.~Christophe\inst{23}
\and
C.~A.~Colesanti\inst{24}\textsuperscript{\textdagger}
\and
M.~Collins\inst{25}\orcidlink{0009-0004-1001-8359}
\and
G.~Columba\inst{26}\orcidlink{0009-0004-2173-7914}
\and
R.~Dunford\inst{21}
\and
D.~W.~Dunham\inst{27}\orcidlink{0000-0001-7527-4207}
\and
J.~Dunham\inst{21}
\and
M.~Emilio\inst{5,28}
\and
W.~G.~Ferrante\inst{3}
\and
T.~George\inst{21}\textsuperscript{\textdagger}
\and
W.~Hanna\inst{21}
\and
G.~Isopi\inst{29,30,31,32}\orcidlink{0000-0002-8458-0588}
\and
R.~Jones\inst{21}
\and
D.~A.~Kenyon\inst{33}
\and
S.~Kerr\inst{17}
\and
V.~Kouprianov\inst{34}
\and
P.~D.~Maley\inst{21,35}
\and
F.~Mallia\inst{30}
\and
J.~Mattei\inst{36}
\and
M.~Meunier\inst{37}
\and
T.~Napoleao\inst{24}
\and
V.~F.~Peixoto\inst{10,4}\orcidlink{0000-0002-3319-4548}
\and
J.~Pollock\inst{38}\textsuperscript{\textdagger}
\and
C.~Snodgrass\inst{22}\orcidlink{0000-0001-9328-2905}
\and
A.~Stechina\inst{39}
\and
W.~Thomas\inst{21}
\and
R.~Venable\inst{21}
\and
G.~R.~Viscome\inst{21,40}
\and
A.~Zapparata\inst{30}\orcidlink{0000-0002-9428-1573}
\and
J.~Bardecker\inst{21}
\and
N.~Castro\inst{41}
\and
C.~Cebral\inst{42}
\and
A.~Chapman\inst{43}
\and
C.~Gao\inst{44}
\and
K.~Green\inst{44,45}
\and
A.~Guimaraes\inst{12}
\and
C.~Jacques\inst{46,47}\orcidlink{0000-0002-3100-463}
\and
E.~Jehin\inst{48}\orcidlink{0000-0001-8923-488X}
\and
M.~Konishi\inst{42}
\and
R.~Leiva\inst{1}\orcidlink{0000-0002-6477-1360}
\and
L.~Liberato\inst{49}
\and
C.~Magliano\inst{42}
\and
L.~A.~Mammana\inst{50,51}
\and
M.~Melita\inst{52}
\and
V.~Moura\inst{12}
\and
Y.~Olivera-Cuello\inst{42}
\and
L.~Peiro\inst{53}
\and
J.~Spagnotto\inst{54}\orcidlink{0009-0007-5363-6298}
\and
P.~C.~Stuart\inst{21}
\and
L.~Vanzi\inst{41}
\and
A.~Wilberger\inst{55}\orcidlink{0009-0001-6346-2902}
\and
M.~Malacarne\inst{56}\orcidlink{0000-0002-0866-1802}
}

\authorrunning{Kilic et al.}

\institute{
Instituto de Astrofísica de Andalucía, IAA-CSIC, Glorieta de la Astronomía s/n, 18008 Granada, Spain \and 
LIRA, CNRS UMR8254, Observatoire de Paris, Meudon, France
\and
Federal University of Technology – Paraná (PPGFA/UTFPR), Curitiba, PR, Brazil
\and
Laboratório Interinstitucional de e-Astronomia (LIneA), Rio de Janeiro, RJ, Brazil
\and
Observatório Nacional (MCTI), Rio de Janeiro, RJ, Brazil
\and
LTE, Observatoire de Paris, Université PSL, Sorbonne Université, Université de Lille, LNE, CNRS, Paris, France
\and
Türkiye National Observatories, TUG, Antalya, Türkiye
\and
The Scientific and Technological Research Council of Türkiye (TÜBİTAK), Ankara, Türkiye
\and
naXys, Department of Mathematics, University of Namur, Namur, Belgium
\and
Federal University of Rio de Janeiro - Observatory of Valongo, Rio de Janeiro, Brazil
\and
Instituto de Física, Universidade Federal de Uberlândia, Uberlândia, MG, Brazil
\and
UNESP – São Paulo State University, Grupo de Dinâmica Orbital e Planetologia, Guaratinguetá, SP, Brazil
\and
Institut Polytechnique des Sciences Avancées (IPSA), Ivry-sur-Seine, France
\and
Deutsches Zentrum für Astrophysik (DZA), Görlitz, Germany
\and
Florida Space Institute, University of Central Florida, Orlando, FL, USA
\and
Department of Space Sciences and Technologies, Akdeniz University, Antalya, Türkiye
\and
Trans-Tasman Occultation Alliance (TTOA), Wellington, New Zealand
\and
Institute of Astronomy and Astrophysics, University of Tübingen, Germany
\and
PADC/DIO, Observatoire de Paris, PSL University, France
\and
Observatoire des Engarouines, Malemort-du-Comtat, France
\and
International Occultation Timing Association (IOTA), Fountain Hills, AZ, USA
\and
Institute for Astronomy, University of Edinburgh, Edinburgh, United Kingdom
\and
Société Astronomique de France (SAF), Paris, France
\and
Grupo Alfa Crucis, Brazil
\and
Lowell Observatory, Flagstaff, AZ, USA
\and
Alma Mater Studiorum – University of Bologna, Department of Physics and Astronomy “A. Righi”, Bologna, Italy
\and
KinetX, Inc., Space Navigation and Flight Dynamics Practice, Simi Valley, California, USA
\and
Universidade Estadual de Ponta Grossa (UEPG), Ponta Grossa, Brazil
\and
Sapienza Università di Roma, Rome, Italy
\and
Campo Catino Astronomical Observatory, Guarcino, Italy
\and
INFN, Sezione Roma1, Rome, Italy
\and
INAF OAC, Via della Scienza, Selargius, Italy
\and
Kenyon Astrophysical Observatory (KAO), Northern California, USA
\and
Skynet Robotic Telescope Network, University of North Carolina, Chapel Hill, NC, USA
\and
Johnson Space Center Astronomical Society, Houston, TX, USA
\and
Independent observer, Brazil
\and
Independent observer, Arbonne-la-Forêt, France
\and
Department of Physics and Astronomy, Appalachian State University, Boone, NC, USA
\and
Centro de Amigos de la Astronomia Reconquista – CAAR, Reconquista, Argentina
\and
Rand Observatory II - MPC W71, Lake Placid, NY, USA
\and
Pontificia Universidad Católica de Chile, Center for Astro Engineering, Santiago, Chile
\and
Asociación Argentina Amigos de la Astronomía, Argentina
\and
Cruz del Sur Private Observatory, El Peral, San Juan Province, Argentina
\and
Westport Astronomical Society, Westport, Connecticut, USA
\and
University of New Haven, West Haven, CT, USA
\and
SONEAR Observatory – CEAMIG, Caeté, Minas Gerais, Brazil
\and
Centro de Estudos Astronômicos de Minas Gerais (CEAMIG), Belo Horizonte, Brazil
\and
Space sciences, Technologies \& Astrophysics Research (STAR) Institute, University of Liège, Liège, Belgium
\and
Université Côte d’Azur, Observatoire de la Côte d’Azur, CNRS, Laboratoire Lagrange, Nice, France
\and
Complejo Astronómico El Leoncito (CASLEO), San Juan, Argentina
\and
Facultad de Ciencias Astronómicas y Geofísicas (UNLP), La Plata, Argentina
\and
Instituto de Astronomía y Física del Espacio, CONICET–Universidad de Buenos Aires, Argentina
\and
Agrupaciones Astronómicas de Madrid (AAM) y Teruel (ACTUEL), Spain
\and
El Catalejo Observatory (MPC I48), Santa Rosa, La Pampa, Argentina
\and
Los Cabezones Observatory (MPC X12), Santa Rosa, La Pampa, Argentina
\and
Federal University of Espírito Santo: Vitória, Espírito Santo, BR
}

\date{Received 4 November, 2025; accepted 12 January, 2026}

\abstract
{Trans-Neptunian objects (TNOs) are among the most primitive remnants of the early Solar System. Determining their sizes, shapes, albedos, and surface properties is essential for understanding their origin and evolution. Stellar occultations provide highly accurate size and shape information for TNOs, while photometry constrains their albedo and surface colours. (28978)~Ixion is one of the largest TNOs and a prominent Plutino, making it a key target for comparative studies.}
{The aim of this work is to constrain Ixion’s projected size, shape, absolute magnitude, geometric albedo, and surface colours, and to search for evidence of an atmosphere or circum-object material.}
{We conducted a series of stellar occultation campaigns by Ixion between 2020 and 2023 as part of the \textit{Lucky Star} collaboration, gathering 51 observations from eight events, including 30 positive detections. Five multi-chord events were used for a global limb fit, enabling an accurate reconstruction of Ixion’s projected shape. Calibrated photometric data, including new and archival measurements, were analysed to derive its absolute magnitude, phase-curve parameters, and broadband colours.}
{The multi-chord occultations reveal a slightly elongated limb, well represented by a single projected ellipse with semi-axes $a = 363.42^{+3.53}_{-3.85}$~km and $b = 333.98^{+7.07}_{-4.96}$~km, yielding $R_\mathrm{equiv} = 348.39^{+5.37}_{-4.43}$~km ($D_\mathrm{equiv} = 696.78^{+10.75}_{-8.87}$~km) and an apparent oblateness $\epsilon' = 0.081^{+0.004}_{-0.010}$. The geometry is consistent with a moderately flattened, nearly spheroidal body that may show slight departures from axial symmetry. The typical radial residuals ($\sim$10~km) support a largely stable shape across the observed epochs, with modest epoch-dependent variations. The phase-curve fit gives $H_V = 3.845 \pm 0.006$, $\beta = 0.1301 \pm 0.0078$~mag\,deg$^{-1}$, and $p_V = 0.106^{+0.003}_{-0.003}$. Multi-band photometry yields $B-V = 1.06 \pm 0.03$, $V-R = 0.61 \pm 0.02$, and $R-I = 0.54 \pm 0.03$, consistent with moderately red TNO surfaces. No atmosphere or circum-object material was detected down to our sensitivity limits. The best-sampled event (13~Oct~2020) also allowed us to measure the angular diameter of the occulted star Gaia~DR3~4056440205544338944, $\theta_\star = 0.670 \pm 0.010$~mas, corresponding to $R_\star = 128 \pm 10\,R_\odot$ at the Gaia distance.}
{}

\keywords{Kuiper belt objects: individual: (28978) Ixion -- Astrometry -- Occultations -- Photometry -- Methods: observational}
\maketitle
\makeatletter
\renewcommand\@makefnmark{}
\makeatother
\footnotetext{\textsuperscript{\textdagger}Deceased.}

\section{Introduction}
\label{sec:intro}

The stellar occultation technique is one of the most precise Earth based methods for determining the physical properties of trans-Neptunian objects (TNOs). Since the first successful observation of a stellar occultation by a TNO other than Pluto or Charon in 2009 \citep{elliot_2010}, this technique has provided highly accurate size measurements with kilometre-level precision, constraints on atmospheric properties down to nanobar levels (e.g., \citealt{sicardy_2024}; \citealt{fribas_2013}), and has enabled the discovery and characterization of ring systems around objects such as Chariklo, Chiron, Haumea, and Quaoar \citep{fribas_2014, ortiz_2015, ort17,sickafoose_2020, ortiz_2023,morgado_2023,pereira_2023,pereira2025_chiron,santos_2025}. Additionally, stellar occultations have revealed topographic features on TNO surfaces, offering insights into surface irregularities and compositions \citep{dias_oliveira_2017,rommel_2023}. Observations of moons, such as Vanth (Orcus/1) and Weywot (Quaoar/1) \citep{sickafoose2019,fribas2025}, further emphasize the technique's power in the study of outer Solar System objects. Further occultation studies have also constrained smaller TNO satellites such as Hi'iaka \citep{fernandezvalenzuela2025}, Namaka \citep{rommel2025_namaka}, Huya’s moon \citep{rommel2025_huya}, 2014WC\textsubscript{510} \citep{leiva2020_wc510}, and 2000YW\textsubscript{134} \citep{varalubiano2023_yw134}. Stellar occultations have also successfully constrained the shapes of Centaurs, such as 2002~GZ\textsubscript{32} \citep{santos_2020}, Chiron \citep{fribas_2023}, Echeclus \citep{pereira_2024}, and Bienor \citep{rizos_2024}, with the shape and spatial orientation of the Centaur Chariklo being a notable example \citep{leiva_2017, morgado_2021}.

Trans-Neptunian objects are icy remnants from the early Solar System beyond Neptune. Their study provides valuable insights into the primordial conditions of Solar System formation and evolution 
\citep{lykawka2008, morbidelli2008}. Among these, (28978) Ixion\footnote{(28978) Ixion had the provisional designation 2001~KX$_{76}$.} is a large Plutino, orbiting the Sun in a 3:2 resonance with Neptune. 
With an estimated diameter in the $\sim$700~km range (see
Table~\ref{tab:ixion-diameters}) and moderately red surface properties, Ixion ranks among the brightest Plutinos and represents a key target for detailed investigation.

\begin{table}[htb]
    \caption{Equivalent diameters of (28978) Ixion from various studies.}
    \label{tab:ixion-diameters}
    \centering
    \resizebox{\columnwidth}{!}{%
        \begin{tabular}{lccc}
        \hline \hline
        Year & Diameter (km) & Method & Reference \\
        \hline
        2002 & $1055 \pm 165$ & Radiometry & \citet{bertoldi_2002} \\
        2003 & $<804$ & Radiometry & \citet{altenhoff2004} \\
        2004 & $<822$ & Optical + Thermal & \citet{grundy2005} \\
        2005 & $475 \pm 75$ & Thermal & \citet{stansberry2005} \\
        2007 & $480^{+152}_{-136}$ & Thermal + Radiometry & \citet{cruikshank2007} \\
        2008 & $650^{+260}_{-220}$ & Thermal (Spitzer) & \citet{stansberry2008} \\
        2008 & $590 \pm 190$ & Thermal & \citet{mikebrown2008} \\
        2013 & $\sim549$ & Thermal (Herschel--Spitzer)& \citet{mommert2013} \\
        2013 & $617^{+19}_{-20}$ & Thermal (Herschel--Spitzer) & \citet{lellouch2013} \\
        2021 & $> 709.6 \pm 0.2$ & Single chord stellar occ. & \citet{levine2021} \\
        \hline
        \end{tabular}
    }
\end{table}

To date, around 259 stellar occultations involving 57 TNOs and 14 Centaurs have been observed, but only approximately 41 have yielded multi-chord observations\footnote{Based on \cite{braga_ribas_2019} and 
the Occultation Portal database: \url{https://occultationportal.org}}. Multi-chord data are essential for accurately determining the shapes, sizes, and densities of these distant bodies. Among these, Ixion stands out due to its unique orbital configuration and physical properties. Ixion was discovered in May 2001 at the Cerro Tololo Inter-American Observatory during the Deep Ecliptic Survey \citep{mpc_2002}. As part of the Plutino group in the 3:2 mean-motion resonance with Neptune, Ixion is a relevant target for studying Neptune’s migration history and its influence on the dynamical evolution of the Kuiper Belt. Such resonance-locked configurations are understood in the framework of the resonance capture mechanism proposed by \citet{malhotra_1995}.

Ixion's orbital parameters closely resemble those of Pluto, with a semi-major axis
$a=39.35~\mathrm{au}$, an eccentricity $e=0.244$, and an inclination $i=19.67^{\circ}$\footnote{JPL Small-Body Database: \url{https://ssd.jpl.nasa.gov/}}. However, Ixion’s slightly higher inclination and distinct surface characteristics make it an intriguing object of study.

The latest equivalent diameter, reported as $709.6 \pm 0.2$~km from a nearly single-chord stellar occultation \citep{levine2021}, was derived under the assumption of a circular limb. As acknowledged in that study, the very small formal uncertainty naturally follows from this modelling choice, given that the event geometry provides only limited information on the limb shape. The multi-chord events analysed here therefore complement that result by providing direct constraints on the limb geometry.

Understanding its physical structure and potential for hydrostatic equilibrium provides critical insights into the processes governing the evolution of such objects \citep{grundy_2019}. The absolute magnitude ($H_V$) of Ixion has been measured as $3.774 \pm 0.021$~mag \citep{alvarezcandal2016}, and its geometric albedo ($p_V$) is calculated to be $0.108 \pm 0.002$ \citep{verbiscer2022}. Photometric measurements indicate a $B{-}V$ color index of $1.03 \pm 0.03$~mag and a $V{-}R$ color index of $0.61 \pm 0.03$~mag, suggesting a moderately red surface \citep{doressoundiram2007}.

Ixion’s rotation period has not been conclusively determined, and studies on this subject have provided varying results. The first attempt to measure Ixion's rotation period was conducted by \citet{ortiz2003}, but no periodicity was identified in the light curve, which had an amplitude of $<0.15$~mag. A subsequent study by \citet{sheppard2003} also failed to detect any significant variability, setting an upper limit of $0.05$~mag for the light curve amplitude. This value is notably lower than the amplitude reported by \citet{ortiz2003}, highlighting its importance. The first successful determination of Ixion's rotation period was achieved in 2010 using the 3.58~m New Technology Telescope (NTT) at the European Southern Observatory. \citet{rousselot2010} derived a period of $15.9 \pm 0.5$~h assuming a single-peaked light curve with an amplitude of $0.06 \pm 0.03$~mag. More recent observations by \citet{galiazzo2016} at the Las Campanas Observatory, using the 1~m Swope Telescope, estimated Ixion's rotation period to be $12.4 \pm 0.3$~h. This determination is one of several published period estimates, and—like the others—should be regarded as a plausible solution rather than a definitive constraint on Ixion’s rotation state. However, Ixion moves in a crowded stellar field near the Galactic plane, and its photometric measurements are frequently affected by contamination from faint background stars, making the determination of a precise rotational period particularly challenging.

Ixion’s surface composition has also been investigated using near-infrared spectroscopy, which suggests a mix of dark organic materials, amorphous carbon, and possibly traces of water ice and silicates \citep{boehnhardt_2004, barkume_2008}. Although no satellite has been observed around Ixion, its surface properties share similarities with Quaoar (which has a satellite called Weywot), another large TNO known for its water ice and tholins.

This article is organised as follows.
Section~\ref{sec:predictions} explains our strategy for predicting the stellar occultations analysed in this work.
Section~\ref{sec:observations} describes the observations obtained from different sites and instruments, and presents the resulting positive chords. Section~\ref{sec:data_analysis} details the analysis of the light curves, leading to constraints on the stellar radius of one of the occulted stars, Ixion’s size and shape, and its physical parameters (absolute magnitude, phase curve, albedo, and colours).
Finally, Section~\ref{sec:discussion} discusses the implications of our results and summarises the main conclusions.

\section{Predictions}\label{sec:predictions}
The remarkable astrometric precision achieved by Gaia Data Releases \citep{gaia_2016,gaia_2016b,gaia_2018,gaia_2023} has markedly improved the reliability of occultation predictions by providing stellar coordinates at the sub-milliarcsecond level \citep{rommel2020}. Under these conditions, the dominant source of uncertainty arises from the TNO’s ephemeris.

To address this, we generated updated orbital solutions for Ixion employing the NIMA algorithm \citep{des15}\footnote{Numerical Integration of the Motion of an Asteroid. For Ixion’s ephemerides, see \url{https://lesia.obspm.fr/lucky-star/obj.php?p=1192}.}, progressively incorporating new astrometric observations and successive Gaia catalogues (DR2, EDR3, DR3) across multiple epochs. The iterative refinement of the orbit, culminating in the NIMAv12 solution, enabled predictions with temporal and spatial accuracy sufficient to design and execute observation campaigns with a high probability of success. In addition, astrometric offsets obtained close to the occultation dates (e.g., from the Observatorio de Sierra Nevada, the Calar Alto Observatory, and the Liverpool Telescope) were incorporated to further improve the predictions.

The predicted mid-times and reference star information for each event are summarised in Table~\ref{tab:j2000_coordinates}, including the NIMA solution applied in each case. The observed occultation chords derived from these predictions are displayed in Figure~\ref{fig:occ_paths}, allowing a direct comparison between the predicted shadow paths and the actual detections.

\begin{table*}[htb]
\caption{J2000 star coordinates, proper motions, parallaxes, magnitudes, and NIMA solutions used for each prediction.}
\label{tab:j2000_coordinates}
\centering
\scriptsize
\begin{tabular}{l c c c c c c c c c c}
\hline\hline
Date & Designation & R.A. & errRA & Decl. & errDEC & pmRA & pmDEC & Plx & G & NIMA \\
 & & (ICRS) & (mas) & (ICRS) & (mas) & (mas yr$^{-1}$) & (mas yr$^{-1}$) & (mas) & (mag) & Version \\
\hline
28 Jul 2023 & 4049162851669604096 & 18$^{\mathrm{h}}$11$^{\mathrm{m}}$3.2654$^{\mathrm{s}}$ & 0.025 & -31$^\circ$15$'$22.477$''$ & 0.020 & 1.748 & -5.399 & 0.108 & 14.53 & NIMAv12 \\
30 Jun 2022 & 4049905056379149824 & 18$^{\mathrm{h}}$06$^{\mathrm{m}}$53.6148$^{\mathrm{s}}$ & 0.033 & -30$^\circ$46$'$56.476$''$ & 0.029 & -4.367 & -8.163 & 0.111 & 15.00 & NIMAv11 \\
2 Jun 2022 & 4049248613603980416 & 18$^{\mathrm{h}}$09$^{\mathrm{m}}$36.9353$^{\mathrm{s}}$ & 0.030 & -30$^\circ$43$'$37.051$''$ & 0.025 & 0.067 & -6.964 & 0.166 & 15.19 & NIMAv11 \\
17 Aug 2021 & 4056220440657653248 & 17$^{\mathrm{h}}$56$^{\mathrm{m}}$18.8133$^{\mathrm{s}}$ & 0.168 & -30$^\circ$17$'$43.765$''$ & 0.131 & -5.521 & -3.726 & -0.074 & 16.93 & NIMAv11 \\
20 May 2021 & 4050180170543871104 & 18$^{\mathrm{h}}$04$^{\mathrm{m}}$5.2362$^{\mathrm{s}}$ & 0.025 & -30$^\circ$13$'$44.615$''$ & 0.021 & -2.538 & -5.993 & 0.284 & 14.18 & NIMAv11 \\
28 Apr 2021 & 4050000576486883840 & 18$^{\mathrm{h}}$05$^{\mathrm{m}}$34.6814$^{\mathrm{s}}$ & 0.034 & -30$^\circ$09$'$09.703$''$ & 0.030 & -3.730 & -7.404 & 0.101 & 14.84 & NIMAv11 \\
13 Oct 2020 & 4056440205544338944 & 17$^{\mathrm{h}}$50$^{\mathrm{m}}$20.3595$^{\mathrm{s}}$ & 0.040 & -29$^\circ$43$'$33.087$''$ & 0.031 & 0.235 & 0.273 & 0.565 & 10.31 & NIMAv10 \\
17 Aug 2020 & 4056438835548480896 & 17$^{\mathrm{h}}$49$^{\mathrm{m}}$53.5830$^{\mathrm{s}}$ & 0.023 & -29$^\circ$48$'$12.589$''$ & 0.018 & 1.007 & -1.704 & 0.293 & 14.52 & NIMAv8 \\
\hline
\end{tabular}
\tablefoot{
RA. and DEC. are barycentric coordinates propagated to the occultation epoch. Errors are standard errors in RA. (corrected by $\cos\delta$) and DEC. Proper motions are also adjusted. Plx: absolute parallax. G: Gaia G-band magnitude. NIMA: Numerical Integration of the Motion of an Asteroid solution used for each event prediction.
}
\end{table*}
\section{Observations}\label{sec:observations}

Stellar occultation campaigns for (28978) Ixion were conducted between 2020 and 2023 as part of the \textit{Lucky Star} collaboration\footnote{\url{https://lesia.obspm.fr/lucky-star}}.  
These campaigns resulted in varying numbers of positive detections, ranging from one to eight chords per event, contributing to a comprehensive multi-chord analysis of Ixion.  
Observational data, including site status and event details, were collected and managed using the \textit{Occultation Portal}\footnote{\url{https://occultationportal.org}}, as described in \citet{kilic_2022}. A summary of these campaigns, including the number of chords, detection outcomes, and the main geometric and observational parameters of each occultation 
(geocentric distance, apparent sky-plane velocity, stellar angular diameter, 
and characteristic spatial sampling), is presented in Table~\ref{tab:ixion_event_summary}.

Across all campaigns, most observers conducted their measurements in Clear mode  (no optical filter) to maximise the signal-to-noise ratio. Weather conditions were generally favourable at the majority of sites, yielding high-quality light curves throughout the 2020–2023 observing period. Detailed information for each 
event — including observing configurations (telescope aperture, camera type, cadence), filter usage, exposure times, and site-specific weather reports — is provided in Table~\ref{tab:obs_ixion_all}.

\begin{table*}[h!]
\centering
\caption{Summary of geometric and observational parameters for all Ixion stellar occultation events.}
\label{tab:ixion_event_summary}
\small
\begin{tabular}{lccccccccc}
\hline\hline
Date (UT) 
& \makecell{$\Delta$ \\ (au)} 
& \makecell{$v$ \\ (km\,s$^{-1}$)} 
& \makecell{$\theta_\star$ \\ (mas)} 
& \makecell{$S^{a}$ \\ (km pt$^{-1}$)} 
& Chords 
& Positive 
& Negative
& Other \\
\hline
28 Jul 2023 06:13:24 & 37.34 & 20.21 & 0.0500 & $\sim$2--20     & 7  & 3 & 3 & 1 \\
30 Jun 2022 00:22:52  & 37.46 & 24.66 & 0.0335 & $\sim$25--100  & 8  & 4 & 1 & 3 \\
2 Jun 2022 18:09:33  & 37.54 & 23.15 & 0.0280 & $\sim$7  & 1  & 1 & 0 & 0 \\
17 Aug 2021 11:36:12 & 38.10 & 12.11 & 0.0183 & $\sim$36    & 1  & 1 & 0 & 0 \\
20 May 2021 06:43:23  & 37.89 & 20.66 & 0.0320 & $\sim$1--40  & 5  & 4 & 1 & 0 \\
28 Apr 2021 07:35:37  & 38.15 & 13.54 & 0.0471 & $\sim$7--70 & 16 & 6 & 5 & 5 \\
13 Oct 2020 01:57:46  & 39.26 & 16.12 & 0.6750 & $\sim$0.5   & 10 & 8 & 2 & 0 \\
17 Aug 2020 01:42:15  & 38.37 & 11.61 & 0.0530 & $\sim$3--6  & 3  & 3 & 0 & 0 \\
\hline
\end{tabular}

\tablefoot{
Nominal geocentric mid-time (Date, UT). Geocentric distances ($\Delta$) and apparent sky-plane velocities ($v$) are evaluated at mid-event. Stellar angular diameters ($\theta_\star$) are those adopted in the light-curve modelling. $S^{a}$ denotes the typical along-chord distance represented by a
single photometric data point, combining exposure time with the apparent shadow velocity. \emph{Chords} refers to the total number of observing stations; \emph{Positive} indicates successful detections; \emph{Negative} indicates stations where no flux drop was recorded within the predicted window; and \emph{Other} includes stations affected by poor weather, technical problems, or incomplete/unreported observations. Detailed ingress and egress timings for all positive detections are listed in Table~\ref{tab:best_instants}. Star identifiers and NIMA solutions used for event predictions appear in Table~\ref{tab:j2000_coordinates}.}
\end{table*}

\subsection{2020 August 17 occultation}

On 2020 August 17, the occultation of the star Gaia DR3~4056438835548480896 ($G=14.52$; see Table~\ref{tab:j2000_coordinates}) was recorded from three stations in the United States (Chester, Naperville, and Lake Placid). The predicted mid-time was 01:42:15~UT, the corresponding shadow path is shown in Fig.~\ref{fig:occ_paths_a}, and the event parameters are summarised in 
Table~\ref{tab:ixion_event_summary}.

Each site reported a positive detection of the occultation. Video and imaging systems with precise timing capabilities were utilised to determine ingress and egress times accurately. The Chester station utilised a WAT-910HX video camera synchronised via Network Time Protocol (NTP), while Naperville and Lake Placid operated QHY174-GPS cameras, providing direct GPS timestamps. Exposure times ranged between 0.3~s and 5~s, depending on the system configuration and observing conditions. All observers used a Luminance (L) filter during the acquisition.
Image sequences were initiated approximately 5 minutes before the predicted event time and continued for at least 5 minutes afterwards.
As seen at Lake Placid (Fig.~\ref{fig:lc_20200817_a}), the disappearance was visually noticed just before the recording was initiated.
Observers also reported meteorological and seeing conditions to support the assessment of the light-curve quality (see Table~\ref{tab:obs_ixion_all}).

\subsection{2020 October 13 occultation}
\label{subsec:20201013_observation}

On 2020 October 13, the occultation was recorded from ten sites across the United States (see Table~\ref{tab:obs_ixion_all}). The target star was Gaia DR3~4056440205544338944 ($G=10.31$; see Table~\ref{tab:j2000_coordinates}). The predicted mid-time was 01:57:46~UT, the shadow geometry is shown in Fig.~\ref{fig:occ_paths_b}, and the corresponding event parameters are summarised in Table~\ref{tab:ixion_event_summary}.

Eight stations reported positive detections, while two sites provided negative chords that contributed essential constraints on the limb geometry. In addition to this extensive campaign, the event was independently observed by \citet{levine2021} using the 4.3-meter Lowell Discovery Telescope (LDT) near Happy Jack, Arizona, and a 0.32-meter Titan Monitor (TiMo) telescope co-mounted with the Lowell’s Mars Hill 0.8-meter telescope in Flagstaff, Arizona. 
The present work incorporates a broader set of light curves and a larger number of observing stations, and the data from \citet{levine2021} were also included to improve consistency across analyses. 
The overall distribution of the observation network and the predicted shadow track for this event are presented in Fig.~\ref{fig:occ_paths_b}.

Most participating observers employed video cameras synchronised via IOTA-VTI GPS 
time inserters, achieving sub-second timing accuracy. All stations acquired their data in Clear mode, consistent with the standard configuration adopted across the campaigns. Exposure times were typically 0.033~s, providing high temporal resolution well matched to the event geometry.

Given the brightness of the occulted star, its projected angular diameter was also a relevant factor limiting the effective resolution. As \citet{levine2021} noted, the star is classified as a mid-M giant (a classification consistent with its location in the Hertzsprung–Russell diagram; \citealt{kilic_2025_HR_diagram}), and its angular diameter was adopted as $0.675 \pm 0.010$~mas, corresponding to $\sim$~19.3~km at Ixion’s distance. This value was used as the initial reference diameter in our analysis and was subsequently cross-checked against our own occultation light curves, as detailed in the analysis section. 

Because of the combination of short exposures, Fresnel diffraction, and the unusually large projected stellar diameter for this event, detailed modelling was required to reproduce the ingress and egress shapes. For stations using 0.033~s exposures, the finite integration time, the Fresnel scale, and the size of the stellar disc 
were explicitly accounted for in the timing fits.

\subsection{2021 April 28 occultation}

This event was remarkable for the wide geographic extent of the observing network, with sixteen stations across Brazil, Argentina, Uruguay, and Chile participating in the campaign (see Table~\ref{tab:obs_ixion_all}). The occultation involved the $G=14.84$\,mag star Gaia DR3~4050000576486883840 (see Table~\ref{tab:j2000_coordinates}). The predicted mid-time was 07:35:37~UT, the shadow geometry is shown in Fig.~\ref{fig:occ_paths_c}, and the corresponding event parameters are summarised in Table~\ref{tab:ixion_event_summary}.

Among the participating sites, six successfully detected the event, while five reported negative chords that were essential for constraining the object’s silhouette. The remaining stations were affected by overcast conditions and did not obtain usable data. The observing setups were highly diverse, ranging from the 1.6~m telescope at Observatório do Pico dos Dias (OPD) to smaller 20–30~cm instruments. Exposure times also spanned more than an order of magnitude, from 0.15~s to 10~s, reflecting both instrumental capabilities and observing strategies. Unlike the earlier campaigns, several sites in this observation employed moderate to long exposure times (5–10~s), which significantly smoothed ingress and egress features. However, the inclusion of faster cadence light curves from larger apertures, such as OPD and CTIO, mitigated these limitations and enabled precise chord fitting. Except for the CASLEO–Cerro Burek ASH station, which used a Luminance (L) filter, all other observers acquired images in Clear mode.

\subsection{2021 May 20 occultation}

The occultation observed on 20 May 2021 involved five stations in the United States (see Table~\ref{tab:obs_ixion_all}). The target star was Gaia DR3~4050180170543871104 ($G=14.18$; see Table~\ref{tab:j2000_coordinates}). The predicted mid-time was 06:43:23~UT, the shadow geometry is shown in Fig.~\ref{fig:occ_paths_d}, and the corresponding event parameters are summarised in Table~\ref{tab:ixion_event_summary}.

Four stations reported positive detections, while one site provided a negative chord.  
The telescopes were of moderate aperture (typically 20--40\,cm), and exposure times ranged from 0.03\,s to 2\,s, yielding spatial samplings of order 1--2\,km along the chords (see Table~\ref{tab:obs_ixion_all} and Table~\ref{tab:ixion_event_summary}). All observers acquired their data in Clear mode to maximise signal-to-noise. Because the observing configurations were relatively homogeneous, the set of positive chords provided a coherent astrometric constraint; however, their tight clustering in cross-track position limited the ability to robustly constrain Ixion’s limb shape.  
In this context, the negative chord from Westport is especially valuable: its proximity to the positive detections sharply restricts the allowed limb extent and significantly improves the silhouette constraint despite the modest network size.

\subsection{2022 June 30 occultation}

The 30 June 2022 occultation was observed predominantly from Brazil, with eight 
stations contributing to the campaign (see Table~\ref{tab:obs_ixion_all}). 
The target star was Gaia DR3~4049905056379149824 ($G=15.0$; see 
Table~\ref{tab:j2000_coordinates}). 
The shadow geometry for this event is shown in Fig.~\ref{fig:occ_paths_g}, 
and the corresponding event parameters are summarised in 
Table~\ref{tab:ixion_event_summary}.

Five stations obtained usable data, with four recording positive detections of the occultation.  The observing configurations were characterized by modest telescope apertures (10--40\,cm), but their geographic distribution provided chords sampling different regions of the limb. Although the network size was limited, the resulting chord pattern yielded meaningful astrometric constraints. The negative chord from the SONEAR Observatory is particularly valuable: its grazing geometry relative to the cluster of positive detections significantly restricts the range of admissible limb solutions and improves the overall silhouette constraint.

Exposure times ranged from 1 to 4.5\,s, corresponding to effective spatial samplings of order 1--2\,km once combined with diffraction and the projected stellar diameter. All observers used Clear mode, consistent with the standard configuration adopted across the campaigns. Weather conditions were favourable at most sites, with only a few stations affected by cloud coverage or technical issues (see Table~\ref{tab:obs_ixion_all}).

\subsection{2023 July 28 occultation}

The 28 July 2023 stellar occultation was monitored from seven stations in Chile, Argentina, and Brazil (see Table~\ref{tab:obs_ixion_all}). The occulted star was Gaia DR3~4049162851669604096 ($G=14.53$; see Table~\ref{tab:j2000_coordinates}). The shadow geometry for this event is shown in Fig.~\ref{fig:occ_paths_h}, and the corresponding event parameters are summarised in Table~\ref{tab:ixion_event_summary}.

Of the seven observing sites, three secured clear positive detections of the occultation. Exposure times ranged from 0.1 to 10~s, and weather conditions varied between excellent transparency and localised cloud cover, affecting the success rate across sites. Notably, the positive chords originated from observatories equipped with relatively large apertures—such as the 4.1\,m SOAR telescope and the 1.54\,m Danish telescope at La Silla permitting high-cadence photometry with exposures as short as 0.1–0.2\,s. All stations except two acquired their data in Clear mode; Danish/ESO used a Red filter, and Astroquinta employed an L--eNhance filter (see Table~\ref{tab:obs_ixion_all}). This combination of large telescopes and fast sampling provided precise timings of ingress and egress, enhancing the accuracy of the limb reconstruction. By contrast, several smaller-aperture sites achieved only partial coverage or negative results. Despite the modest number of positive chords, their proximity in the plane of the sky offered useful constraints on Ixion’s profile, albeit limiting the ability to model shape asymmetries comprehensively.

The presence of negative chords, particularly those from CASLEO - Cerro Burek HSH and OPD observatories, proved valuable for bounding the possible limb solutions. These non-detections passed close enough to the positive chords to significantly reduce the uncertainty in Ixion’s size and centre position.

\subsection{Single-chord stellar occultations}\label{subsec:single_chords}

Two single-chord stellar occultations were successfully recorded in 2021 and 2022, providing useful constraints on Ixion’s size and astrometric position despite their limited geometrical coverage. Both shadow paths crossed southern Australia (see Fig.~\ref{fig:occ_paths_e} and Fig.~\ref{fig:occ_paths_f}), enabling observations from well-equipped amateur and professional facilities. The identifiers of the occulted stars are listed in Table~\ref{tab:j2000_coordinates}; event-specific geometric parameters are summarised in Table~\ref{tab:ixion_event_summary}, and the corresponding observing configurations are given in Table~\ref{tab:obs_ixion_all}.

17 Aug 2021: The occulted star was Gaia DR3~4056220440657653248 ($G=16.9$; see 
Table~\ref{tab:j2000_coordinates}). The event was recorded from Heaven’s Mirror Observatory (New South Wales) with a 50.8\,cm telescope equipped with a QHY174M–GPS camera (Table~\ref{tab:obs_ixion_all}). A continuous sequence of 3.0\,s exposures captured a clear flux drop, yielding a well-defined single chord under stable atmospheric conditions.

2 Jun 2022: The target star was Gaia DR3~4049248613603980416 ($G=15.3$; see Table~\ref{tab:j2000_coordinates}). The event was recorded from Glenlee Observatory (Australia) using a 30\,cm telescope equipped with a Watec~910BD camera operating at a 0.32\,s cadence. Conditions were cloudless with moderate seeing, and timing was provided by an IOTA-VTI GPS inserter (see Table~\ref{tab:obs_ixion_all} for full site details). These observations yielded a nearly grazing chord of $104.3 \pm 2.5$\,km. The grazing geometry significantly reduced the apparent stellar motion relative to the observer, requiring a dedicated determination of the effective limb-crossing speed to accurately derive the ingress and egress times.

Although single-chord events cannot constrain Ixion’s full limb profile, they provide independent astrometric measurements and robust bounds on its size. Continued multi-chord campaigns remain essential to improve constraints on Ixion’s shape and refine its ephemerides.

\section{Data analysis and results}
\label{sec:data_analysis}

The stellar occultation observations analyzed here were obtained during international campaigns between 2020 and 2023, involving both professional and amateur observatories equipped with telescopes from small portable instruments to 4-meter-class facilities. 
Details of the observing sites and acquisition parameters are provided in Appendix~\ref{sec:obs_details}, 
while Table~\ref{tab:archive_obs} summarises the archival photometric dataset used to determine Ixion’s absolute magnitude ($H_V$), phase coefficient, and geometric albedo ($p_V$). 
Data reduction and analysis were carried out using independent pipelines and photometric tools developed within the collaboration.

\subsection{Data acquisition and reduction}
The observational data sets collected for Ixion span a wide range of formats and image quality, including both Flexible Image Transport System (FITS)\footnote{\url{https://fits.gsfc.nasa.gov/}} images and video-based recordings (\texttt{.avi}, \texttt{.ser}, \texttt{.adv}, \texttt{.aav}, etc.). For FITS data accompanied by calibration frames (bias, dark, and flat), a standard pre-reduction routine was applied, which performed bias subtraction, dark correction, and flat-field normalisation using well-established procedures. When video files were provided, they were first converted to FITS format through a dedicated module in the \textit{Occultation Portal} built upon \texttt{PyMovie} using \texttt{stacker.py}. The details of this conversion pipeline are extensively described in \citet{kilic_2022} and \citet{anderson2019}.

All resulting light curves were generated directly within the \textit{Occultation Portal}. 
To ensure the reliability of these results, all data sets were independently analysed with the \texttt{PRAIA}\footnote{\url{https://ov.ufrj.br/praia-photometry-task/}} (Package for the Reduction of Astronomical Images Automatically) software \citep{assafin2023}, providing a cross-check with the portal-generated light curves. No significant discrepancies were found.

\subsection{Light-curve modelling}
\label{sec:light_curve_modelling}

All positive light curves were modelled using the \texttt{SORA}\footnote{\url{https://sora.readthedocs.io/latest/overview.html}} (Stellar Occultation Reduction Analysis) package \citep{altair2022}. For each station, the model includes Fresnel diffraction, the finite angular size of the occulted star, the integration time of the detector, and the effective CCD bandwidth. Ingress and egress times were obtained by fitting the SORA model to
each light curve, and the resulting timings with their associated $1\sigma$ uncertainties are listed in Table~\ref{tab:best_instants}, making use of the stellar parameters and event geometries presented in Tables~\ref{tab:j2000_coordinates} and \ref{tab:ixion_event_summary}.

Across all events, Fresnel diffraction and the projected stellar diameter are both in the 1.3--1.6\,km range, comparable to one another and to the finest spatial sampling achieved in the data. The Fresnel scale is computed as $F=\sqrt{\lambda\Delta/2}$ for $\lambda=600$\,nm, yielding $F\simeq1.3$--1.4\,km over the geocentric distance interval $\Delta\simeq37.8$--39.3\,au. Typical exposure times of 0.1--0.4\,s correspond to along-chord resolutions of $\sim$2.5--4.5\,km for Ixion's sky-plane velocities ($v\simeq10$--25\,km\,s$^{-1}$), while the fastest observations (0.033\,s cadence) reach $\lesssim1$\,km sampling, allowing diffraction and finite-star effects to be directly resolved.

Except for the 17 August 2021 event, all occulted stars lie in the giant regime of the Hertzsprung--Russell diagram (see \citealt{kilic_2025_HR_diagram}). Their angular diameters were therefore derived using the giant-star calibration of the
\citet{van99} surface-brightness relations, ensuring a uniform and internally consistent treatment across campaigns. The 17 August 2021 star lies on the main sequence and was modelled using the corresponding van Belle coefficients. In all cases, the resulting angular diameters were converted into projected sizes at Ixion’s geocentric distance and used as fixed inputs in the SORA light-curve fits.

\subsubsection*{Refined stellar diameter estimates from the 13 October 2020 event}
\label{sec:13102020_ixion_results}

As described in Section~\ref{subsec:20201013_observation}, the 13 October 2020 event involved Ixion occulting a relatively bright star ($G=10.31$\,mag), Gaia DR3~4056440205544338944. Because of the star’s brightness and Ixion’s geocentric distance (39.26~au), the projected stellar diameter (hereafter simply ``stellar diameter'') was a key factor to account for during the modelling process. The \textit{Occultation Portal} data were obtained using various instruments, primarily small-aperture telescopes, resulting in light curves with different S/N levels. Consequently, the stellar diameter derived from each chord showed minor variations but remained consistent within the uncertainties.

To determine the stellar diameter for each chord, we applied a $\chi^{2}$ minimisation approach (Eq.~\ref{eq:chi2_lc}; \citealt{altair2022}).

\begin{equation}
\label{eq:chi2_lc}
\chi^{2}=\sum_{i=1}^{N}\left(\frac{(\phi_{i,obs}-\phi_{i,cal})^{2}}{\sigma_{i}^{2}}\right).
\end{equation}

The parameter scan was performed over the range $9.00~\mathrm{km} \leq d_{\star} \leq 30.00~\mathrm{km}$ with a step size of $\Delta d_{\star} = 0.25~\mathrm{km}$, and the values satisfying $\chi^{2} < \chi^{2}_{\mathrm{min}} + \Delta \chi^{2}$ at the 3$\sigma$ confidence level were selected.

From the set of stellar diameters $d_{i}$ obtained for each chord (each with its associated uncertainty $\sigma_{i}$), we derived the weighted mean stellar diameter using:

\begin{equation}
\label{eq:mean_star_diam}
d_{\star}=\frac{\sum_{i=1}^{n}{\frac{d_{i}}{\sigma _{i}^{2}}}}{\sum_{i=1}^{n}{\frac{1}{\sigma _{i}^{2}}}},
\end{equation}

Where $n$ is the total number of chords. The corresponding uncertainty of the weighted mean was calculated as:

\begin{equation}
\label{eq:mean_star_diam_err}
\sigma _{\star}=\left( \sum_{i=1}^{n}{\frac{1}{\sigma _{i}^{2}}} \right)^{-\frac{1}{2}}.
\end{equation}

\begin{figure}[h]
    \centering
    \begin{minipage}{0.5\textwidth}
        \includegraphics[width=\textwidth]{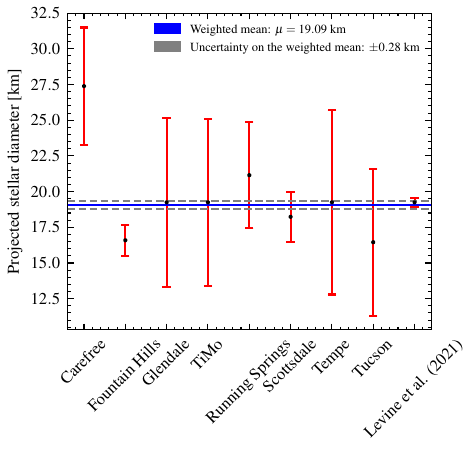}
        \caption{
            Stellar diameter estimates for the occulted star Gaia DR3~4056440205544338944, obtained by scanning models across the observed light curves of the October 13, 2020, occultation.
            The weighted mean of all results is $\mu = 19.09 \pm 0.28$~km, shown with a 3$\sigma$ confidence interval.
            Data from \citet{levine2021} were also included in the final estimate.  The error bars represent the uncertainties derived from model fits to each chord.
        }
        \label{fig:star_diameters_3sigma}
    \end{minipage}
\end{figure}
As illustrated in Fig.~\ref{fig:star_diameters_3sigma}, the light curve with the lowest uncertainty was obtained by \citet{levine2021} using the 4.3\,m Lowell Discovery Telescope (LDT). The relative flux dispersion in this light curve is $<0.1$, making it the most precise measurement and thus the largest contributor to the weighted mean. Most other light curves were obtained with small telescopes (20–30~cm apertures) under the \textit{Lucky Star} campaign (see Table~\ref{tab:obs_ixion_all}).

The final weighted mean angular diameter of the occulted star Gaia~DR3~4056440205544338944 is 
$\theta_\star = 0.670 \pm 0.010$\,mas, corresponding to a projected linear diameter of 
$19.09 \pm 0.28$~km at Ixion’s distance of 39.26~au. 
This represents a remarkably precise angular size determination from a trans-Neptunian object occultation and is entirely consistent with the value derived by \citet{levine2021} ($19.25 \pm 0.30$~km). 
Combining our result with the star’s Gaia parallax ($0.5645 \pm 0.0417$\,mas) yields a physical stellar radius of $128 \pm 10\,R_{\odot}$, also consistent with \citet{levine2021} ($130^{+20}_{-17}\,R_{\odot}$).

Gaia~DR3 provides stellar radii inferred by the GSP-Phot module\footnote{General Stellar Parametrizer from Photometry} \citep{creevey2023}, which uses multiple stellar-atmosphere libraries within the Aeneas algorithm \citep[see][]{andrae2023}. 
In the main Gaia~DR3 catalogue, only a single "best" solution is published for each source. For Gaia~DR3~4056440205544338944, this default solution corresponds to the OB-library model,
which yields a radius of $\sim56\,R_{\odot}$ (which corresponds to an angular diameter of $\theta_\star \simeq 0.29$\,mas), i.e. more than a factor of two smaller than our occultation-based estimate.
However, the Gaia~DR3 supplementary GSP-Phot results for the same source provide alternative solutions derived from different stellar libraries. In particular, the MARCS-library solution returns a radius of $\sim116\,R_{\odot}$ ($\theta_\star \simeq 0.61$\,mas), much closer to the $0.670 \pm 0.010$\,mas angular diameter that we obtain from the occultation. This indicates that the apparent tension arises primarily from the automatic library selection in the GSP-Phot module (Aeneas algorithm) rather than from an intrinsic inconsistency between Gaia and the occultation result. For cool, luminous late-type giants, the module may select as ``best'' a solution based on a library that is not optimal for this spectral regime, while an alternative library (e.g. MARCS) yields a physically more plausible radius \citep[Section~2.6 of][]{andrae2023}.

\subsection{Limb fitting and astrometry}\label{sec:ixion_limb_fit}

The ingress and egress times derived from all positive detections (see Table~\ref{tab:best_instants}) were projected onto the sky plane to reconstruct the silhouette of (28978) Ixion during each event. These timings were analysed with the \texttt{SORA} package \citep{altair2022}, using a sharp-edge occultation model that accounts for the finite stellar diameter, Fresnel diffraction, exposure-time smearing, and CCD bandwidth. Negative chords, when available, were incorporated via the \texttt{filter\_negative\_chord} function to constrain the limb geometry further. When negative chords are present, they naturally restrict the family of 
acceptable limb solutions, which may clip the outer portions of the $1\sigma$ confidence region in Fig.~\ref{fig:ellipse_fits}, as occurs for events with constraining non-detections (e.g., 28~Apr~2021, 20~May~2021, 30~Jun~2022).

For the multi-chord events (August 17, 2020; October 13, 2020; April 28, 2021; May 20, 2021; June 30, 2022; and July 28, 2023), the observed limb was modelled as an ellipse defined by five free parameters: the center offsets $(f, g)$ with respect to the NIMA v13 prediction ephemeris, the apparent equatorial radius $R_{\mathrm{equatorial}} = a$, the apparent oblateness $\epsilon' = (a-b)/a$, and the position angle of the minor axis, denoted by $\phi$ and defined as the angle measured eastward from celestial north. The statistical significance of the fits was evaluated through the reduced chi-square $\chi^2_{\mathrm{pdf}} = \chi^2/(N-M)$, where $N$ is the number of fitted points and $M = 5$ the number of free parameters. The 1$\sigma$ uncertainties for each parameter were obtained by varying the parameter around its nominal value until $\chi^2 = \chi^2_{\mathrm{min}} + 1$.

Given that several events have only a limited number of chords or exhibit significant timing uncertainties, a purely free fit may yield poorly constrained position angles. To mitigate this, we adopted a strategy inspired by \citet{fribas_2013,fribas_2014,BenedettiRossi2016, ort17,dias_oliveira_2017,rizos2025}, where $\phi$ is allowed to vary freely but within a physically plausible interval guided by the best-constrained events. In particular, the October 13, 2020, April 28, 2021, and June 30, 2022 occultations, which provide the most reliable constraints, yielded $\phi$ values of $34.1^\circ \pm 15.7^\circ$, $35.3^\circ \pm 17.5^\circ$, and $29.1^\circ \pm 11.7^\circ$, respectively. These values are mutually consistent, defining a preferred range of approximately $25^\circ$–$40^\circ$ for Ixion’s apparent limb orientation. For other events, our fits were guided by this range, ensuring convergence toward physically meaningful solutions without imposing a strict fixed $\phi$.

The projection of single chords onto the sky plane yields two symmetrical solutions for the centre position along the direction perpendicular to the chord, leaving the limb shape unconstrained. For this reason, the single-chord events (17~Aug~2021 and 02~Jun~2022) are used only to derive the astrometric offsets $(f,g)$, while the limb parameters $(a,b,\phi)$ are fixed to the global multi-chord solution. Although the August~17,~2021 chord is notably long ($\sim$770~km), single-chord 
geometries do not permit a reliable size estimate, and we therefore do not derive any diameter constraints from these events.

The results of all ellipse fits are summarised in Table~\ref{tab:ixion_limb_solutions_full}, which lists the derived $(f, g)$ offsets, equatorial and polar radii, apparent oblateness, $\phi$, equivalent radius, radial dispersion, and $\chi^2_{\mathrm{pdf}}$. Visual representations of the fitted limbs are shown in Fig.~\ref{fig:ellipse_fits} (panels a–h).

\begin{table*}[h!]
\centering
\caption{Best-fitted limb solutions for (28978) Ixion (1$\sigma$ uncertainties) derived from stellar occultations.}
\label{tab:ixion_limb_solutions_full}
\footnotesize
\begin{tabular}{@{\extracolsep{\fill}}lcccccccc@{}}
\toprule\midrule
Parameter & 17 Aug 2020 & 13 Oct 2020 & 28 Apr 2021 & 20 May 2021 & 17 Aug 2021 & 2 Jun 2022 & 30 Jun 2022 & 28 Jul 2023 \\\midrule
$f$ (km) & $1.7 \pm 7.6$   & $-8.0 \pm 1.4$    & $-41.6 \pm 7.6$   & $-71.7 \pm 31.2$ & $-8.3 \pm 18.4$   & $30.3 \pm 75.6$ & $-38.5 \pm 16.6$  & $-132.0 \pm 9.8$ \\
$g$ (km)              & $5.4 \pm 7.2$  & $4.6 \pm 5.8$     & $1.8 \pm 4.4$      & $48.7 \pm 110.8$ & $19.5 \pm 47.6$   & $-31.6 \pm 15.7$ & $-27.0 \pm 7.2$  & $-47.6 \pm 55.1$ \\
$R_{\mathrm{equatorial}}$ (km) & $375.1 \pm 11.7$ & $366.8 \pm 6.2$ & $391.9 \pm 9.4$ & $375.9 \pm 20.8$ & -- & -- & $389.5 \pm 16.8$  & $371.1 \pm 33.5$ \\
$R_{\mathrm{polar}}$ (km)   & $330.2 \pm 20.1$  & $346.1 \pm 7.4$ & $322.2 \pm 17.2$ & $338.2 \pm 41.6$ & -- & -- & $325.6 \pm 31.5$  & $334.0 \pm 47.8$\\
$\epsilon^{'}$        & $0.120 \pm 0.046$ & $0.056 \pm 0.012$ & $0.231 \pm 0.102$ & $0.100 \pm 0.099$ & -- & -- & $0.16 \pm 0.072$  & $0.100 \pm 0.100$ \\
$\phi$ ($^\circ$)         & $14.86 \pm 8.10$ & $34.11 \pm 15.70$   & $35.25 \pm 17.50$    & $-4.23 \pm 63.65$  & --    & -- & $31.35 \pm 13.51$     & $-0.04 \pm 89.81$ \\
$R_{\mathrm{equiv}}$ (km)      & $351.9 \pm 12.0$         & $356.3 \pm 4.9$        & $386.3 \pm 59.8$         & $356.6 \pm 24.1$         & --        &--         & $356.1 \pm 18.9$          & $352.0 \pm 29.8$         \\
$R_{\mathrm{dispersion}}$ (km) & $-0.23 \pm 5.89$ & $-1.87 \pm 2.83$ & $-6.79 \pm 8.83$ & $-21.1 \pm 6.18$ &-- & -- & $-3.77 \pm 10.85$  & $-12.2 \pm 8.6$ \\
$\chi^2_{\mathrm{pdf}}$        & $0.816$         & $0.926$         & $0.393$         & $0.303$         & --         & --         & $0.098$         & $0.938$ \\
Ellipse fit   & Fig.~\ref{fig:ellipse_fits_a} & Fig.~\ref{fig:ellipse_fits_b} & Fig.~\ref{fig:ellipse_fits_c} & Fig.~\ref{fig:ellipse_fits_d} & Fig.~\ref{fig:ellipse_fits_e} & Fig.~\ref{fig:ellipse_fits_f} & Fig.~\ref{fig:ellipse_fits_g} & Fig.~\ref{fig:ellipse_fits_h} \\
\bottomrule
\end{tabular}
\begin{flushleft}
\footnotesize
\textbf{Notes.} 
For the single-chord events (17 Aug 2021 and 2 Jun 2022), only the astrometric offsets $(f,g)$ were fitted; the limb shape parameters were fixed to the global multi-chord solution.
\end{flushleft}
\end{table*}

\subsection{Global limb fit}
\label{sec:ixion_global_fit}

We derived a global limb solution for (28978)~Ixion by jointly fitting all multi-chord stellar occultations observed between 2020 and 2023 using a per-event centre minimization approach (PECM). In this framework, Ixion’s apparent limb is described by a single projected ellipse, while the astrometric centre of each occultation event is optimized independently.

The global shape parameters, defined by the semi-axes $(a,b)$ and the position angle $\phi$ (measured eastward from celestial north, modulo $180^\circ$), were explored using a Monte Carlo sampling of the three-dimensional $(a,b,\phi)$ parameter space, centred on physically motivated initial values. For each Monte Carlo realization, the astrometric centres of all events were re-optimized by minimizing the event-wise contributions to the total $\chi^2$. The global goodness of fit was then computed as the sum of these event-wise $\chi^2$ values.

Allowing independent astrometric centres for each event prevents ephemeris or astrometric offsets from biasing the inferred global limb parameters, ensuring that the shape solution is driven by the relative geometry of the chord extremities rather than by systematic offsets between observing epochs. In addition, a small extra model uncertainty of $\sigma_{\mathrm{model}} = 4$~km was added in quadrature to the uncertainties of the chord extremities in the projected $f$–$g$ plane, in order to account for possible unmodelled limb effects or small-scale topography \citep{rommel_2023}.

The final confidence intervals were derived from the ensemble of accepted Monte Carlo solutions within a $\Delta\chi^2$ threshold of $\Delta\chi^2 = 3.53$, appropriate for three global shape parameters. The total number of free parameters in the fit is therefore $M = 3 + 2N_{\mathrm{events}}$, corresponding to the global ellipse parameters and the per-event astrometric centres. The resulting fit involves $N_{\mathrm{data}} = 56$ chord extremities and $\nu = 41$ degrees of freedom, yielding a reduced $\chi^2_\nu = 1.17$, consistent with the expected statistical distribution given the adopted observational and model uncertainties.

The best-fitting global solution is
\begin{equation}
a = 363.42^{+3.53}_{-3.85}~\mathrm{km}, \quad
b = 333.98^{+7.07}_{-4.96}~\mathrm{km}, \quad
\phi = 110.97^{+7.00}_{-6.54}~\mathrm{deg},
\end{equation}
where the quoted uncertainties correspond to asymmetric $1\sigma$ confidence intervals derived from the $\chi^2$ surface.

From these values, we derive an equivalent radius based on the projected area,
\begin{equation}
R_{\mathrm{equiv}} = \sqrt{ab} = 348.39^{+5.37}_{-4.43}~\mathrm{km},
\end{equation}
corresponding to an equivalent diameter
\begin{equation}
D_{\mathrm{equiv}} = 696.78^{+10.75}_{-8.87}~\mathrm{km}.
\end{equation}
The apparent oblateness is
\begin{equation}
\epsilon' = \frac{a-b}{a} = 0.081^{+0.004}_{-0.010},
\end{equation}
indicating a moderately flattened projected shape.

\begin{figure}
\centering
\includegraphics[width=9.2cm]{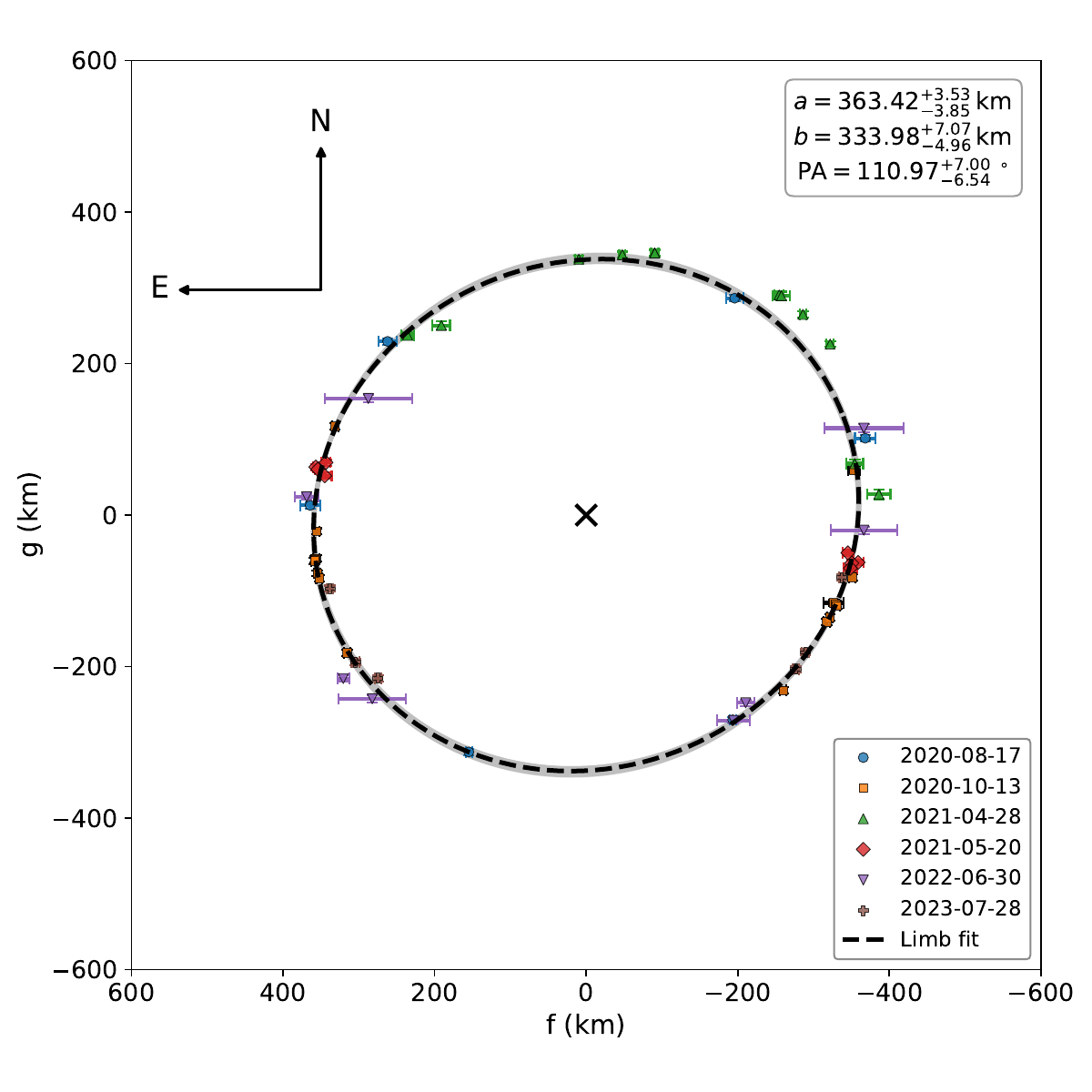}
\caption{Global limb fit of (28978)~Ixion obtained by combining the chord extremities from all multi-chord stellar occultations observed between 2020 and 2023. 
Each colour and marker shape corresponds to a different event (see legend). 
The black dashed curve shows the best-fitting ellipse derived from the global $\chi^2$ minimization, while the grey shaded region represents the $1\sigma$ uncertainty envelope of the global limb solution. Error bars represent the projected uncertainties of the chord extremities in the $f$–$g$ plane (in kilometres). The black \textsf{x} marks the limb-centred reference frame, and the arrows indicate the celestial North (N) and East (E) directions. The data used in this figure are available in \citep{kilic_2025_globalfitdata}.
}
\label{fig:ixion_all_dates_fit}
\end{figure}

The resulting global limb solution is illustrated in Fig.~\ref{fig:ixion_all_dates_fit}, where the best-fitting ellipse and its $1\sigma$ uncertainty envelope are shown together with the chord extremities from all observing epochs; the corresponding orthogonal residuals are presented in Appendix~\ref{fig:global_fit_residuals}.

\subsection{Astrometry}

The astrometric solutions for (28978) Ixion derived from eight stellar occultations between 2020 and 2023 are summarised in Table\,\ref{tab:ixion_astrometric_solutions}. For each event, the ellipse fit provides the centre coordinates (offsets \(X, Y\)) relative to the predicted position, defined by the object’s ephemeris and the star coordinates. These offsets were used to determine the precise ICRS equatorial coordinates of Ixion at the occultation mid-time (see Table\,\ref{tab:ixion_astrometric_solutions}).

The sub-milliarcsecond precision achieved in most events is a direct consequence of the accurate timing of the occultation light curves combined with the Gaia DR3 catalogue positions of the occulted stars. These eight astrometric measurements will be incorporated into the orbital refinement of Ixion, improving its ephemeris and the accuracy of future stellar occultation predictions.

\begin{table*}[h!]
\centering
\caption{Astrometric solutions for (28978) Ixion derived from stellar occultations.}
\label{tab:ixion_astrometric_solutions}
\footnotesize
\begin{tabular}{@{\extracolsep{\fill}}lccccc@{}}
\toprule\midrule
Date (UT) & Time (UT) & RA (hh mm ss.s) & DEC (dd mm ss.s) & Offset X (km) & Offset Y (km) \\\midrule
2020-08-17 & 01:37:11.820 & $17^{h}49^{m}53.5853^{s} \pm 0.284$ mas & $-29^{\circ}48'12.3881^{s} \pm 0.282$ mas & $1.7 \pm 7.6$ & $5.4 \pm 7.2$ \\
2020-10-13 & 01:57:46.320 & $17^{h}50^{m}20.3585^{s} \pm 0.141$ mas & $-29^{\circ}43'32.9060^{s} \pm 0.274$ mas & $-8.0 \pm 1.4$ & $4.6 \pm 5.8$ \\
2021-04-28 & 07:36:45.400 & $18^{h}05^{m}34.6798^{s} \pm 0.567$ mas & $-30^{\circ}09'09.7432^{s} \pm 0.240$ mas & $-41.6 \pm 7.6$ & $1.8 \pm 4.4$ \\
2021-05-20 & 06:43:21.420 & $18^{h}04^{m}05.2324^{s} \pm 1.200$ mas & $-30^{\circ}13'44.4493^{s} \pm 3.868$ mas & $-71.7 \pm 31.2$ & $48.7 \pm 110.8$ \\
2021-08-17 & 11:36:10.400 & $17^{h}56^{m}18.8131^{s} \pm 1.030$ mas & $-30^{\circ}17'43.7814^{s} \pm 2.120$ mas & $-8.3 \pm 18.4$ & $19.5 \pm 47.6$ \\
2022-06-02 & 18:09:30.140 & $18^{h}09^{m}36.9353^{s} \pm 2.783$ mas & $-30^{\circ}43'37.0537^{s} \pm 0.616$ mas & $30.3 \pm 75.6$ & $-31.6 \pm 15.7$ \\
2022-06-30 & 00:22:49.920 & $18^{h}06^{m}53.6124^{s} \pm 0.642$ mas & $-30^{\circ}46'56.5116^{s} \pm 0.351$ mas & $-38.5 \pm 16.6$ & $-27.0 \pm 7.2$ \\
2023-07-28 & 06:13:19.080 & $18^{h}11^{m}03.2660^{s} \pm 0.399$ mas & $-31^{\circ}15'22.5483^{s} \pm 2.046$ mas & $-132.0 \pm 9.8$ & $-47.6 \pm 55.1$ \\
\bottomrule
\end{tabular}
\end{table*}

\subsection{Archival photometry: $H_V$, $G$, and $p_V$}
We analysed the photometric properties of (28978) Ixion using archival observations. A total of 39 raw V-band images obtained with the NTT/EMMI were retrieved from the ESO archive\footnote{ESO programme ID 075.C-0431(A)}, together with two V-band images from VLT/FORS1\footnote{ESO programme ID 178.C-0036(N)}. In addition, 14 unique measurements of Ixion were extracted from the 138 available data of Ixion in the \textit{Gaia} DR3 database \citep{gaia_2023}, and one additional V-band magnitude reported by \citep{boehnhardt_2004} was included to improve the phase-angle coverage (see Table~ \ref{tab:archive_obs}).

From the 39 NTT frames, 28 images were selected based on focus, tracking quality, S/N ratio and minimal contamination from background sources. The ESO data were first processed through standard pre-reduction procedures, and apparent magnitudes were derived using the \textit{PhoPS} algorithm \citep{erece2023}. For the \textit{Gaia} data, the $G$ magnitudes were transformed into $V$ magnitudes using the equations published by \citet{gaia_dr3_doc}. Reduced magnitudes, $V(1,\alpha)$, normalized to $r = \Delta = 1$ AU, were then computed to Eq.~\eqref{redmag}, where $r$ and $\Delta$ denote the heliocentric and geocentric distances, respectively.

\begin{equation}\label{redmag}
V(1,\alpha) = V - 5\log(r\Delta)
\end{equation}

\begin{figure}
\centering
\includegraphics[width=9.2cm]{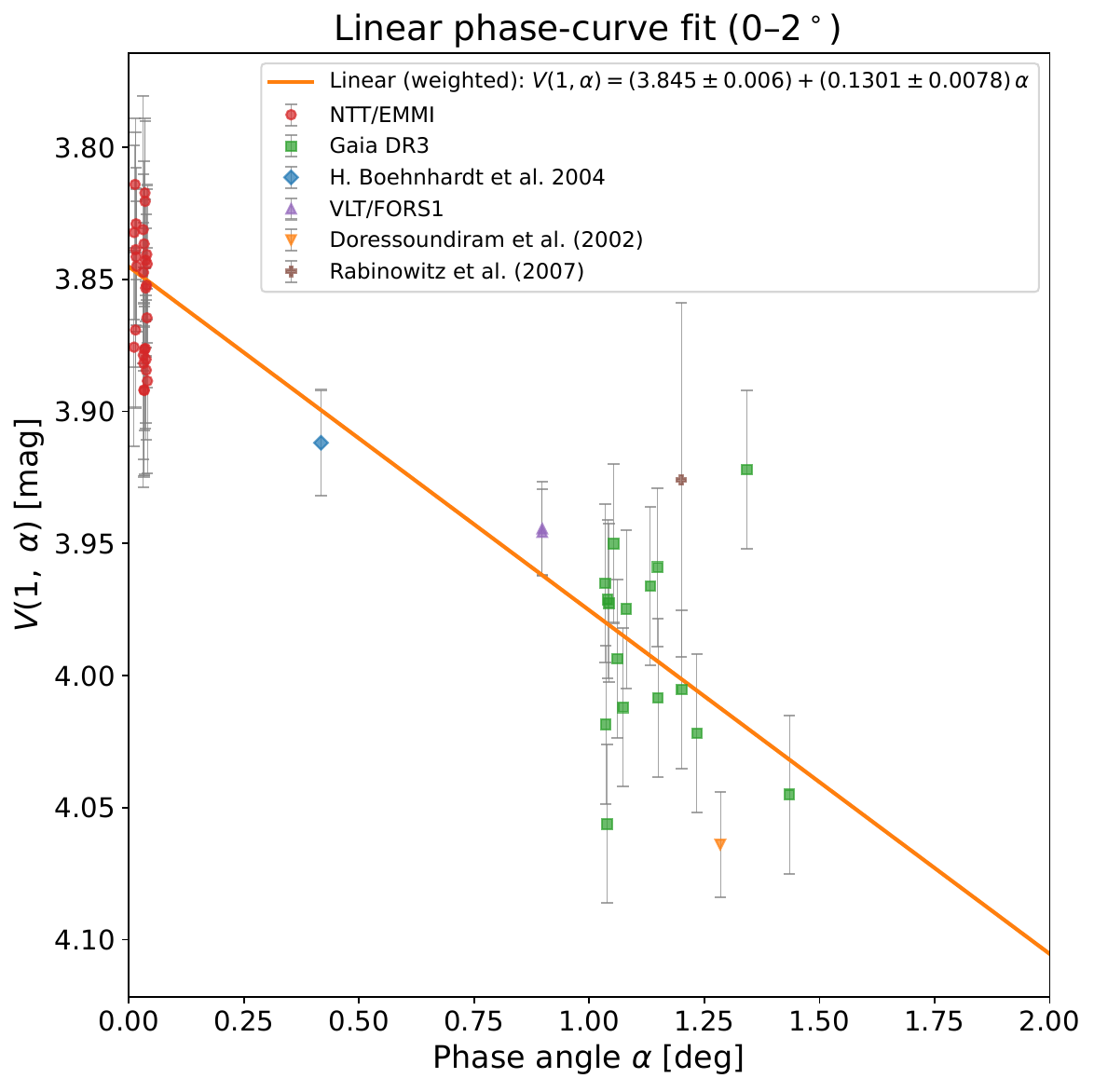}
\caption{Linear phase-curve fit for Ixion in the restricted range 
$0$–$2^\circ$ in phase angle (datasets listed in Table~\ref{tab:archive_obs}).}
\label{fig:linear_fitting}
\end{figure}

Because such distant objects cannot be observed over a wide range of phase angles, we modelled the phase curve in the restricted range $0$–$2^\circ$ with a linear relation,
$V(1,\alpha)=H + \beta\,\alpha$. 
The fit yields $H = 3.845 \pm 0.006$ and a slope $\beta = 0.1301 \pm 0.0078$~mag\,deg$^{-1}$ (Fig.~\ref{fig:linear_fitting}).

We also retrieved B, R, and I magnitudes from the NTT/EMMI data to derive colour indices for Ixion. To minimise effects related to rotation, phase angle, or distance, we used measurements obtained closest in time. We derived $B-V = 1.06 \pm 0.03$, $V-R = 0.61 \pm 0.02$, and $R-I = 0.54 \pm 0.03$, values consistent with the red spectral slope typically observed among TNOs, and in good agreement with previous results published by  \citet{boehnhardt_2004}.

Using the equivalent radius derived from the global limb fit (Sect.~\ref{sec:ixion_global_fit}), we re-estimated Ixion’s geometric albedo in the visible band.The best-fitting ellipse obtained from the combined multi-chord occultations (2020--2023)
yields an equivalent radius of \( R_{\mathrm{equiv}} = 348.39^{+5.37}_{-4.43}~\mathrm{km} \), corresponding to an equivalent diameter of \( D_{\mathrm{equiv}} = 696.78^{+10.75}_{-8.87}~\mathrm{km} \). Combining this size with our determination of the absolute magnitude, \( H_V = 3.845 \pm 0.006 \), we derived the geometric albedo \(p_V\) using the classical relation introduced by \citet{russell1916}, expressed with a modern calibration constant \(C\) (e.g., \(C = 1330 \pm 18~\mathrm{km}\);~\citealt{masiero2021}):

\begin{equation}\label{eq:pv_general}
p_V = \left(\frac{C}{D_{\mathrm{equiv}}}\right)^{2} \, 10^{-0.4H_V}.
\end{equation}

This calculation yields a visible geometric albedo of \( p_V = 0.106^{+0.003}_{-0.003} \), indicating that Ixion’s surface is slightly darker than previously estimated. The result remains fully consistent within uncertainties with earlier determinations, such as the \( p_V = 0.108 \pm 0.002 \) value reported by \citet{verbiscer2022}, and confirms that Ixion lies among the brighter members of the Plutino population.

\begin{table*}[h!]
\centering
\caption{Observational data used in this study.}
\label{tab:archive_obs}
\small 
\begin{tabular}{lccccccc}
\hline\hline
Telescope/Inst. & Date (UT) & Filter & \makecell{$N_{\mathrm{obs}}$ \\ used/total} 
& \makecell{Exp. \\ time (s)} & \makecell{Apparent \\ mag (V)} 
& \makecell{Phase angle \\ ($^\circ$)} & Notes \\
\hline
NTT/EMMI  & 2005-05-29 & B & 3/28 & 360 & 21.15 & 0.9 & Prog. ID 075.C-0431(A) \\
NTT/EMMI  & 2005-05-29/30/31 & V & 28/39 & 120 & 20.08 & 0.01--0.04 & Prog. ID 075.C-0431(A) \\
NTT/EMMI  & 2005-05-29 & R & 12/20 & 120 & 19.47 & 0.9 & Prog. ID 075.C-0431(A) \\
NTT/EMMI  & 2005-05-29 & I & 3/12 & 60 & 18.91 & 0.9 & Prog. ID 075.C-0431(A) \\
VLT/FORS1 & 2007-07-14 & V & 2/2  & 40 & 20.14 & 0.90 & Prog. ID 178.C-0036(N) \\
Gaia DR3  & 2014-2017 & G & 14/14 & --  & 19.61--19.81 & 0.5--1.4 & \citet{gaia_2023} \\
CFHT      & 2002-08-12 & V & 1/1 & -- & 20.39 & 1.28 & \citet{doressoundiram2002} \\
VLT/FORS1 & 2002-05-11 & V & 1/1 & -- & 20.21 & 0.42 & \citet{boehnhardt_2004} \\
SMARTS/ANDICAM & 2003-03-25 & V & 1/1 & -- & -- & 1.20 & \citet{rabinowitz2007} \\
\hline
\end{tabular}
\tablefoot{Apparent magnitudes correspond to average values per dataset. Phase angle refers to the Sun–Ixion–Earth angle at the time of observation.}
\end{table*}

\section{Discussion and conclusions}
\label{sec:discussion}

Our analysis indicates a moderately elongated body with an area-equivalent diameter of $D_{\mathrm{equiv}} = 696.78^{+10.75}_{-8.87}$~km and a visible geometric albedo of $p_V =0.106^{+0.003}_{-0.003}$. According to the colour–albedo distribution presented by \citet[][their Fig.~2]{lacerda2014}, Ixion lies between the dark-neutral and bright-red groups, ranking among the brighter members of the Plutino population. The absolute magnitude we derive ($H_V = 3.845 \pm 0.006$) is slightly fainter than the value reported by \citet{alvarezcandal2016} ($H_V = 3.774 \pm 0.021$), likely reflecting the broader temporal coverage of our dataset and the consequent mitigation of rotational-phase bias. Although no secure rotational period could be established, the scatter in both the light curves and the residuals of the phase-curve fit suggests slight departures from axial symmetry and/or heterogeneous surface properties, as expected for objects of this size \citep[e.g.][]{duffard2009}. 

The global limb fit (Sect.~\ref{sec:ixion_global_fit}) shows that the chord extremities from the 2020–2023 occultation campaigns are well reproduced, to first order, by a single projected ellipse. No evidence for systematic trends is found in the residuals, and the reduced $\chi^2_\nu$ close to unity indicates a statistically acceptable fit given the adopted observational and model uncertainties. The derived apparent oblateness, $\epsilon' = 0.081^{+0.004}_{-0.010}$, is consistent with a moderately flattened projected figure, providing an adequate first-order description of Ixion’s global shape.

While the majority of occultation epochs are mutually consistent with this global limb solution, the events of 28 Apr 2021 and 30 Jun 2022 exhibit noticeable offsets relative to the best-fitting projected ellipse. These deviations are not indicative of shortcomings in the modelling approach, but may reflect changes in the projected figure of the body sampled at different rotational phases or aspect angles. In the absence of a well-constrained rotational period, such epoch-dependent variations are naturally expected for a body that departs slightly from axial symmetry, consistent with a mildly triaxial figure. The lack of any systematic trend in the residuals for the remaining well-constrained events (Appendix~\ref{fig:global_fit_residuals}), together with their limited scatter (typically at the $\sim$10~km level), supports the interpretation that a modestly flattened figure provides an adequate global description of Ixion’s shape, while allowing for small departures from perfect axisymmetry.

A unique outcome of this work is the precise measurement of the angular diameter of the occulted star 
Gaia~DR3~4056440205544338944, yielding $\theta_\star = 0.670 \pm 0.010$\,mas, which corresponds to a physical stellar radius 
of $128 \pm 10\,R_{\odot}$. 
This rare, direct constraint on the size of an M~III giant at a distance of $\sim$1.7\,kpc is remarkable, as such measurements are typically limited to much closer giants observed with interferometry or lunar occultations \citep[e.g.][]{white1987,nordgren1999,baines2016,baines2018}. 
Our result also reveals a significant discrepancy with the Gaia~DR3 GSP-Phot ``best'' estimate obtained from the OB-library model ($\sim56\,R_{\odot}$, $\theta \simeq 0.29$\,mas), 
while the alternative MARCS-library solution ($\sim116\,R_{\odot}$, $\theta \simeq 0.61$\,mas) is in much better agreement with our occultation-based value. 
This confirms that the apparent mismatch stems primarily from the automatic library selection in the GSP-Phot module rather than from an intrinsic inconsistency between Gaia and the occultation result. 
Consequently, multi-chord stellar occultations by trans-Neptunian objects can serve not only to constrain the physical properties of the occulting bodies, but also to independently validate Gaia-derived stellar radii in the cool-giant regime.

In addition, our analysis of detection limits (see Appendix~\ref{appendix:detection_limits}) shows no evidence for rings or other circum-object material around Ixion. 
The sensitivity of our light curves, particularly the April 2021 event observed from OPD, rules out opaque structures wider than a few hundred meters in the sky plane and constrains tenuous material with optical depths $\tau > 0.1$ at kilometre scales. 
These limits are stringent enough that the dense rings of Chariklo, as well as those observed around Haumea and Quaoar, would have been detected if present. 
Thus, our results indicate that Ixion lacks ring systems or significant debris within the probed radial ranges, providing an essential negative constraint in the comparative study of ring-bearing small bodies.

Another key outcome of this work was the derivation of precise astrometric positions of Ixion. The eight multi-chord events provided sub-milliarcsecond accuracy, thanks to the combination of accurate timing and Gaia DR3 star catalogues (Table~\ref{tab:ixion_astrometric_solutions}). These astrometric constraints offer valuable input for orbit refinement and will enhance the accuracy of future occultation predictions, underscoring the dual contribution of these observations to both physical and dynamical studies of trans-Neptunian objects.

Future progress will rely on improved photometric coverage around opposition, denser occultation campaigns, and complementary thermal radiometry. In particular, determining a reliable rotational period remains a key open issue, as our re-analysis of all available datasets revealed no statistically significant periodicity. Such data will allow us to constrain Ixion’s phase curve, spin state, and 3D shape more accurately, and to refine surface property estimates within the context of the broader trans-Neptunian population.

\begin{acknowledgements}
We dedicate this paper to the memory of C. A. Colesanti\textsuperscript{\textdagger}, J. Pollock\textsuperscript{\textdagger}, and T. George\textsuperscript{\textdagger}, whose commitment, passion, and careful work in the field of occultation astronomy remain an inspiration to us all. Their contributions have enriched this study and the broader community, and their legacy will continue to guide future observations. We thank the anonymous referee for their careful review and constructive suggestions, which have helped us to improve the clarity and overall quality of the manuscript. Multiple funding agencies and institutions supported this work. It was partly funded by the Spanish projects PID2020-112789GB-I00 (AEI) and Proyecto de Excelencia de la Junta de Andalucía PY20-01309. This study was financed in part by the Coordenação de Aperfeiçoamento de Pessoal de Nível Superior – Brasil (CAPES) – Finance Code 001. J.L.O., P.S.-S., N.M., A.A.C, R.D., Y.K., J.L.R., and J.M.L.G. acknowledge financial support from the Severo Ochoa grant CEX2021-001131-S (MCIN/AEI/10.13039/501100011033). F.B.-R. acknowledges CNPq (grant 316604/2023-2) and the financial support of the NAPI “Fenômenos Extremos do Universo” of Fundação de Apoio à Ciência, Tecnologia e Inovação do Paraná. P.S.-S. and Y.K. also acknowledge support from the Spanish I+D+i project PID2022-139555NB-I00 (TNO-JWST) funded by MCIN/AEI/10.13039/501100011033. This work has been supported by the French ANR project Roche, number ANR-23-CE49-0012. J.L.R. acknowledges financial support from grant PID2021-126365NB-C21. AAC acknowledges financial support from the project PID2023-153123NB-I00 funded by MCIN/AEI. J.M.G.L. acknowledges funding by the Spanish Ministry of Universities through the university training programme FPU2022/00492. BEM thanks CAPES Grant 23079.212658/2024-30. ARGJ thanks the financial support of FAPEMIG APQ-02987-24. This work is partly based on observations collected at the Centro Astronómico Hispano en Andalucía (CAHA), Observatorio de Sierra Nevada (IAA-CSIC), and the Liverpool Telescope at the Roque de los Muchachos Observatory (IAC). LMC thanks CAPES for the financial support, Finance Code 001, and 88881.981125/2024-01. Based on data acquired at Complejo Astronómico El Leoncito, operated under agreement between the Consejo Nacional de Investigaciones Científicas y Técnicas de la República Argentina and the National Universities of La Plata, Córdoba and San Juan. We also thank T. Linder for his valuable contribution to the 2021 April 28 observation of Ixion, performed with the CTIO 1-meter telescope, which provided important constraints for this study. Based on observations collected at the La Silla European Southern Observatory. TRAPPIST is funded by the Belgian F.R.S.-FNRS under grant PDR T.0120.21. EJ is Director of Research at the Belgian F.R.S.-FNRS. This research has made use of data from the European Space Agency (ESA) mission Gaia (\url{https://www.cosmos.esa.int/gaia}), processed by the Gaia Data Processing and Analysis Consortium (\url{https://www.cosmos.esa.int/web/gaia/dpac/consortium}), with funding provided by institutions participating in the Gaia Multilateral Agreement. CLP thanks the FAPERJ/DSC-10 E-26/204.141/2022, FAPERJ/PDR-10 E-26/200.107/2025, and FAPERJ 200.108/ 2025. This work was supported by the project Gaia Moons of the Agence Nationale de Recherche (France), grant ANR-22-CE49-0002.
\end{acknowledgements}

\bibliographystyle{aa}
\bibliography{ixion}

\appendix
\onecolumn

\section{Ingress and egress times from light-curve modelling}
\label{appendix:timing_table}
\begin{table*}[htbp]
    \footnotesize  
    \centering
    \caption{Ingress and egress times obtained from light curve modelling of all positive detections. The $1\sigma$ error bars are in seconds.}
    \label{tab:best_instants}
    \begin{tabular}{l c c c c c c} \hline \hline
    \multirow{2}{*}{Event}  &   \multirow{2}{*}{Site}   
                            & Ingress time (UT)     
                            & Egress time (UT)  
                            & Chord length 
                            & Light curve \\ 
                            &                           
                            &  (hh:mm:ss.s)             
                            &  (hh:mm:ss.s)         
                            &   (km)     
                            & (Fig.) \\ 
    \hline
    \multirow{3}{*}{17 Aug 2020}    
        & Lake Placid    & 01:36:59.90 (0.90)  & 01:37:38.54 (0.92)  & 461.59 (15.4) & \ref{fig:lc_20200817_a} \\
        & Naperville     & 01:38:21.50 (1.10)  & 01:39:22.24 (1.06)  & 738.02 (18.3) & \ref{fig:lc_20200817_b} \\
        & Chester        & 01:38:02.48 (0.10)  & 01:38:31.71 (0.09)  & 350.82 (1.6)  & \ref{fig:lc_20200817_c} \\ 
    \hline
    \multirow{8}{*}{13 Oct 2020}   
        & Running Springs & 02:00:21.52 (0.01)  & 02:01:05.06 (0.39)  & 687.3 (6.2)   & \ref{fig:lc_20201013_a} \\
        & Glendale        & 02:00:45.99 (0.02)  & 02:01:29.34 (0.76)  & 685.3 (12.4)  & \ref{fig:lc_20201013_b} \\
        & Lowell Obs. (TiMo)\textsuperscript{a} & - & 02:01:30.22 (0.02) & - & \ref{fig:lc_20201013_c} \\        
        & Tucson          & 02:00:57.68 (0.12)  & 02:01:34.16 (0.09)  & 578.0 (2.3)   & \ref{fig:lc_20201013_d} \\
        & Carefree        & 02:00:47.00 (0.04)  & 02:01:30.59 (0.01)  & 689.1 (0.7)   & \ref{fig:lc_20201013_e} \\
        & Fountain Hills  & 02:00:48.66 (0.01)  & 02:01:31.61 (0.01)  & 679.1 (0.4)   & \ref{fig:lc_20201013_f} \\
        & Scottsdale      & 02:00:47.33 (0.02)  & 02:01:31.79 (0.02)  & 691.6 (0.4)   & \ref{fig:lc_20201013_g} \\
        & Tempe           & 02:00:48.17 (0.03)  & 02:01:30.69 (0.02)  & 672.0 (0.5)   & \ref{fig:lc_20201013_h} \\ 
    \hline
    \multirow{6}{*}{28 Apr 2021}     
        & OPD  & 07:34:38.43 (0.005) & 07:34:56.47 (0.022) & 251.1 ($0.3$)   & \ref{fig:lc_20210428_a} \\ 
        & Reconquista     & 07:36:21.60 (0.78)  & 07:37:02.93 (0.81)  & 575.4 ($15.7$)  & \ref{fig:lc_20210428_b} \\ 
        & La Canelilla    & 07:37:44.56 (0.19)  & 07:38:09.72 (0.19)  & 350.1 ($3.7$)   & \ref{fig:lc_20210428_c} \\ 
        & CTIO One-Meter  & 07:37:49.83 (0.83)  & 07:38:02.49 (0.32)  & 176.3 ($12.4$)  & \ref{fig:lc_20210428_d} \\ 
        & PROMPT-6        & 07:37:50.05 (0.35)  & 07:38:02.51 (0.01)  & 173.5 ($4.9$)   & \ref{fig:lc_20210428_e} \\ 
        & CASLEO - ASH           & 07:37:28.12 (1.12)  & 07:38:15.29 (0.57)  & 656.4 ($17.5$)  & \ref{fig:lc_20210428_f} \\ 
    \hline
    \multirow{4}{*}{20 May 2021}       
        & Bartlett Lake   & 06:45:14.26 (0.26)  & 06:45:48.96 (0.13)  & 726.7 ($5.5$)   & \ref{fig:lc_20210520_a} \\ 
        & Scottsdale      & 06:45:14.84 (0.03)  & 06:45:48.97 (0.03)  & 715.4 ($1.1$)   & \ref{fig:lc_20210520_b} \\
        & Clear Lake Shores & 06:44:19.36 (0.27)  & 06:44:52.59 (0.21)  & 699.0 ($7.3$)  & \ref{fig:lc_20210520_c} \\
        & Fountain Hills  & 06:45:14.85 (0.39)  & 06:45:48.62 (0.40)  & 704.6 ($11.8$)  & \ref{fig:lc_20210520_d} \\
    \hline
    17 Aug 2021                     
        & Heaven's Mirror  & 11:33:12.16 (1.08)  & 11:34:13.79 (1.06)  & 769.2 (26.8)  & \ref{fig:lc_20210817_a} \\
    \hline
    2 Jun 2022                        
        & Glenlee         & 18:06:43.11 (0.05)  & 18:06:47.55 (0.06)  & 104.3 (2.5)   & \ref{fig:lc_20220602_a} \\
    \hline
    \multirow{4}{*}{30 Jun 2022}      
        & UTFPR           & 00:25:00.41 (0.84)  & 00:25:19.50 (1.78)  & 477.4 (49.2)  & \ref{fig:lc_20220630_a} \\
        & UEPG            & 00:25:01.67 (0.23)  & 00:25:22.52 (0.24)  & 532.1 (12.4)  & \ref{fig:lc_20220630_b} \\
        & Orion           & 00:24:48.17 (2.46)  & 00:25:15.84 (2.63)  & 736.7 (46.3)  & \ref{fig:lc_20220630_c} \\
        & YPO             & 00:24:44.11 (1.37)  & 00:25:10.19 (1.47)  & 655.1 (77.9) & \ref{fig:lc_20220630_d} \\
    \hline
    \multirow{3}{*}{28 Jul 2023}      
        & SOAR            & 06:09:16.73 (0.13)  & 06:09:45.77 (0.14)  & 594.1 (4.0)   & \ref{fig:lc_20230728_a} \\ 
        & CAO     & 06:09:17.98 (0.03)  & 06:09:44.95 (0.05)  & 551.6 (2.5)   & \ref{fig:lc_20230728_b} \\
        & Danish/ESO          & 06:09:12.23 (0.04)  & 06:09:45.28 (0.02)  & 675.9 (0.5)   & \ref{fig:lc_20230728_c} \\
    \hline
    \multicolumn{6}{l}{\footnotesize \textsuperscript{a} 5× binned to improve S/N.}\\
    \end{tabular}
\end{table*}

\section{Detection limits on additional material around Ixion}
\label{appendix:detection_limits}

We analysed the light curves from all the events studied in this work to search for additional material around Ixion. Following the procedures outlined in \citet{morgado_2023, pereira_2023, fribas_2023, pereira_2024}, we determined the upper limits for the apparent equivalent width and optical depth obtained from $\mathrm{E'}_{p} = [1 - \phi(i)]\Delta r(i)$ and $\tau = \tau'/2$ (where $\tau' = -\ln(1-p'_{3\sigma})$), respectively. Table \ref{tab:detection_limits} presents the results obtained in each event. The limits were determined using the original spatial resolution $\delta_{r}$ in the regions external to the occultation by the central body (when a positive detection), covering a certain distance in the sky plane, not necessarily centred on the body's position. Using as an example the light curve obtained at the Observatório do Pico dos Dias (OPD) for the April 28, 2021 event, the $3\sigma$ limit for the apparent equivalent width is $E'_{p} = 400$~meters. This indicates that an opaque structure with a width in the sky plane of $W_{\perp} > 400$~meters would cause a flux drop in the light curve above our $3\sigma$ cut. On the other hand, a structure with an apparent width $W_{\perp} = 2.1$~km ($\delta_r$) would be detected if its optical depth were $\tau > 0.1$. 
We can compare the detection limits obtained for Ixion with the physical properties of known rings around other small bodies. In our example using the April 2021 light curve from OPD, Chariklo's densest ring (C1R) would be detected ($\mathrm{E}_{p} \sim 2$~km). Additionally, the Q1R ring around Quaoar and Haumea's ring would also be detected if present around Ixion.

\begin{table*}[h!]
    \centering
    \small
    \caption{Detection limits on additional material derived from all light curves. The limits are at the $3\sigma$ level.}
    \begin{tabular}{l l c c c c} \hline \hline
    \multirow{2}{*}{Date} & \multirow{2}{*}{Site} & $\delta_{r}$ & $\mathrm{E'}_{p}$ & \multirow{2}{*}{$\tau$} & Sky Cover \\ 
    & & [km] & [km] & & [km] \\ 
    \hline 
    2020-08-17 & Lake Placid                    & 35.8 & 29.4 & 0.5 & 7\,913  \\
    2020-08-17 & Naperville                     & 59.8 & 57.3 & 0.6 & 12\,255 \\
    2020-08-17 & Chester                        & 3.6  & 4.4  & 0.8 & 1\,581  \\
    \hline
    2020-10-13 & Running Springs                & 0.5 & 0.6 & 0.7 & 1\,371    \\
    2020-10-13 & Lowell Observatory (TiMo)             & 3.9 & 2.4 & 0.3 & 3\,648    \\
    2020-10-13 & Glendale                       & 2.1 & 1.6 & 0.4 & 3\,833    \\
    2020-10-13 & Carefree                       & 0.5 & 1.0 & 1.5 & 2\,298    \\
    2020-10-13 & Scottsdale                     & 0.5 & 0.6 & 0.7 & 4\,047    \\
    2020-10-13 & Fountain Hills                 & 0.5 & 0.5 & 0.5 & 3\,798    \\
    2020-10-13 & Tempe                          & 2.1 & 1.5 & 0.4 & 3\,884    \\
    2020-10-13 & Tucson                         & 1.0 & 1.5 & 0.9 & 5\,552    \\
    2020-10-13 & Clear Lake Shores              & 5.3 & 5.9 & 0.7 & 14\,421   \\
    2020-10-13 & Gardnerville                   & 5.3 & 3.9 & 0.4 & 2\,898    \\
    \hline
    2021-04-28 & PROMPT-6                        & 13.9 & 4.4 & 0.2 & 16\,330   \\
    2021-04-28 & CTIO One-meter                 & 13.9 & 1.4 & 0.05 & 16\,860  \\
    2021-04-28 & Observatório do Pico dos Dias  & 2.1 & 0.4 & 0.1 & 14\,801    \\
    2021-04-28 & La Canelilla                   & 7.0 & 13.6 & 1.6 & 6\,699    \\
    2021-04-28 & Reconquista                    & 69.6 & 32.6 & 0.3 & 26\,124  \\
    2021-04-28 & CASLEO - Cerro Burek ASH                & 139.2 & 68.4 & 0.3 & 20\,835 \\ 
    2021-04-28 & A.A.A.A.                       & 69.5 & 107 & 1.0 & 17\,936   \\
    2021-04-28 & Cruz del Sur                   & 83.4 & 0.4 & 0.6 & 4\,023    \\
    2021-04-28 & Trappist-South                 & 34.8 & 4.3 & 0.06 & 15\,433  \\
    2021-04-28 & Observatório Los Cabezones                  & 69.5 & 54.9 & 0.5 & 14\,208  \\
    2021-04-28 & Observatório El Catalejo                    & 27.8 & 35 & 0.8 & 14\,183    \\
    \hline
    2021-05-20 & Westport Astronomical Society                       & 21.0 & 42.4 & 1.7 & 7\,595   \\
    2021-05-20 & Clear Lake Shores                 & 22.5 & 30.5 & 0.9 & 12\,590  \\
    2021-05-20 & Barltlett Lake Turnoff                      & 22.1 & 16.0 & 0.4 & 4\,811   \\
    2021-05-20 & Scottsdale                   & 33.6 & 22.9 & 0.4 & 5\,514   \\
    2021-05-20 & Fountain Hills                 & 41.9 & 29.6 & 0.4 & 11\,330  \\
    \hline
    2021-08-17 & Heaven's Mirror                 & 37.4 & 18.8 & 0.3 & 12\,994  \\
    \hline
    2022-06-02 & Glenlee                        & 7.5  & 3.9  & 0.3 & 5\,563   \\
    \hline
    2022-06-30 & YPO     & 50.0 & 46.2& 0.6 & 9\,416   \\
    2022-06-30 & Orion Observatory              & 75.1 & 32.3 & 0.2 & 17\,290  \\
    2022-06-30 & UEPG                           & 25.0 & 21.5 & 0.5 & 17\,689  \\
    2022-06-30 & UTFPR-Neoville                 & 97.5 & 96.6 & 0.6 & 14\,021  \\
    2022-06-30 & SONEAR-CEAMIG                  & 50.1 & 14.0 & 0.1 & 14\,546  \\
    \hline
    2023-07-28 & Danish/ESO                     & 2.0  & 0.6  & 0.1 & 6\,136   \\
    2023-07-28 & SOAR                           & 4.1  & 1.8  & 0.08& 19\,999  \\
    2023-07-28 & Campocatino Austral Observatory (CAO) & 20.4 & 4.9  & 0.1 & 17\,570  \\
    2023-07-28 & CASLEO - Cerro Burek HSH                         &122.6 & 20.6 & 0.09 & 19\,876 \\
    \hline
    \end{tabular}
    \label{tab:detection_limits}
\end{table*}

\clearpage
\section{Summary of Occultation Campaigns}
\label{sec:summary_occultation_paths}

\begin{figure*}[h!]
\centering
\resizebox{\hsize}{10cm}{
\begin{tabular}{cc}
    \includegraphics[width=0.48\textwidth]{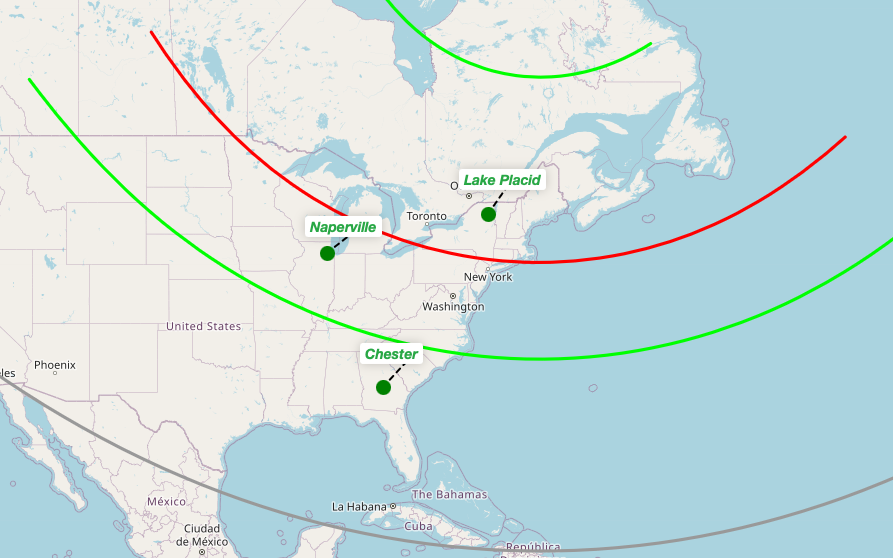}\phantomsubcaption\label{fig:occ_paths_a} &
    \includegraphics[width=0.48\textwidth]{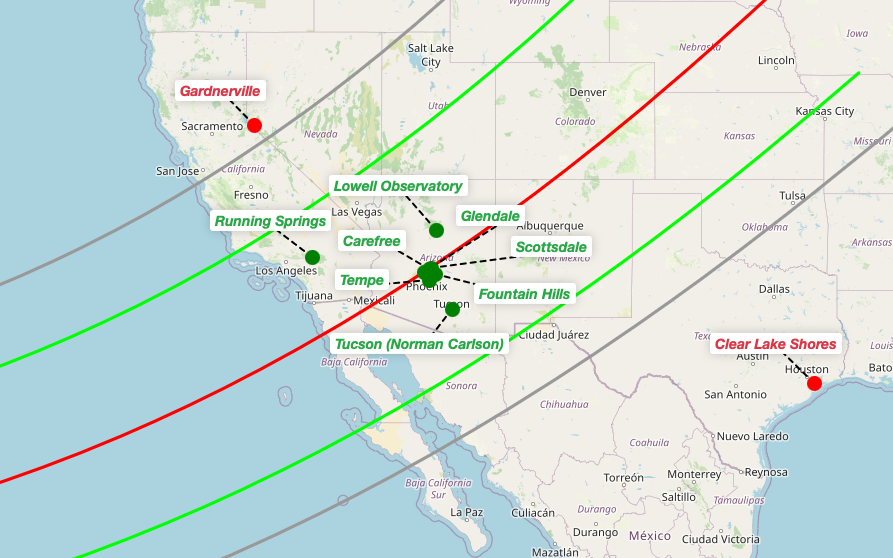}\phantomsubcaption\label{fig:occ_paths_b} \\
    \textbf{(a)} 17 August 2020 & \textbf{(b)} 13 October 2020 \\

    \includegraphics[width=0.48\textwidth]{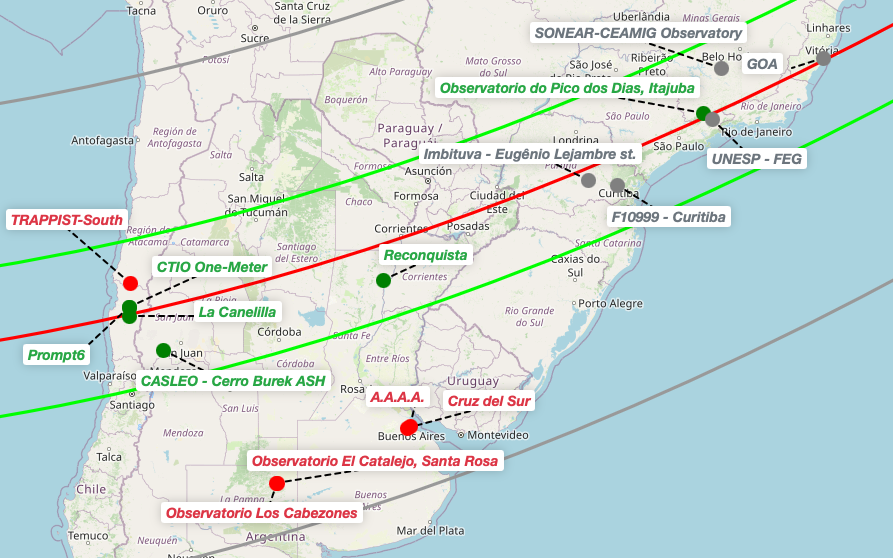}\phantomsubcaption\label{fig:occ_paths_c} &
    \includegraphics[width=0.48\textwidth]{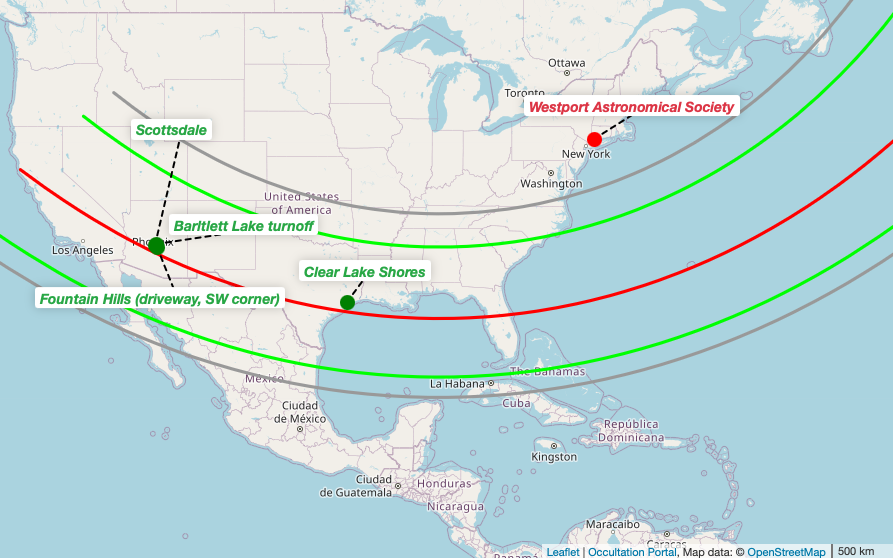}\phantomsubcaption\label{fig:occ_paths_d} \\
    \textbf{(c)} 28 April 2021 & \textbf{(d)} 20 May 2021 \\

    \includegraphics[width=0.48\textwidth]{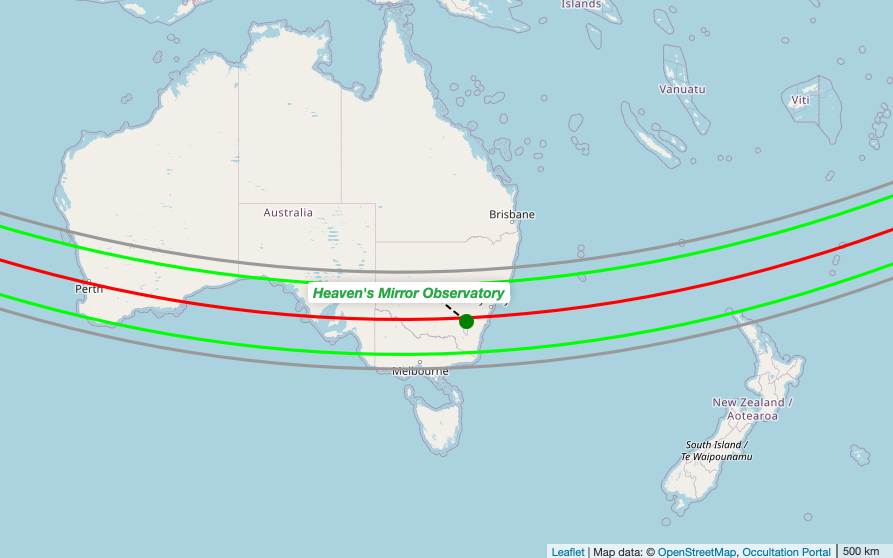}\phantomsubcaption\label{fig:occ_paths_e} &
    \includegraphics[width=0.48\textwidth]{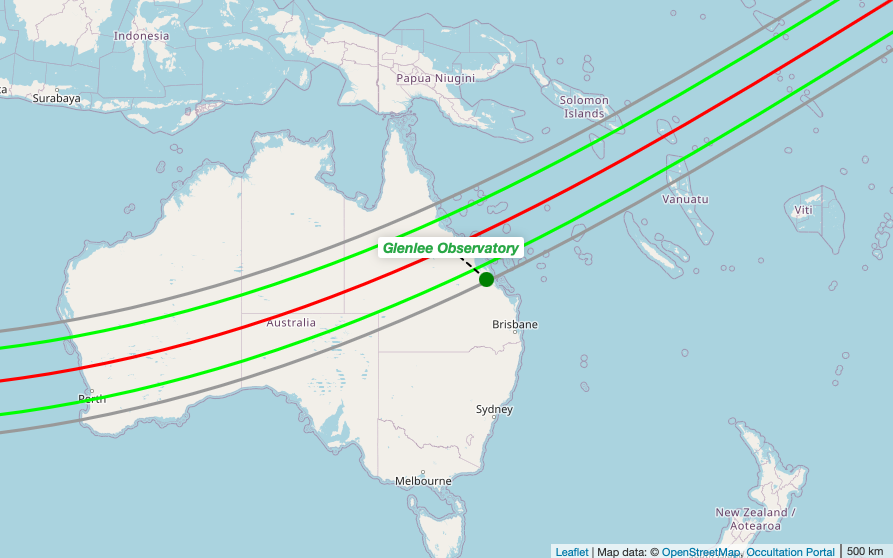}\phantomsubcaption\label{fig:occ_paths_f} \\
    \textbf{(e)} 17 August 2021 & \textbf{(f)} 2 June 2022 \\

    \includegraphics[width=0.48\textwidth]{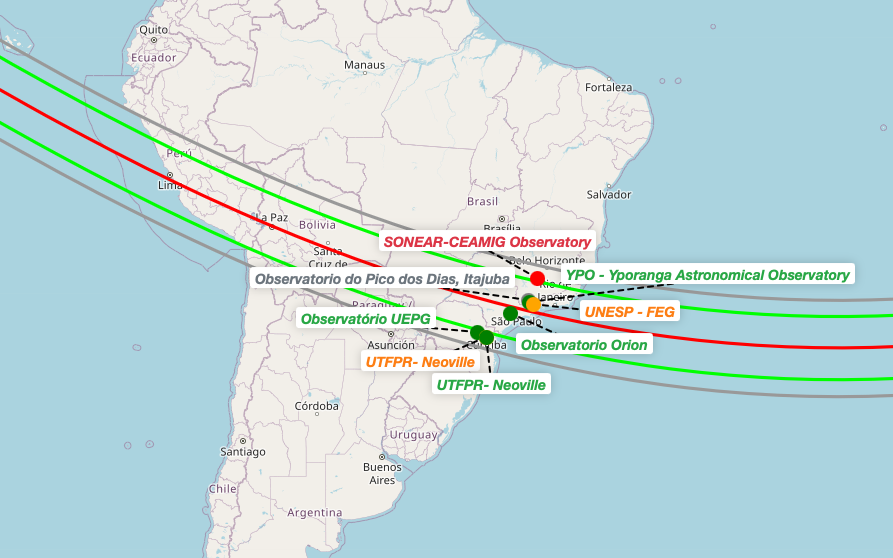}\phantomsubcaption\label{fig:occ_paths_g} &
    \includegraphics[width=0.48\textwidth]{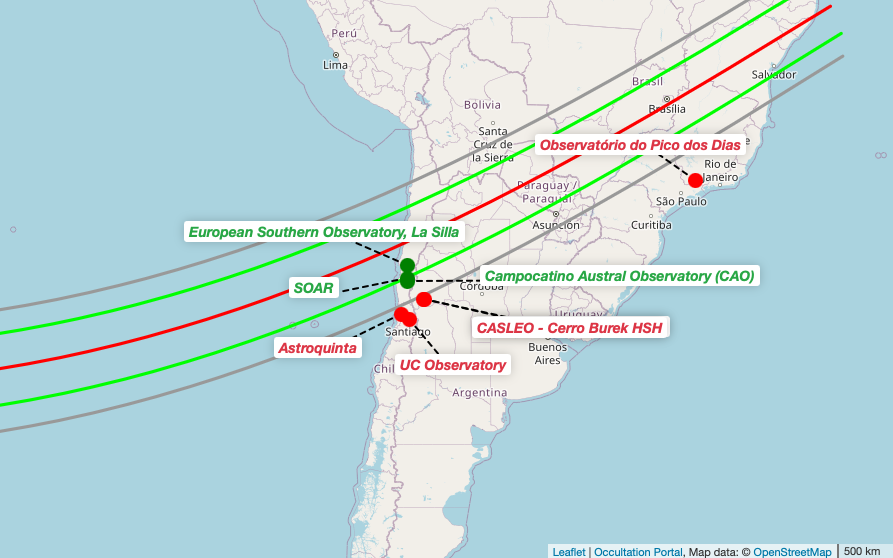}\phantomsubcaption\label{fig:occ_paths_h} \\
    \textbf{(g)} 30 June 2022 & \textbf{(h)} 28 July 2023 \\
\end{tabular}
}
\caption{
Occultation paths for (28978)~Ixion events. Each panel corresponds to a specific observation date.  
Green circles indicate positive detections, red circles indicate negative detections, grey circles indicate overcast conditions, and orange circles indicate technical failures.  
The green lines show the maximum and minimum shadow diameter limits of Ixion's NIMA prediction, the red line represents the prediction centre, and the grey band indicates the $1\sigma$ uncertainty.  
Data compiled via the \textit{Occultation Portal}.
}
\label{fig:occ_paths}
\end{figure*}

\clearpage
\section{Observation Details}
\label{sec:obs_details}

The observing circumstances and notation conventions are described in the table footnotes.

\begin{small}
\setlength{\LTcapwidth}{\textwidth}
\renewcommand{\arraystretch}{1.08}
\setlength\tabcolsep{3pt}

\begin{longtable}{@{\extracolsep{\fill}}rllllllllll@{}}
\caption{Observation details for the occultations by (28978)~Ixion.}
\label{tab:obs_ixion_all}\\
\toprule\midrule
\# & Site (Country) & Latitude (dms) & Transparency & Telescope (cm) & Method & Observation \\
  & Observer(s)     & Longitude (dms) & Wind        & Camera         & ExpTime & DeadTime \\
  &                 & Elevation (m)   & Seeing      & Filter         & TimeSrc &           \\
\midrule
\endfirsthead

\caption[]{Observation details for the occultations by (28978)~Ixion (continued).}\\
\toprule\midrule
\# & Site (Country) & Latitude (dms) & Transparency & Telescope (cm) & Method & Observation \\
  & Observer(s)     & Longitude (dms) & Wind        & Camera         & ExpTime & DeadTime \\
  &                 & Elevation (m)   & Seeing      & Filter         & TimeSrc &           \\
\midrule
\endhead

\midrule
\multicolumn{11}{r}{\emph{Continued on next page}}\\
\endfoot

\bottomrule
\multicolumn{11}{p{\textwidth}}{\footnotesize
Site latitude, longitude (format dms) and elevation (AMSL in m) are given in the WGS84 datum.
\textbf{Telescope}: Tx refers to the telescope aperture in cm.
\textbf{Method} is the recording method: \textbf{IMG} means digital (CCD, CMOS) sequential imaging, while \textbf{VID} means analogue video recording.
\textbf{TimeSrc} refers to the used timing source and method: \textbf{GPS}: 1-PPS (one pulse per second) driven video-time-insertion (VID) or camera-internal GPS timestamps (IMG). \textbf{NTP}: Network Time Protocol computer system clock synchronization. \textbf{CamGPS}: Camera synchronized directly to a GPS signal for timestamping accuracy. \textbf{ComGPS}: Computer synchronized to a GPS signal. \textbf{ComNTP}: Computer synchronized using the Network Time Protocol. \textbf{TimeBox}: A TimeBox is used for precise timestamping during observations. \textbf{IOTA-VTI}: A hardware device developed by IOTA (International Occultation Timing Association) for video time insertion. \textbf{Other}: Any other synchronization method not listed above.
\textbf{Transparency} qualitatively describes sky clarity during the observation (e.g., clear, hazy, cloudy, overcast).
\textbf{Seeing} qualitatively refers to atmospheric stability affecting image sharpness (e.g., good, moderate, poor).
Observation can be either positive (occultation detected/recorded) or negative (occultation not detected).
\textbf{ExpTime} represents the exposure time in seconds, while \textbf{DeadTime} refers to the interval between subsequent images in seconds. The sampling cadence is the sum of ExpTime and DeadTime.
}\\
\endlastfoot

\multicolumn{11}{c}{\textbf{2020-08-17}}\\[-0.35em]\midrule

1 & Chester (USA) & 32$^\circ$ 22$^\prime$ 15.0996$^{\prime\prime}$ N& Clear & T35.6& VID & Positive \\
&  & 83$^\circ$ 12$^\prime$ 7.2$^{\prime\prime}$ W & <2~km/h & WAT-910HX & 0.3~s & (0~s)\\
& \multicolumn{1}{l}{\parbox[t]{5.5cm}{\raggedright\em Roger Venable}} & 104.00 & Moderate & Luminance & ComNTP &\\\midrule
2 & Naperville (USA) & 41$^\circ$ 45$^\prime$ 32.4$^{\prime\prime}$ N& Clear & T35.6& IMG & Positive \\
&  & 88$^\circ$ 7$^\prime$ 0.0084$^{\prime\prime}$ W & <2~km/h & QHY174-GPS & 5~s & (0~s)\\
& \multicolumn{1}{l}{\parbox[t]{5.5cm}{\raggedright\em Robert Dunford}} & 230.00 & Moderate & Luminance & CamGPS &\\\midrule
3 & Lake Placid (USA) & 44$^\circ$ 15$^\prime$ 18.3996$^{\prime\prime}$ N & Clear& T30.5& IMG & Positive \\
&  & 74$^\circ$ 1$^\prime$ 59.4192$^{\prime\prime}$ W & <2~km/h & QHY174-GPS & 3.00088~s & (0~s)\\
& \multicolumn{1}{l}{\parbox[t]{5.5cm}{\raggedright\em George R. Viscome}} & 608.00 & Good & Luminance & CamGPS &\\

\midrule
\multicolumn{11}{c}{\textbf{2020-10-13}}\\[-0.35em]\midrule

1 & Lowell Observatory (TiMo) (USA)& 35$^\circ$ 12$^\prime$ 10.3644$^{\prime\prime}$ N & Clear & T31.8& VID & Positive \\
&  & 111$^\circ$ 40$^\prime$ 1.4628$^{\prime\prime}$ W & <2~km/h & Point Grey BFLY & 0.050~s & (0~s)\\
& \multicolumn{1}{l}{\parbox[t]{5.5cm}{\raggedright\em Michael Collins}} & 2198.00 & Good  & Clear & ComGPS &\\\midrule
2 & Scottsdale (USA) & 33$^\circ$ 49$^\prime$ 0.0984$^{\prime\prime}$ N& Clear & T30.0& VID & Positive \\
&  & 111$^\circ$ 52$^\prime$ 7.2984$^{\prime\prime}$ W & <2~km/h & Watec 910HX & 0.033~s & (0~s)\\
& \multicolumn{1}{l}{\parbox[t]{5.5cm}{\raggedright\em Tony George}} & 843.00& Moderate  & Clear & IOTA-VTI &\\\midrule
3 & Tempe (USA) & 33$^\circ$ 23$^\prime$ 24.3348$^{\prime\prime}$ N& Clear & T28.0& VID & Positive \\
&  & 111$^\circ$ 57$^\prime$ 20.574$^{\prime\prime}$ W& <2~km/h  & RunCam Astro & 0.134~s & (0~s)\\
& \multicolumn{1}{l}{\parbox[t]{5.5cm}{\raggedright\em Wayne Thomas}} & 358.00& Good   & Clear & IOTA-VTI &\\\midrule
4 & Glendale & 33$^\circ$ 41$^\prime$ 39.0012$^{\prime\prime}$ N& Clear & T25.4& VID & Positive \\
&  & 112$^\circ$ 12$^\prime$ 46.0008$^{\prime\prime}$ W& <2~km/h  & RunCam Astro & 0.134~s & (0~s)\\
& \multicolumn{1}{l}{\parbox[t]{5.5cm}{\raggedright\em David A. Kenyon}} & 400.00 & Good  & Clear & IOTA-VTI &\\\midrule
5 & Tucson (USA) & 32$^\circ$ 18$^\prime$ 28.314$^{\prime\prime}$ N& Clear & T23.5& VID & Positive \\
&  & 110$^\circ$ 57$^\prime$ 46.62$^{\prime\prime}$ W& <2~km/h  & RunCam Astro & 0.067~s & (0~s)\\
& \multicolumn{1}{l}{\parbox[t]{5.5cm}{\raggedright\em Norman Carlson}}  & 750.00& Moderate & Clear & IOTA-VTI &\\\midrule
6 & Running Springs (USA) & 34$^\circ$ 13$^\prime$ 7.8996$^{\prime\prime}$ N& Clear & T20.3& VID & Positive \\
&  & 117$^\circ$ 7$^\prime$ 52.3992$^{\prime\prime}$ W & <2~km/h & Watec 910HX & 0.033~s & (0~s)\\
& \multicolumn{1}{l}{\parbox[t]{5.5cm}{\raggedright\em Robert Jones}} & 1874.00& Good  & Clear & IOTA-VTI &\\\midrule
7 & Carefree (USA) & 33$^\circ$ 48$^\prime$ 42.858$^{\prime\prime}$ N& Clear & T20.0& VID & Positive \\
&  & 111$^\circ$ 57$^\prime$ 7.974$^{\prime\prime}$ W& <2~km/h  & Watec 910HX & 0.033~s & (0~s)\\
& \multicolumn{1}{l}{\parbox[t]{5.5cm}{\raggedright\em Paul D. Maley}} & 654.00& Best  & Clear & IOTA-VTI &\\\midrule
8 & Fountain Hills (USA) & 33$^\circ$ 37$^\prime$ 21$^{\prime\prime}$ N& Clear & T20.0& VID & Positive \\
&  & 111$^\circ$ 43$^\prime$ 35.0004$^{\prime\prime}$ W& <2~km/h  & Watec 910HX & 0.033~s & (0~s)\\
& \multicolumn{1}{l}{\parbox[t]{5.5cm}{\raggedright\em Ted Blank}}& 515.00 & Best  & Clear & IOTA-VTI &\\\midrule
9 & Gardnerville (USA) & 38$^\circ$ 53$^\prime$ 23.4996$^{\prime\prime}$ N& Clear & T30.5& VID & Negative \\
& & 119$^\circ$ 40$^\prime$ 20.3016$^{\prime\prime}$ W  & <2~km/h & Watec 910HX & 0.334~s & (0~s)\\
& \multicolumn{1}{l}{\parbox[t]{5.5cm}{\raggedright\em Jerry Bardecker}} & 1524.00& Good  & Clear & IOTA-VTI &\\\midrule
10 & Clear Lake Shores (USA) & 29$^\circ$ 32$^\prime$ 55.5101$^{\prime\prime}$ N& Clear & T20.3& VID & Negative \\
&  & 95$^\circ$ 2$^\prime$ 6.57996$^{\prime\prime}$ W& <2~km/h  & Watec 910BD & 0.334~s & (0~s)\\
& \multicolumn{1}{l}{\parbox[t]{5.5cm}{\raggedright\em Phil C. Stuart}} & 15.80& Moderate  & Clear & IOTA-VTI &\\

\midrule
\multicolumn{11}{c}{\textbf{2021-04-28}}\\[-0.35em]\midrule
1 & Observatorio do Pico dos Dias (Brazil) & 22$^\circ$ 32$^\prime$ 7.7532$^{\prime\prime}$ S& Clear & T157.4& IMG & Positive \\
& Itajuba & 45$^\circ$ 34$^\prime$ 57.54$^{\prime\prime}$ W& <2~km/h  & IXon & 0.15~s & (0~s)\\
& \multicolumn{1}{l}{\parbox[t]{5.5cm}{\raggedright\em G. Rossi, G. Margoti, V. Peixoto}} & 1810.71& Good  & Clear & CamGPS &\\\pagebreak
2 & CTIO One-Meter (Chile) & 30$^\circ$ 10$^\prime$ 7.7916$^{\prime\prime}$ S& Clear & T100.0& IMG & Positive \\
&  & 70$^\circ$ 48$^\prime$ 21.69$^{\prime\prime}$ W & <2~km/h & FLI PL23042 & 1.0~s & (0~s)\\
& \multicolumn{1}{l}{\parbox[t]{5.5cm}{\raggedright\em J. Pollock, T. Linder }} & 2201.00 & Good  & Clear & ComNTP &\\\midrule
3 & La Canelilla (Chile) & 30$^\circ$ 32$^\prime$ 3$^{\prime\prime}$ S& Clear & T52.0& IMG & Positive \\
&  & 70$^\circ$ 47$^\prime$ 45$^{\prime\prime}$ W& <2~km/h  & ZWO1600 & 0.50112~s & (0~s)\\
& \multicolumn{1}{l}{\parbox[t]{5.5cm}{\raggedright\em M. Meunier, B. Christophe, L. Bernasconi}} & 1548.00 & Good  & Clear & ComNTP &\\\midrule
4 & CASLEO - Cerro Burek ASH (Argentina) & 31$^\circ$ 47$^\prime$ 13.2$^{\prime\prime}$ S & Clear& T45.7& IMG & Positive \\
&  & 69$^\circ$ 18$^\prime$ 23.868$^{\prime\prime}$ W& <2~km/h  & SBIG STL11000 & 10.0~s & (2~s)\\
& \multicolumn{1}{l}{\parbox[t]{5.5cm}{\raggedright\em N. Morales, J.L. Ortiz}}& 2591.00 & Good  & Luminance & ComNTP &\\\midrule
5 & PROMPT-6 (Chile) & 30$^\circ$ 10$^\prime$ 3.72$^{\prime\prime}$ S& Clear & T40.0& IMG & Positive \\
&  & 70$^\circ$ 48$^\prime$ 18.8208$^{\prime\prime}$ W & <2~km/h  & FLI PL23042 & 1.0~s & (0~s)\\
& \multicolumn{1}{l}{\parbox[t]{5.5cm}{\raggedright\em J. Pollock , Vladimir Kouprianov}} & 2166.00& Good  & Clear & ComNTP &\\\midrule
6 & Reconquista (Argentina) & 29$^\circ$ 8$^\prime$ 25.1826$^{\prime\prime}$ S& Clear & T30.5& IMG & Positive \\
&  & 59$^\circ$ 38$^\prime$ 36.6086$^{\prime\prime}$ W & <2~km/h  & QHY174M & 5.0~s & (0~s)\\
& \multicolumn{1}{l}{\parbox[t]{5.5cm}{\raggedright\em Ariel Stechina}} & 50.00 & Good  & Clear & ComNTP &\\\midrule
7 & TRAPPIST-South (Chile) & 29$^\circ$ 15$^\prime$ 16.56$^{\prime\prime}$ S & Clear & T60.0& IMG & Negative \\
&  & 70$^\circ$ 44$^\prime$ 21.84$^{\prime\prime}$ W& 20-28~km/h & FLI PL3041-BB & 2.5~s & (1.2~s)\\
& \multicolumn{1}{l}{\parbox[t]{5.5cm}{\raggedright\em Emmanuel Jehin}} & 2315.00 & Good & Clear & ComNTP &\\\midrule
8 & A.A.A.A. (Argentina) & 34$^\circ$ 36$^\prime$ 18.6984$^{\prime\prime}$ S& High humidity & T25.0& IMG & Negative \\
& {\raggedright\em C.Cebral, M. Konishi} & 58$^\circ$ 26$^\prime$ 4.6$^{\prime\prime}$ W & <2~km/h & SBIG ST9e & 5~s & (0~s)\\
& \multicolumn{1}{l}{\parbox[t]{5.5cm}{\raggedright\em C. Magliano, Y.O. Cuello}} & 39.40 & Good  & No filter & Other &\\\midrule
9 & Observatorio El Catalejo, Santa Rosa (Argentina) & 36$^\circ$ 38$^\prime$ 15.99$^{\prime\prime}$ S& Clear & T20.0& IMG & Negative \\
&  & 64$^\circ$ 19$^\prime$ 27.48$^{\prime\prime}$ W & <2~km/h  & QHY174M GPS & 2~s & (0~s)\\
& \multicolumn{1}{l}{\parbox[t]{5.5cm}{\raggedright\em Julio Spagnotto}} & 182.00 & Good  & Clear & CamGPS &\\\midrule
10 & Observatorio Los Cabezones (Argentina) & 36$^\circ$ 38$^\prime$ 8.28996$^{\prime\prime}$ S& Clear & T20.0& IMG & Negative \\
&  & 64$^\circ$ 17$^\prime$ 16.8299$^{\prime\prime}$ W & <2~km/h  & QHY174M-GPS & 5~s & (0~s)\\
& \multicolumn{1}{l}{\parbox[t]{5.5cm}{\raggedright\em A. Wilberger}} & 180.00 & Good  & Clear & CamGPS &\\\midrule
11 & Cruz del Sur (Argentina) & 34$^\circ$ 40$^\prime$ 10.56$^{\prime\prime}$ S & Clear & T20.0& IMG & Negative \\
&  & 58$^\circ$ 34$^\prime$ 24.276$^{\prime\prime}$ W & <2~km/h & QHY174M-GPS & 6~s & (0~s)\\
& \multicolumn{1}{l}{\parbox[t]{5.5cm}{\raggedright\em Andres Chapman}} & 39.00 & Bad  & Clear & CamGPS &\\\midrule
12 & SONEAR-CEAMIG Observatory (Brazil) & 20$^\circ$ 42$^\prime$ 54.27$^{\prime\prime}$ S& Overcast & T45.0& IMG & Overcast \\
&  & 44$^\circ$ 47$^\prime$ 5.82$^{\prime\prime}$ W& -  & QHY600 & -~s & (1.0~s)\\
& \multicolumn{1}{l}{\parbox[t]{5.5cm}{\raggedright\em C. Jacques}} & 1113.00 & -  & Clear & ComNTP &\\\midrule
13 & UNESP - FEG (Brazil) & 22$^\circ$ 48$^\prime$ 6.0012$^{\prime\prime}$ S & Overcast & T40.6& IMG & Overcast \\
&  & 45$^\circ$ 11$^\prime$ 25.5984$^{\prime\prime}$ W & -  & Merlin 247 Raptor & -~s & (0~s)\\
& \multicolumn{1}{l}{\parbox[t]{5.5cm}{\raggedright\em R. Sfair, A. R. Gomes, Jr }} & 540.00& -  & Clear & ComGPS &\\\midrule
14 & Imbituva - Eugênio Lejambre St. (Brazil) & 25$^\circ$ 13$^\prime$ 39.6527$^{\prime\prime}$ S& Overcast & T30.5& IMG & Overcast \\
& & 50$^\circ$ 36$^\prime$ 41.7111$^{\prime\prime}$ W& -  & Raptor & -~s & (0~s)\\
& \multicolumn{1}{l}{\parbox[t]{5.5cm}{\raggedright\em Chrystian L. Pereira}} & 968.00& -  & Clear & CamGPS &\\\midrule
15 & GOA (Brazil) & 20$^\circ$ 18$^\prime$ 1.99998$^{\prime\prime}$ S& Overcast & T30.4& IMG & Overcast \\
&  & 40$^\circ$ 19$^\prime$ 2$^{\prime\prime}$ W& -  & Canon 600D & -~s & (0~s)\\
& \multicolumn{1}{l}{\parbox[t]{5.5cm}{\raggedright\em M. Malacarne}} & 24.00& -  & Clean & ComNTP &\\\midrule
16 & F10999 - Curitiba (Brazil) & 25$^\circ$ 26$^\prime$ 10.3926$^{\prime\prime}$ S& Overcast & T25.4& IMG & Overcast \\
&  & 49$^\circ$ 20$^\prime$ 23.73$^{\prime\prime}$ W & -  & QHY174M-GPS & -~s & (0~s)\\
& \multicolumn{1}{l}{\parbox[t]{5.5cm}{\raggedright\em F. B. Ribas}} & 1024.00 & -  & Clear & CamGPS &\\

\midrule
\multicolumn{11}{c}{\textbf{2021-05-20}}\\[-0.35em]\midrule
1 & Fountain Hills (USA)& 33$^\circ$ 37$^\prime$ 27.8868$^{\prime\prime}$ N & Clear & T40.6& IMG & Positive \\
& & 111$^\circ$ 43$^\prime$ 39.5879$^{\prime\prime}$ W& 12-19~km/h & QHY174M-GPS & 2.0~s & (0~s)\\
& \multicolumn{1}{l}{\parbox[t]{5.5cm}{\raggedright\em David W. Dunham, Joan Dunham}} & 518.00& Moderate  & Empty & CamGPS &\\\pagebreak
2 & Scottsdale (USA) & 33$^\circ$ 49$^\prime$ 0.0984$^{\prime\prime}$ N& Clear & T30.0& VID & Positive \\
& AZ & 111$^\circ$ 52$^\prime$ 7.2984$^{\prime\prime}$ W & <2~km/h  & WAT910HX & 0.533860~s & (0~s)\\
& \multicolumn{1}{l}{\parbox[t]{5.5cm}{\raggedright\em Tony George}} & 843.00& Moderate  & Clear & CamGPS &\\\midrule
3 & Clear Lake Shores (USA) & 29$^\circ$ 32$^\prime$ 55.5101$^{\prime\prime}$ N& Clear & T20.3& VID & Positive \\
&  & 95$^\circ$ 2$^\prime$ 6.57996$^{\prime\prime}$ W& 2-5~km/h  & Watec 910BD EIA & 1.068~s & (0~s)\\
& \multicolumn{1}{l}{\parbox[t]{5.5cm}{\raggedright\em P. Stuart}} & 15.80 & Moderate  & None & IOTA-VTI &\\\midrule
4 & Barltlett Lake Turnoff (USA) & 33$^\circ$ 50$^\prime$ 54.8412$^{\prime\prime}$ N& Clear & T20.0& VID & Positive \\
&  & 111$^\circ$ 49$^\prime$ 58.08$^{\prime\prime}$ W & 6-11~km/h  & Watec910HX & 0.0333~s & (0~s)\\
& \multicolumn{1}{l}{\parbox[t]{5.5cm}{\raggedright\em Paul D. Maley}} & 996.00 & Good  & None & CamGPS &\\\midrule
5 & Westport Astronomical Society (USA) & 41$^\circ$ 10$^\prime$ 15.8999$^{\prime\prime}$ N & Partly cloudy & T35.6& IMG & Negative \\
&  & 73$^\circ$ 19$^\prime$ 39.2999$^{\prime\prime}$ W & <2~km/h  & QHY174M-GPS & 1.00~s & (0~s)\\
& \multicolumn{1}{l}{\parbox[t]{5.5cm}{\raggedright\em Kevin Green, Chang Gao}} & 88.00 & Good  & None & CamGPS &\\

\midrule
\multicolumn{11}{c}{\textbf{2021-08-17}}\\[-0.35em]\midrule
1 & Heaven's Mirror Observatory (Australia) & 34$^\circ$ 51$^\prime$ 50.89$^{\prime\prime}$ S& Clear & T50.8& IMG & Positive \\
&  & 148$^\circ$ 58$^\prime$ 35.0602$^{\prime\prime}$ E & <2~km/h & QHY174M-GPS & 3.0~s & (0~s)\\
& \multicolumn{1}{l}{\parbox[t]{5.5cm}{\raggedright\em W. Hanna}} & 536.00 & Moderate  & None & CamGPS &\\

\midrule
\multicolumn{11}{c}{\textbf{2022-06-02}}\\[-0.35em]\midrule
1 & Glenlee Observatory (Australia) & 23$^\circ$ 16$^\prime$ 10.0597$^{\prime\prime}$ S& Clear & T30.4& VID & Positive \\
&  & 150$^\circ$ 30$^\prime$ 1.61841$^{\prime\prime}$ E & <2~km/h  & Watec 910BD & 0.32~s & (0~s)\\
& \multicolumn{1}{l}{\parbox[t]{5.5cm}{\raggedright\em Stephen Kerr}} & 53.40 & Moderate  & None & IOTA-VTI &\\

\midrule
\multicolumn{11}{c}{\textbf{2022-06-30}}\\[-0.35em]\midrule
1 & Observatório UEPG (Brazil) & 25$^\circ$ 5$^\prime$ 22.671$^{\prime\prime}$ S& Clear & T40.6& IMG & Positive \\
&  & 50$^\circ$ 5$^\prime$ 56.7901$^{\prime\prime}$ W& <2~km/h  & Raptor & 1.0~s & (0~s)\\
& \multicolumn{1}{l}{\parbox[t]{5.5cm}{\raggedright\em C. L. Pereira, M. Emilio}} & 923.00 & Moderate  & Clear & CamGPS &\\\midrule
2 & Observatorio Orion (Brazil) & 23$^\circ$ 34$^\prime$ 10.2036$^{\prime\prime}$ S& Clear & T35.6& IMG & Positive \\
&  & 47$^\circ$ 12$^\prime$ 40.5144$^{\prime\prime}$ W& 2-5~km/h  & ATIK One 9.0 & 3.0~s & (3.0~s)\\
& \multicolumn{1}{l}{\parbox[t]{5.5cm}{\raggedright\em Tasso Napoleao, Carlos Colesanti}} & 884.00& Good  & No filter & ComNTP &\\\midrule
3 & Yporanga Astronomical (Brazil) & 22$^\circ$ 41$^\prime$ 49$^{\prime\prime}$ S& Clear & T14.0& IMG & Positive \\
& Observatory (YPO) & 45$^\circ$ 31$^\prime$ 5.99988$^{\prime\prime}$ W& <2~km/h  & ZWO ASI 6200M & 2.0~s & (1.2~s)\\
& \multicolumn{1}{l}{\parbox[t]{5.5cm}{\raggedright\em J. Mattei}}& 1640.00 & Good  & LRGB & ComGPS &\\\midrule
4 & UTFPR- Neoville (Brazil) & 25$^\circ$ 30$^\prime$ 31.7373$^{\prime\prime}$ S& Clear & T11.4& IMG & Positive \\
&  & 49$^\circ$ 18$^\prime$ 57.1841$^{\prime\prime}$ W & <2~km/h  & Sony IMX224LQR & 3.971~s & (0.016~s)\\
& \multicolumn{1}{l}{\parbox[t]{5.5cm}{\raggedright\em F. B. Ribas , W. G. Ferrante}} & 937.67& Good  & Clear & ComNTP &\\\midrule
5 & SONEAR-CEAMIG Observatory (Brazil) & 20$^\circ$ 42$^\prime$ 54.27$^{\prime\prime}$ S& Clear & T45.0& IMG & Negative \\
&  & 44$^\circ$ 47$^\prime$ 5.82$^{\prime\prime}$ W& <2~km/h  & QHY600 & 2~s & (1.0~s)\\
& \multicolumn{1}{l}{\parbox[t]{5.5cm}{\raggedright\em C. Jacques}} & 1113.00 & Good  & Clear & ComNTP &\\\midrule
6 & Observatorio do Pico dos Dias (Brazil) & 22$^\circ$ 32$^\prime$ 7.7532$^{\prime\prime}$ S& Overcast & T160.0& IMG & Overcast \\
&  & 45$^\circ$ 34$^\prime$ 57.54$^{\prime\prime}$ W& -  & Andor - Ixon & -~s & (0~s)\\
& \multicolumn{1}{l}{\parbox[t]{5.5cm}{\raggedright\em J. I. B. Camargo }} & 1810.71& -  & Clear & CamGPS &\\\midrule
7 & UNESP - FEG (Brazil) & 22$^\circ$ 48$^\prime$ 6.0012$^{\prime\prime}$ S& Clear & T40.6& IMG & Technical failure \\
&  & 45$^\circ$ 11$^\prime$ 25.5984$^{\prime\prime}$ W & <2~km/h  & Merlin 247 Raptor & -~s & (0~s)\\
& \multicolumn{1}{l}{\parbox[t]{5.5cm}{\raggedright\em G. Rossi, R. Sfair}} & 540.00& Good  & Clear & ComGPS &\\\midrule
8 & UTFPR - Neoville (Brazil) & 25$^\circ$ 30$^\prime$ 31.7373$^{\prime\prime}$ S& Clear & T25.4& IMG & Technical failure \\
&  & 49$^\circ$ 18$^\prime$ 57.1841$^{\prime\prime}$ W& <2~km/h  & QHY174M-GPS & -~s & (0~s)\\
& \multicolumn{1}{l}{\parbox[t]{5.5cm}{\raggedright\em F. B. Ribas }}& 937.67 & Good  & Clear & CamGPS &\\

\midrule
\multicolumn{11}{c}{\textbf{2023-07-28}}\\[-0.35em]\midrule
1 & SOAR (Chile) & 30$^\circ$ 14$^\prime$ 16.89$^{\prime\prime}$ S& Clear & T410.0& IMG & Positive \\
&  & 70$^\circ$ 44$^\prime$ 21.12$^{\prime\prime}$ W& 2-5~km/h  & Raptor & 0.200~s & (0~s)\\
& \multicolumn{1}{l}{\parbox[t]{5.5cm}{\raggedright\em J. I. B. Camargo, C. L. Pereira}} & 2693.95& Good  & Clear & CamGPS &\\\pagebreak
2 & Danish/ESO  (Chile)& 29$^\circ$ 15$^\prime$ 31.7592$^{\prime\prime}$ S & Clear & T154.0& IMG & Positive \\
&  & 70$^\circ$ 44$^\prime$ 1.464$^{\prime\prime}$ W& $\sim$20~km/h   & Andor Ixion Ultra & 0.1~s & (0.0~s)\\
& \multicolumn{1}{l}{\parbox[t]{5.5cm}{\raggedright\em C. Snodgrass, M. Bonavita, G. Columba}} & 2345.44& Good  & red & CamGPS &\\\midrule
3 & Campocatino Austral Observatory (Chile) & 30$^\circ$ 28$^\prime$ 15.2905$^{\prime\prime}$ S& Clear & T61.0& IMG & Positive \\
& & 70$^\circ$ 45$^\prime$ 53.9015$^{\prime\prime}$ W & <2~km/h & Moravian C3-PRO & 1.0~s & (0~s)\\
& \multicolumn{1}{l}{\parbox[t]{5.5cm}{\raggedright\em A. Zapparata, F. Mallia, G. Isopi}} & 1525.00& Good  & Clear & CamGPS &\\\midrule
4 & Observatório do Pico dos Dias (Brazil) & 22$^\circ$ 32$^\prime$ 49.5658$^{\prime\prime}$ S& Clear & T60.0& IMG & Negative \\
& {\raggedright\em L. Liberato, V. Moura} & 45$^\circ$ 26$^\prime$ 43.8041$^{\prime\prime}$ W & 12-19~km/h  & Ixon 4269 & 0.2877~s & (0.013~s)\\
& \multicolumn{1}{l}{\parbox[t]{5.5cm}{\raggedright\em A.L. Guimaraes, J. Arcas, Silva }} & 1112.15 & Moderate  & Empty & CamGPS &\\\midrule
5 & CASLEO - Cerro Burek HSH (Argentina) & 31$^\circ$ 47$^\prime$ 13.2$^{\prime\prime}$ S& Partly cloudy & T60.0& IMG & Negative \\
&  & 69$^\circ$ 18$^\prime$ 24.12$^{\prime\prime}$ W & 20-28~km/h  & SBig & 6.0~s & (3~s)\\
& \multicolumn{1}{l}{\parbox[t]{5.5cm}{\raggedright\em Mario Melita, Luis Mammana}} & 2591.00& Moderate  & Clear & Other &\\\midrule
6 & UC Observatory (Chile) & 33$^\circ$ 16$^\prime$ 8.9364$^{\prime\prime}$ S & Clear & T50.0& IMG & Negative \\
&& 70$^\circ$ 32$^\prime$ 4.1352$^{\prime\prime}$ W & <2~km/h & Raptor Merlin & 5~s & (0~s)\\
& \multicolumn{1}{l}{\parbox[t]{5.5cm}{\raggedright\em N. Castro, L. Vanzi, R. Leiva}} & 1475.20 & Good  & Clear & ComGPS &\\\midrule
7 & Astroquinta (Chile) & 32$^\circ$ 53$^\prime$ 0.2796$^{\prime\prime}$ S& Clear & T27.9& IMG & Negative \\
&  & 71$^\circ$ 14$^\prime$ 55.7412$^{\prime\prime}$ W & <2~km/h  & QHY168C & 10~s & (2.7~s)\\
& \multicolumn{1}{l}{\parbox[t]{5.5cm}{\raggedright\em Leo Peiro}} & 154.14 & Good  & L-eNhance & ComNTP &\\

\end{longtable}
\end{small}

\clearpage
\section{Light curves, limb fits, and radial residuals}
\label{appendix:lightcurves}

\begin{figure*}[h!]
\centering
\lcpanel[fig:lc_20200817_a]{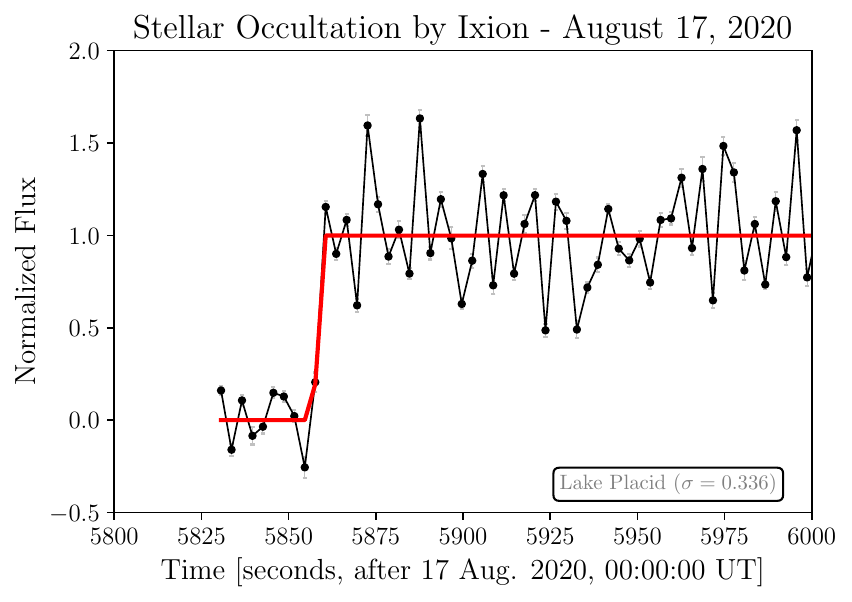}{Lake Placid}\hfill
\lcpanel[fig:lc_20200817_b]{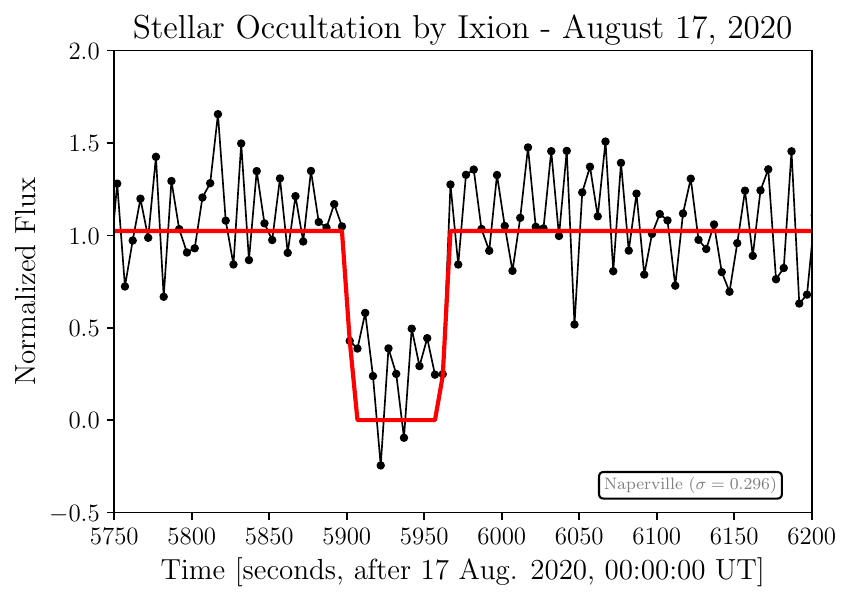}{Naperville}\hfill
\lcpanel[fig:lc_20200817_c]{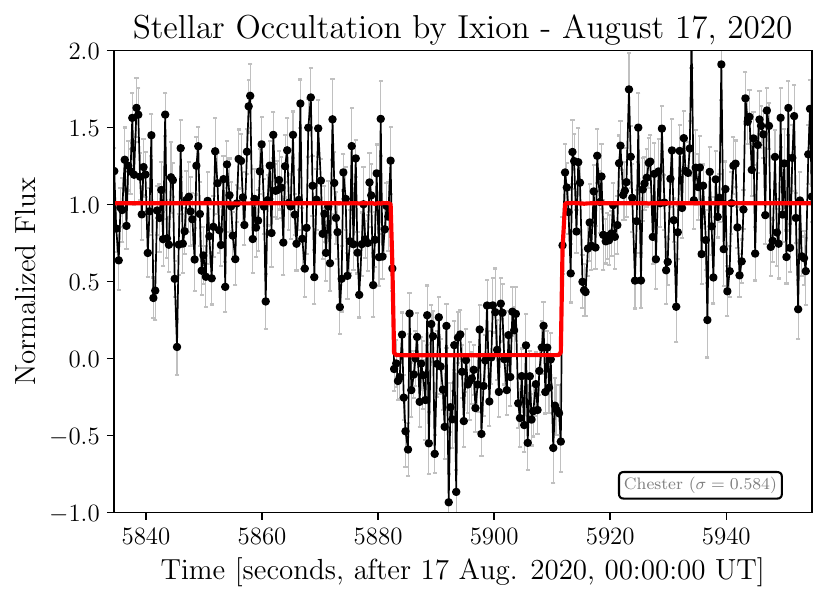}{Chester}\hfill
\lcpanel[fig:lc_20201013_a]{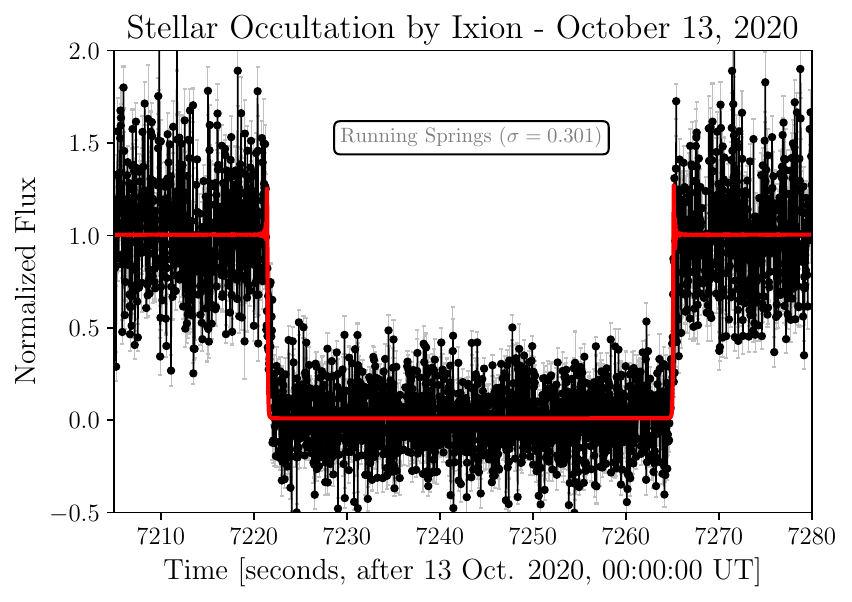}{Running Springs}\\[0.6em]

\lcpanel[fig:lc_20201013_b]{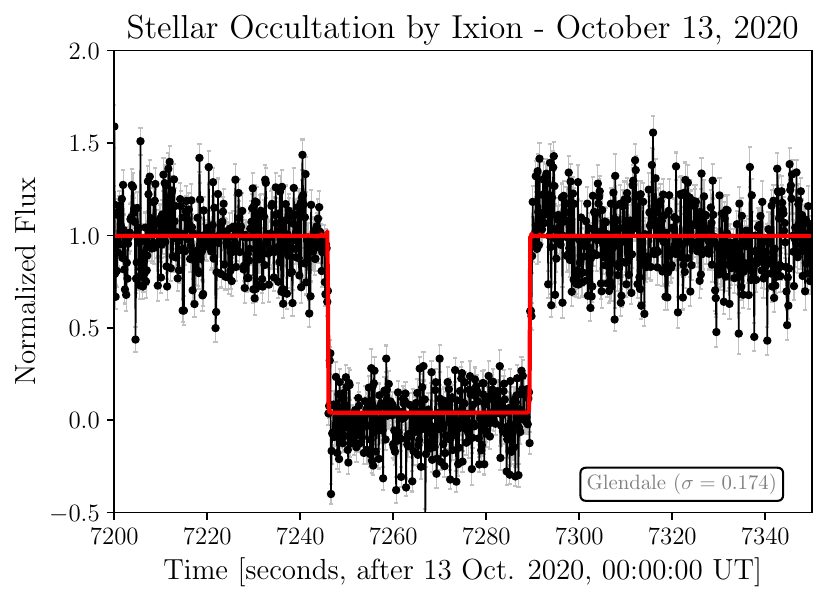}{Glendale}\hfill
\lcpanel[fig:lc_20201013_c]{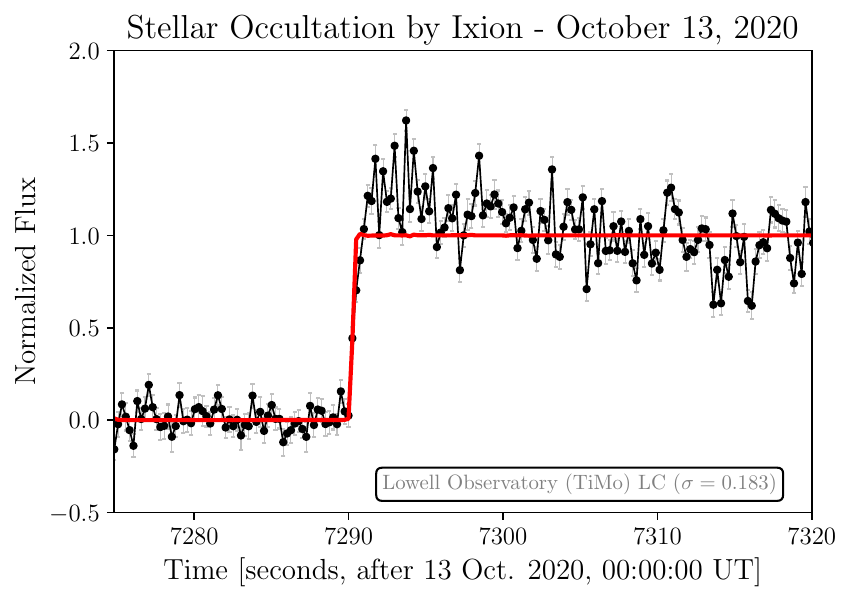}{Lowell Observatory (TiMo)}\hfill
\lcpanel[fig:lc_20201013_d]{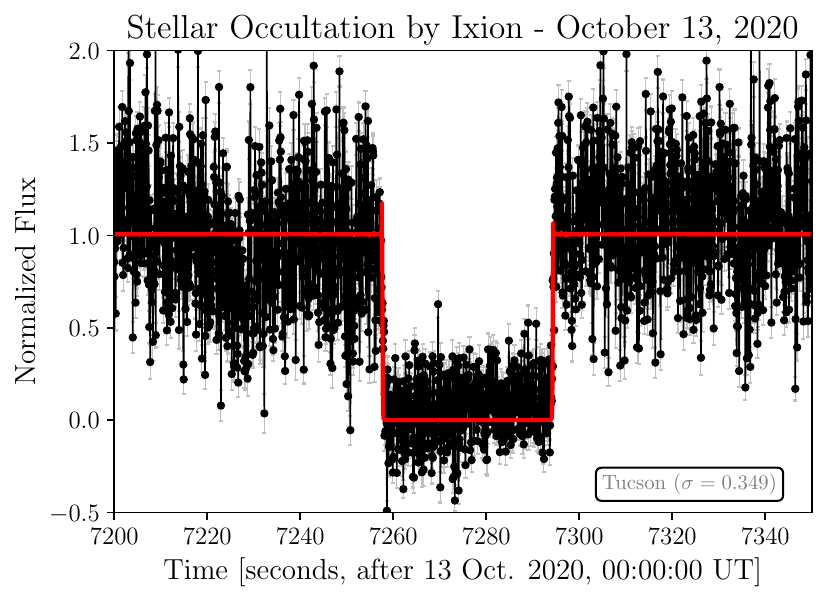}{Tucson}\hfill
\lcpanel[fig:lc_20201013_e]{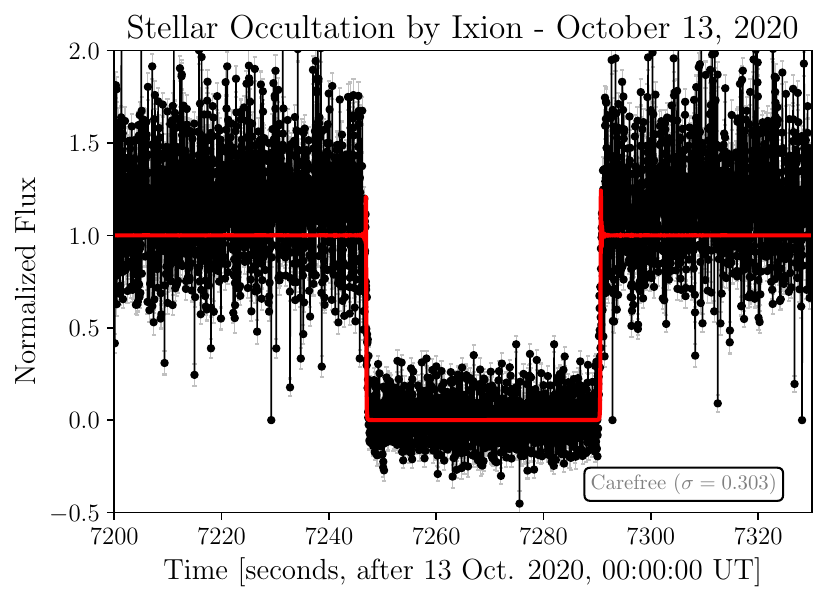}{Carefree}\\[0.6em]

\lcpanel[fig:lc_20201013_f]{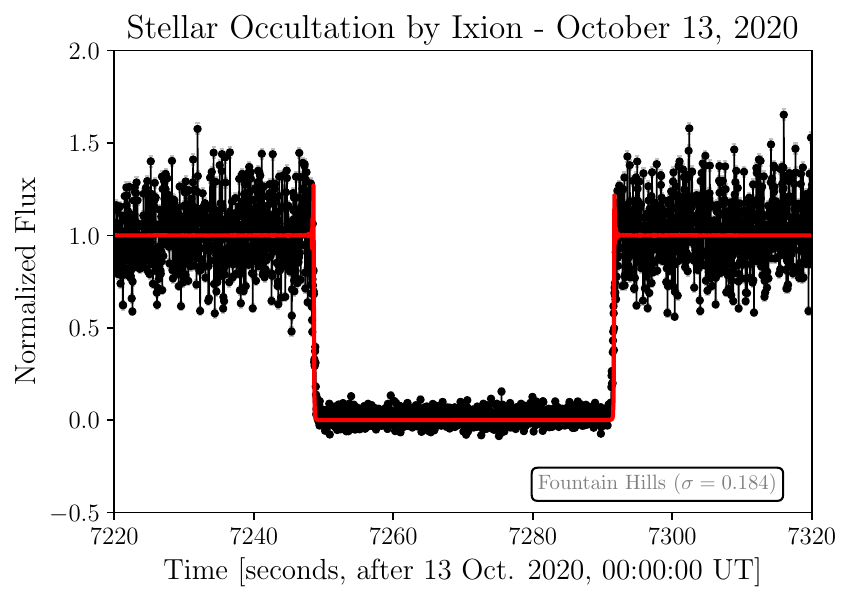}{Fountain Hills}\hfill
\lcpanel[fig:lc_20201013_g]{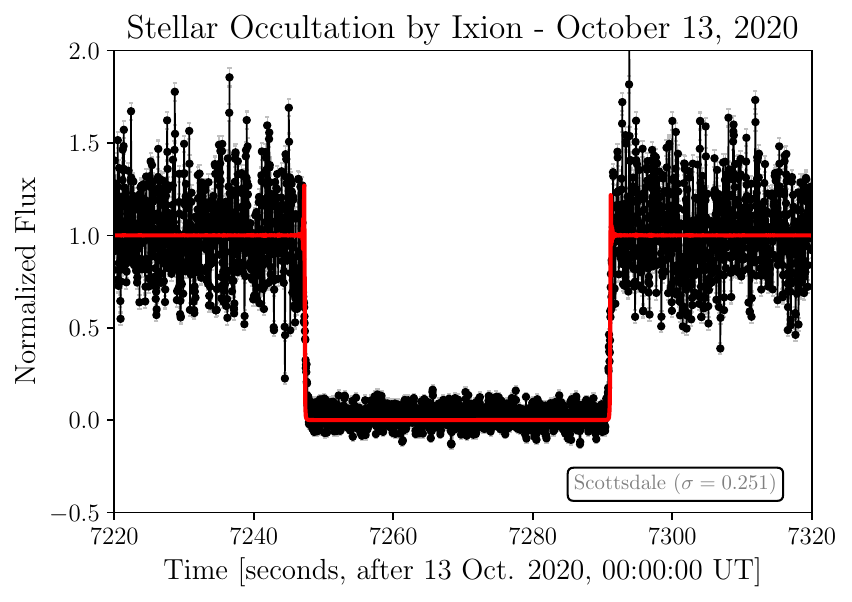}{Scottsdale}\hfill
\lcpanel[fig:lc_20201013_h]{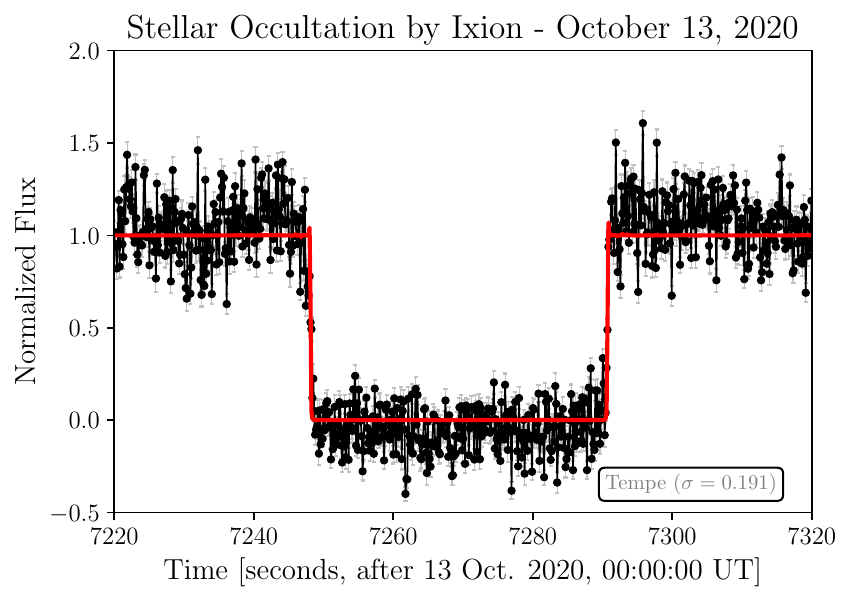}{Tempe}\hfill
\lcpanel[fig:lc_20210428_a]{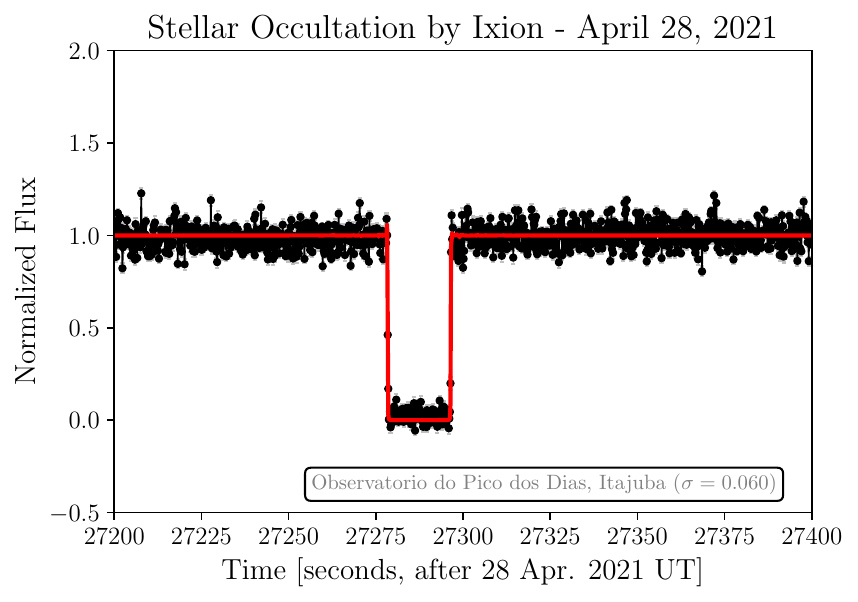}{OPD}\\[0.6em]

\lcpanel[fig:lc_20210428_b]{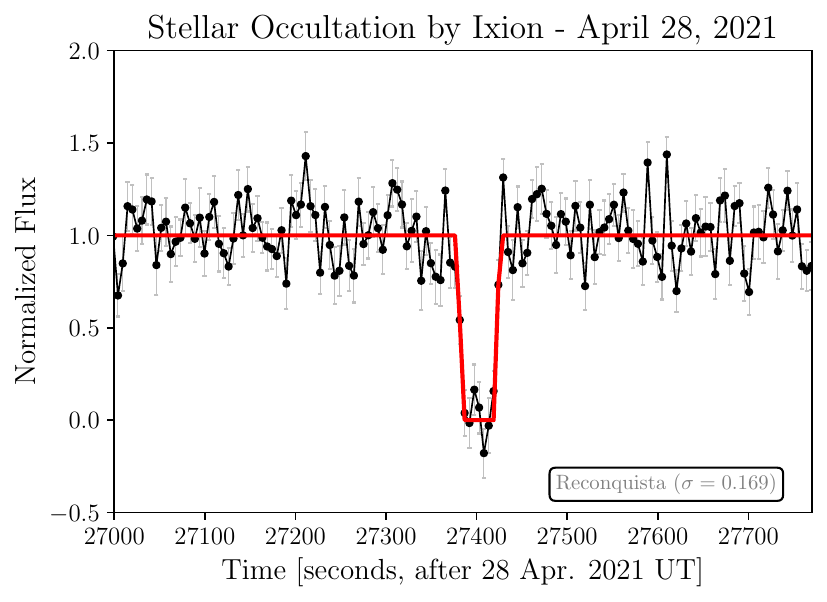}{Reconquista}\hfill
\lcpanel[fig:lc_20210428_c]{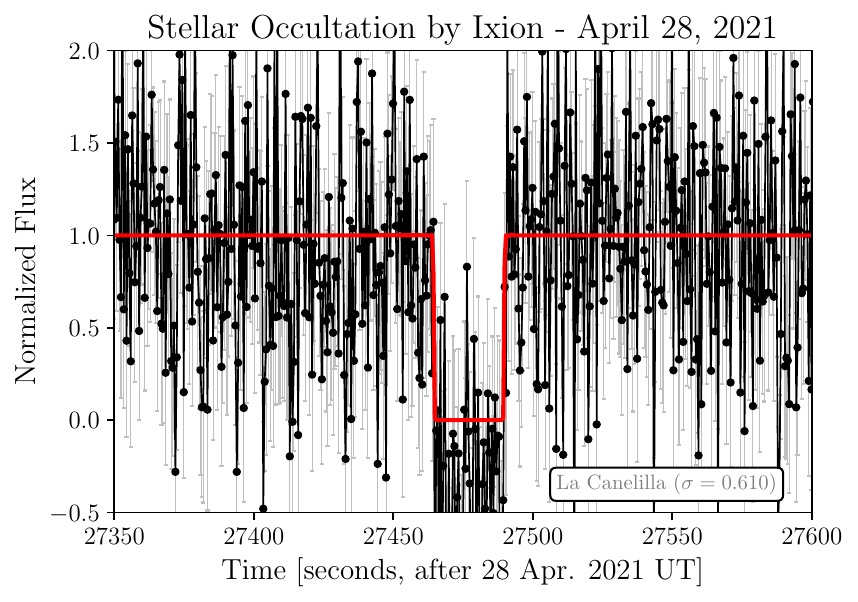}{La Canelilla}\hfill
\lcpanel[fig:lc_20210428_d]{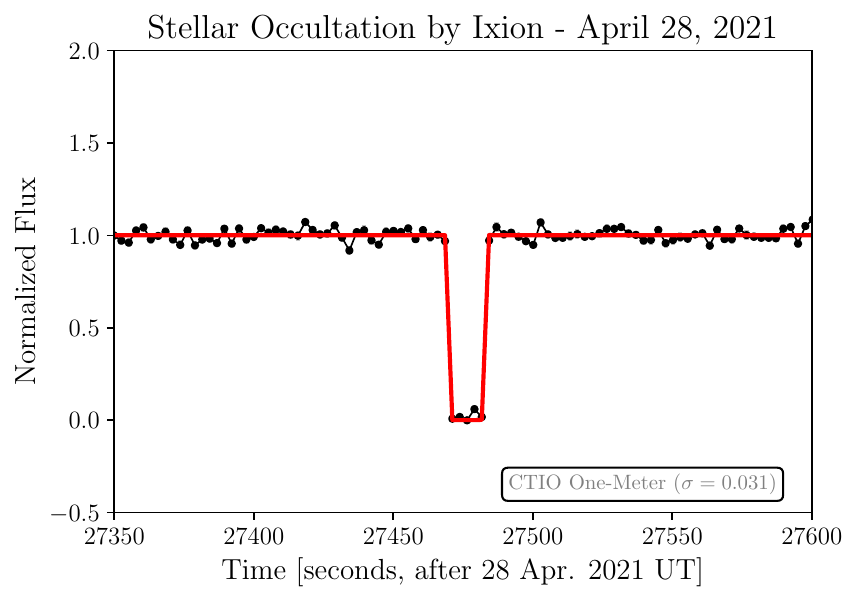}{CTIO 1-m}\hfill
\lcpanel[fig:lc_20210428_e]{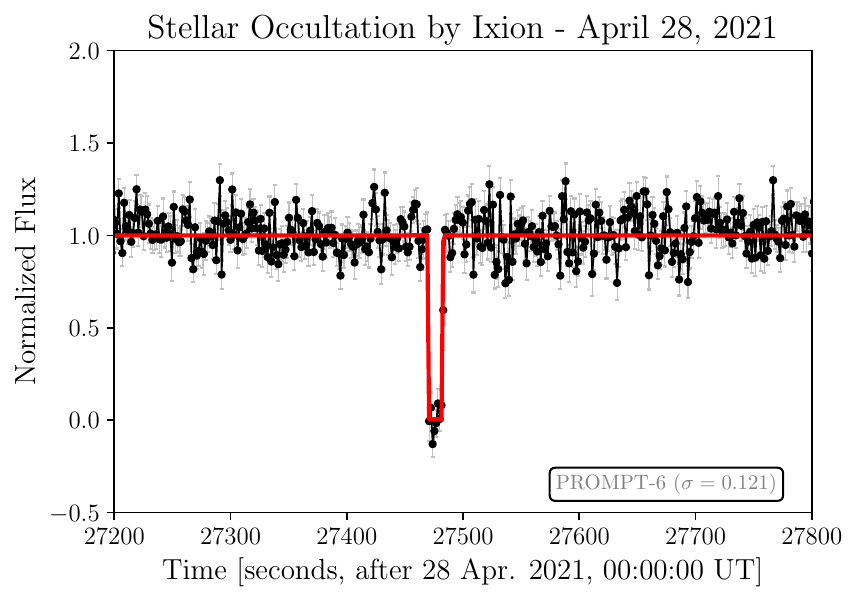}{PROMPT-6}\\[0.6em]

\lcpanel[fig:lc_20210428_f]{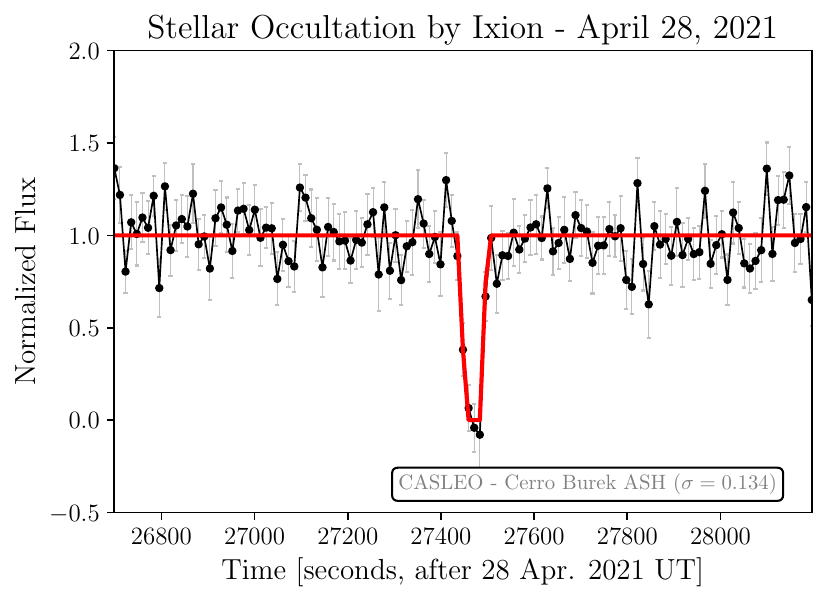}{CASLEO -- Cerro Burek ASH}\hfill
\lcpanel[fig:lc_20210520_a]{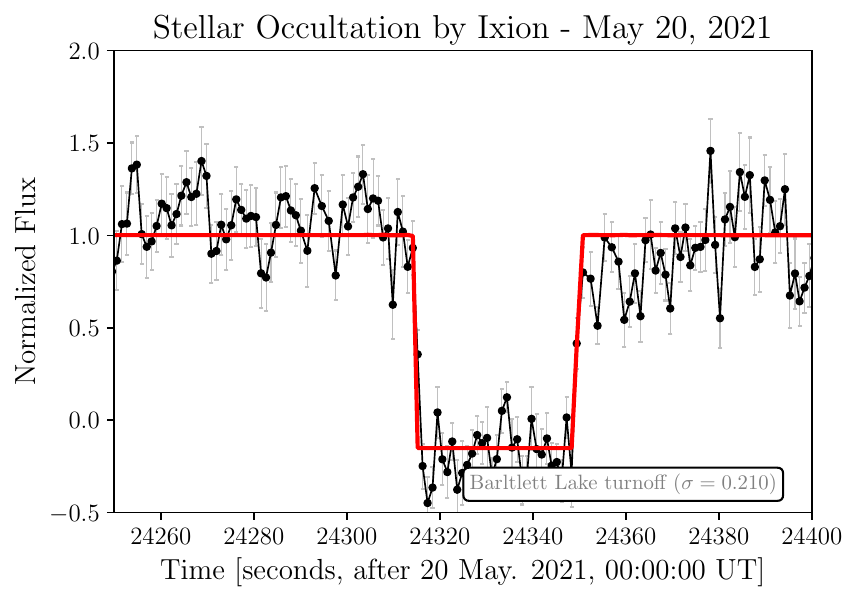}{Bartlett Lake Turnoff}\hfill
\lcpanel[fig:lc_20210520_b]{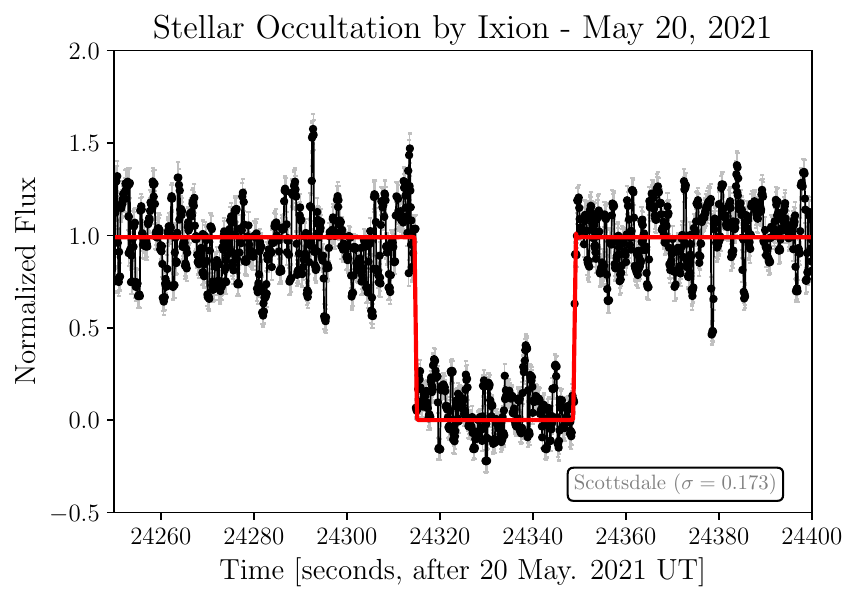}{Scottsdale}\hfill
\lcpanel[fig:lc_20210520_c]{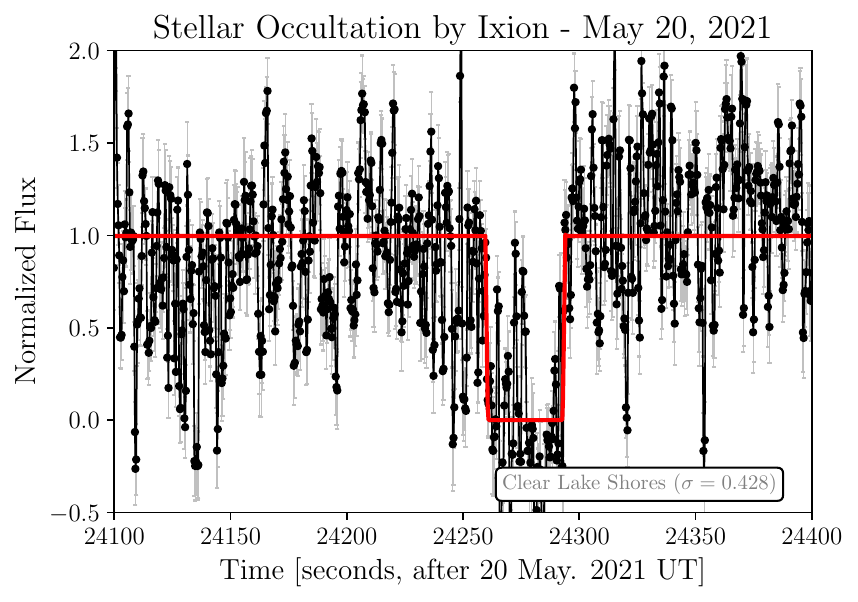}{Clear Lake Shores}\\[0.6em]

\caption{Normalised flux light curves of the stellar occultations by (28978)~Ixion. The black points represent the observed flux with uncertainties, and the red line shows the best-fit model. The x-axis represents the time in seconds relative to 00:00:00~UT on the observation date, and the y-axis shows the normalised flux.}
\label{fig:ixion_lcs_all}
\end{figure*}

\begin{figure*}[t]
\ContinuedFloat
\centering
\setcounter{subfigure}{20}
\lcpanel[fig:lc_20210520_d]{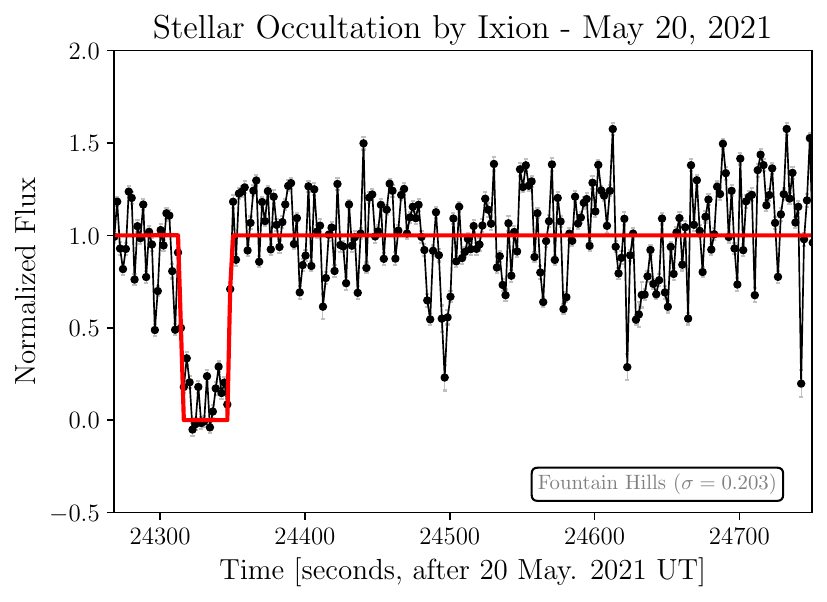}{Fountain Hills}\hfill
\lcpanel[fig:lc_20210817_a]{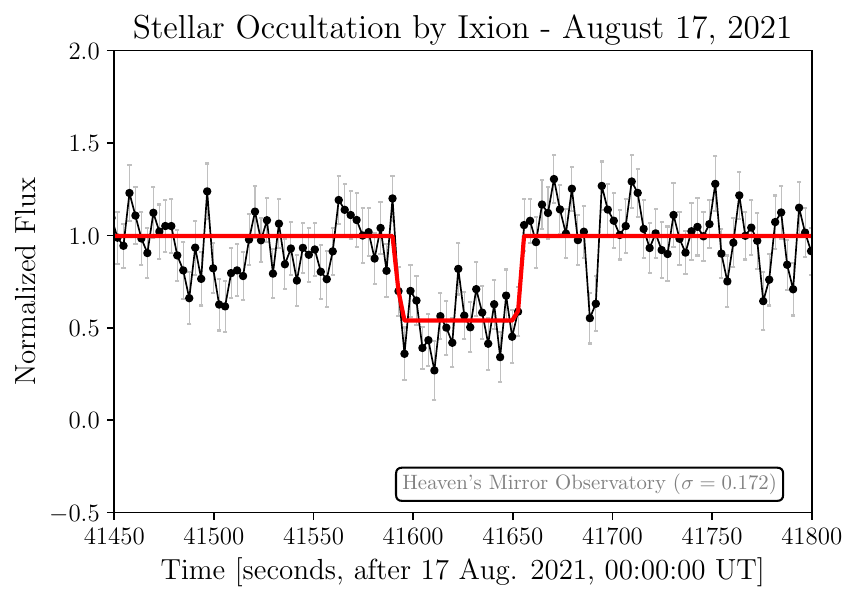}{Heaven's Mirror Obs.}\hfill
\lcpanel[fig:lc_20220602_a]{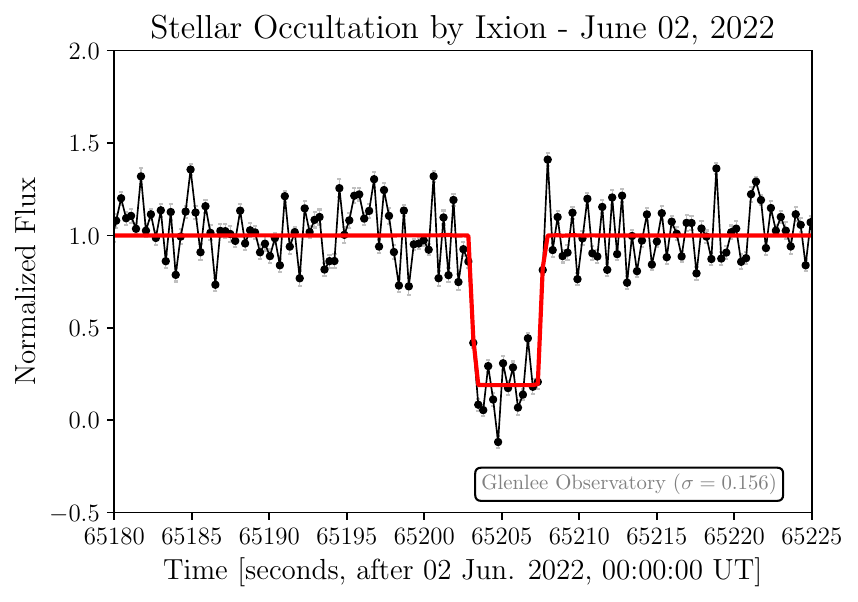}{Glenlee Observatory}\hfill
\lcpanel[fig:lc_20220630_a]{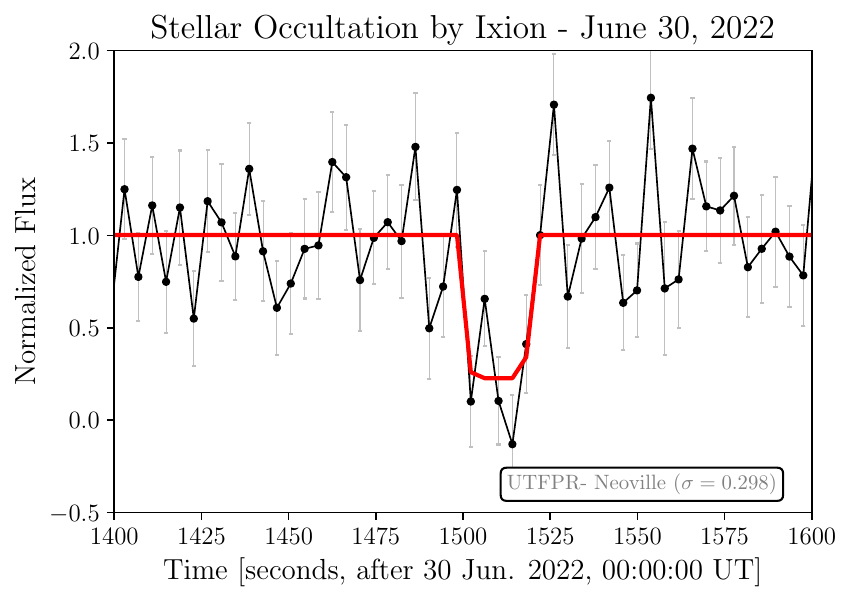}{UTFPR--Neoville}\\[0.6em]

\lcpanel[fig:lc_20220630_b]{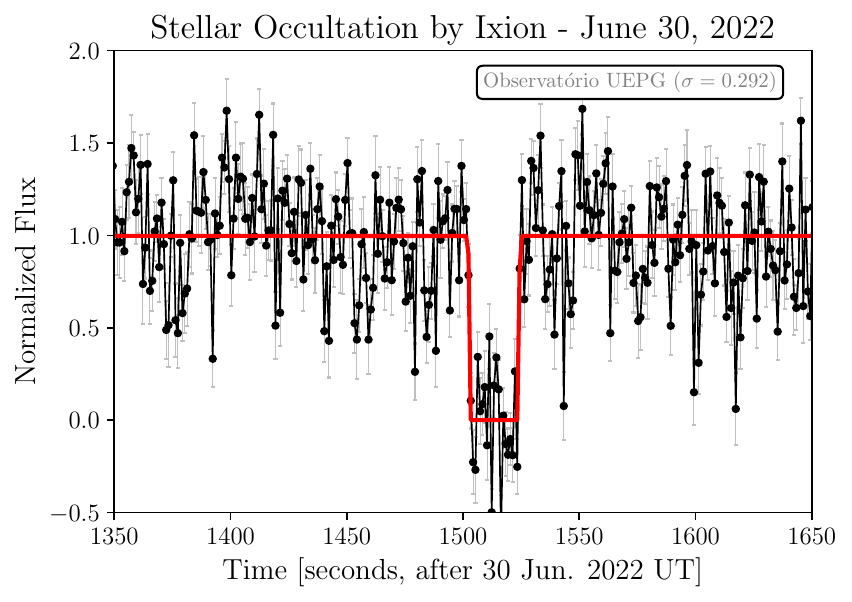}{Observatório UEPG}\hfill
\lcpanel[fig:lc_20220630_c]{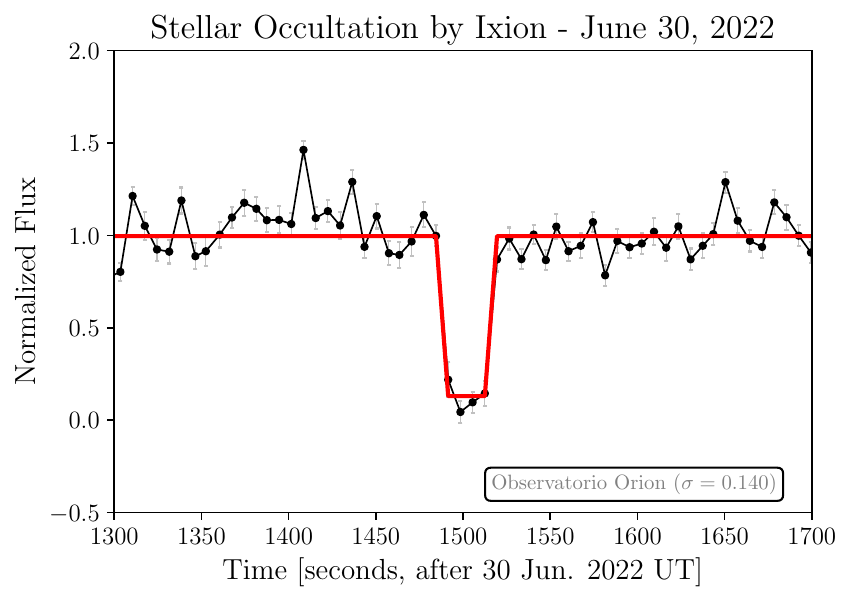}{Observatorio Orion}\hfill
\lcpanel[fig:lc_20220630_d]{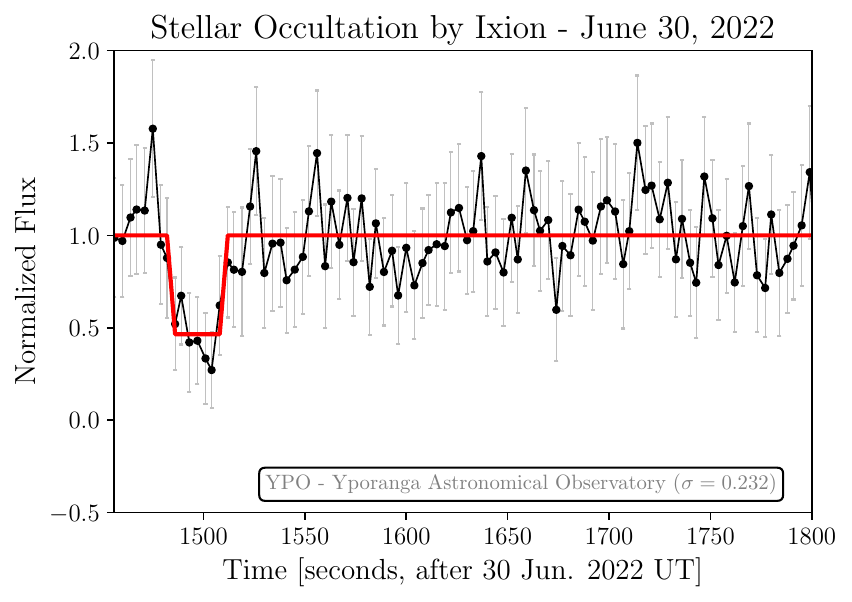}{YPO}\hfill
\lcpanel[fig:lc_20230728_a]{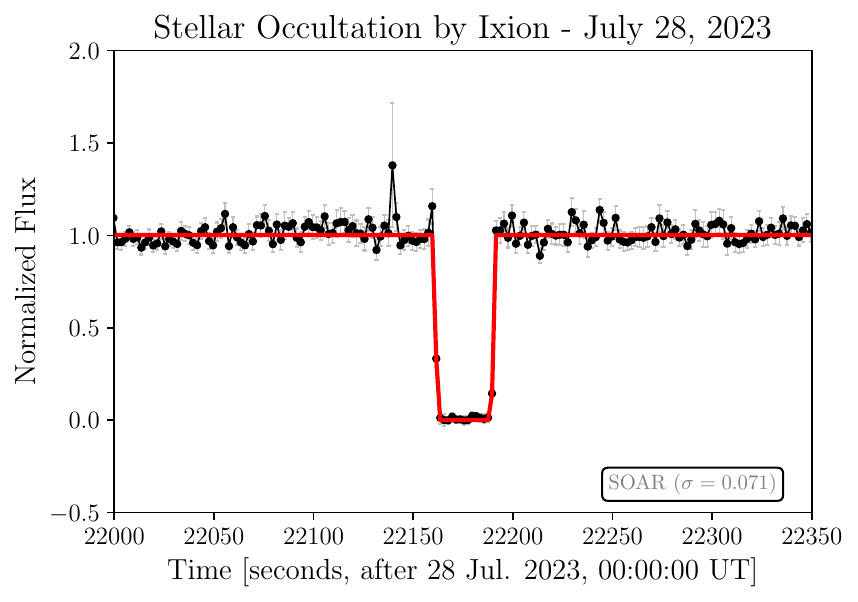}{SOAR}\\[0.6em]

\lcpanel[fig:lc_20230728_b]{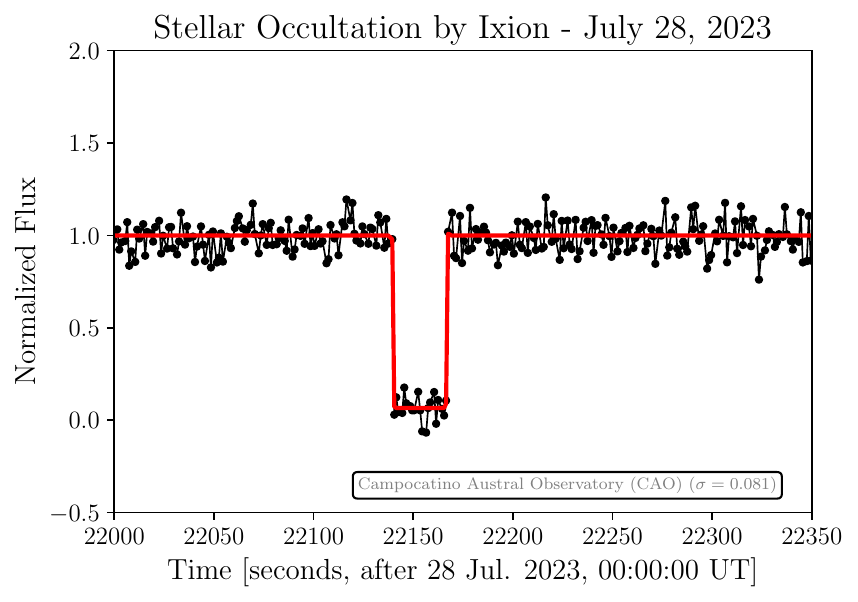}{CAO}\hfill
\lcpanel[fig:lc_20230728_c]{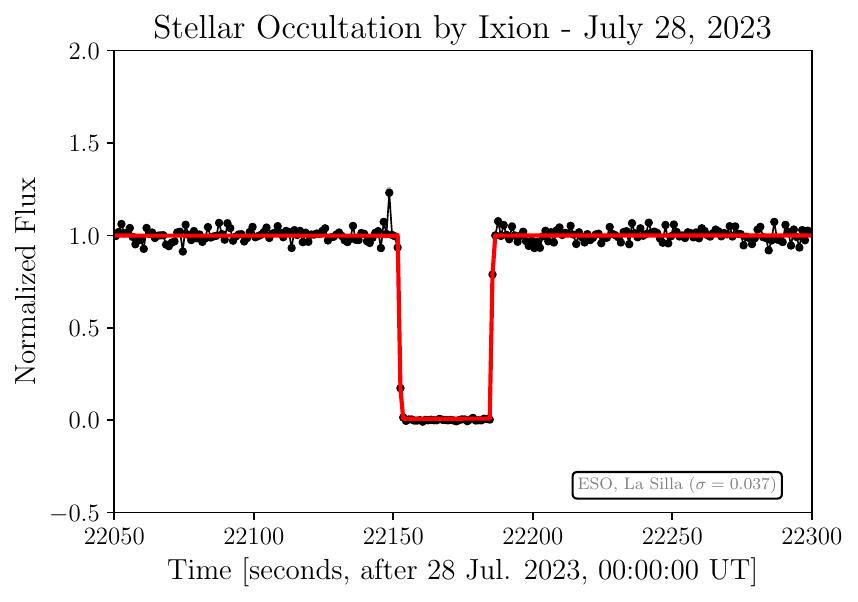}{Danish/ESO}\hfill

\caption[]{\emph{(continued)}}
\end{figure*}

\begin{figure*}[h!]
\centering
\begin{subfigure}[t]{0.24\textwidth}
  \includegraphics[width=\linewidth]{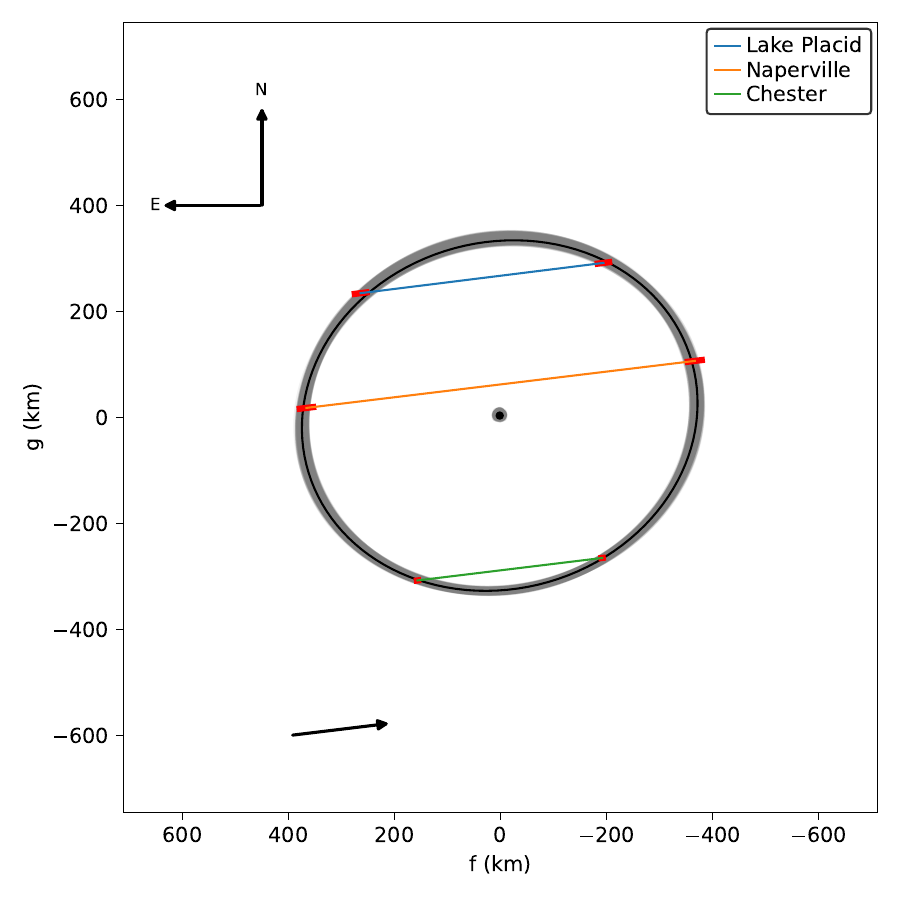}
  \caption{17 August 2020}\label{fig:ellipse_fits_a}
\end{subfigure}\hfill
\begin{subfigure}[t]{0.24\textwidth}
  \includegraphics[width=\linewidth]{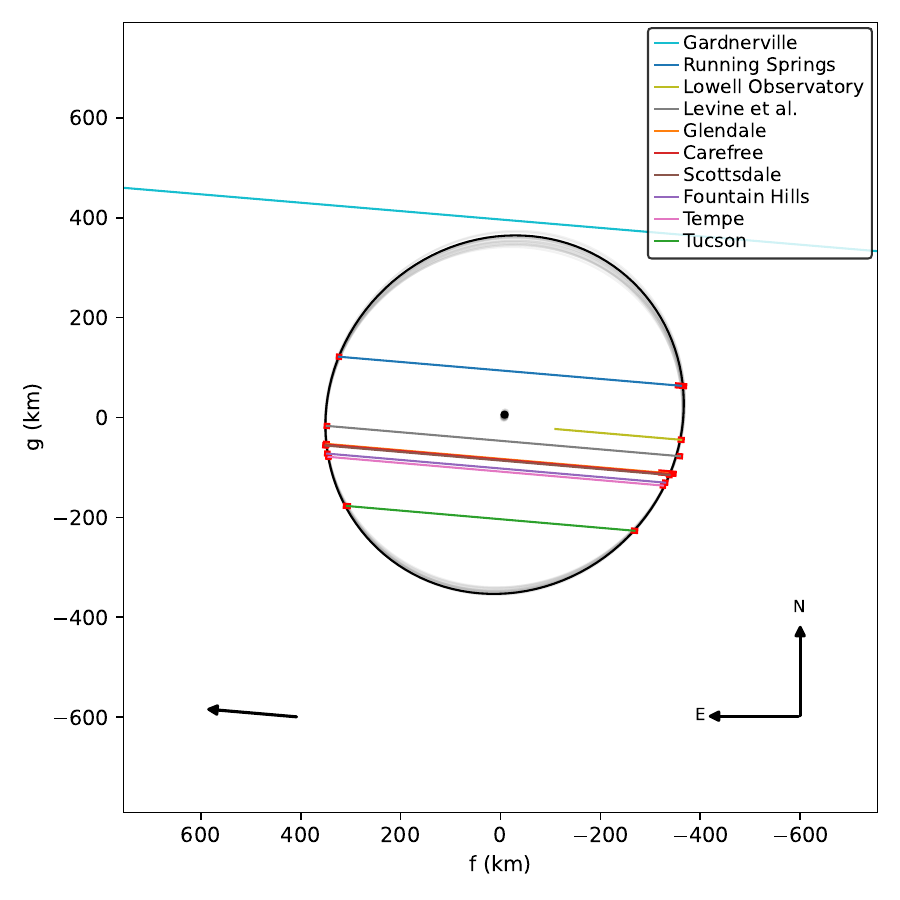}
  \caption{13 October 2020}\label{fig:ellipse_fits_b}
\end{subfigure}\hfill
\begin{subfigure}[t]{0.24\textwidth}
  \includegraphics[width=\linewidth]{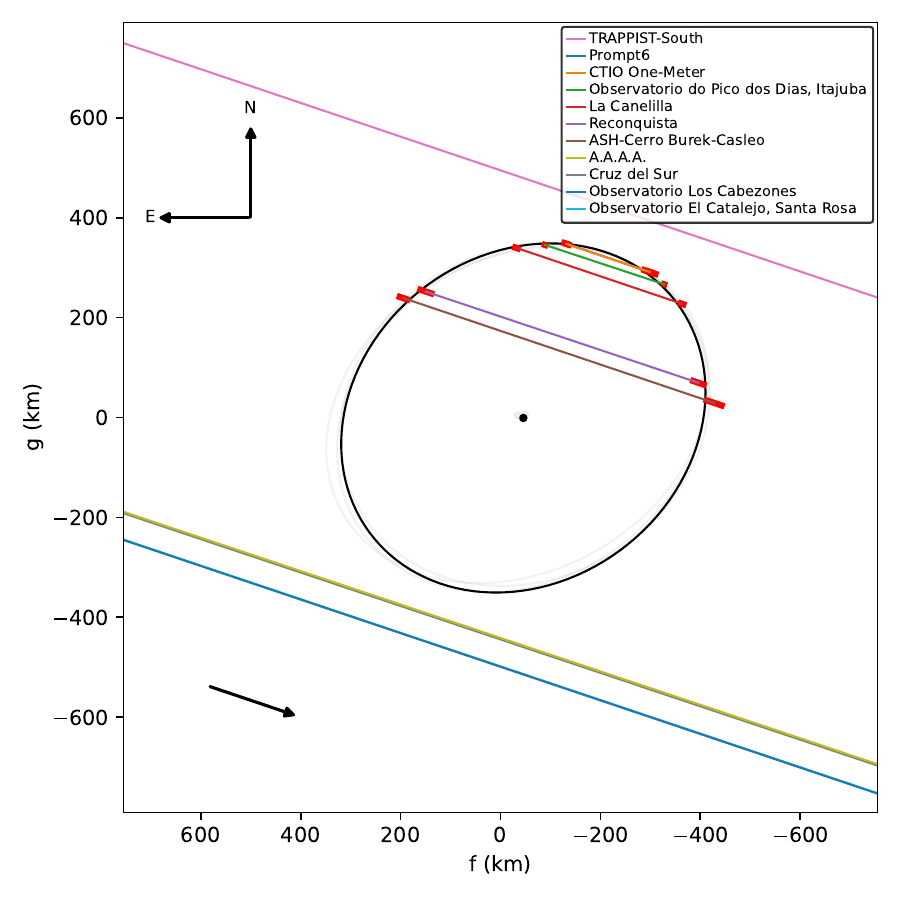}
  \caption{28 April 2021}\label{fig:ellipse_fits_c}
\end{subfigure}\hfill
\begin{subfigure}[t]{0.24\textwidth}
  \includegraphics[width=\linewidth]{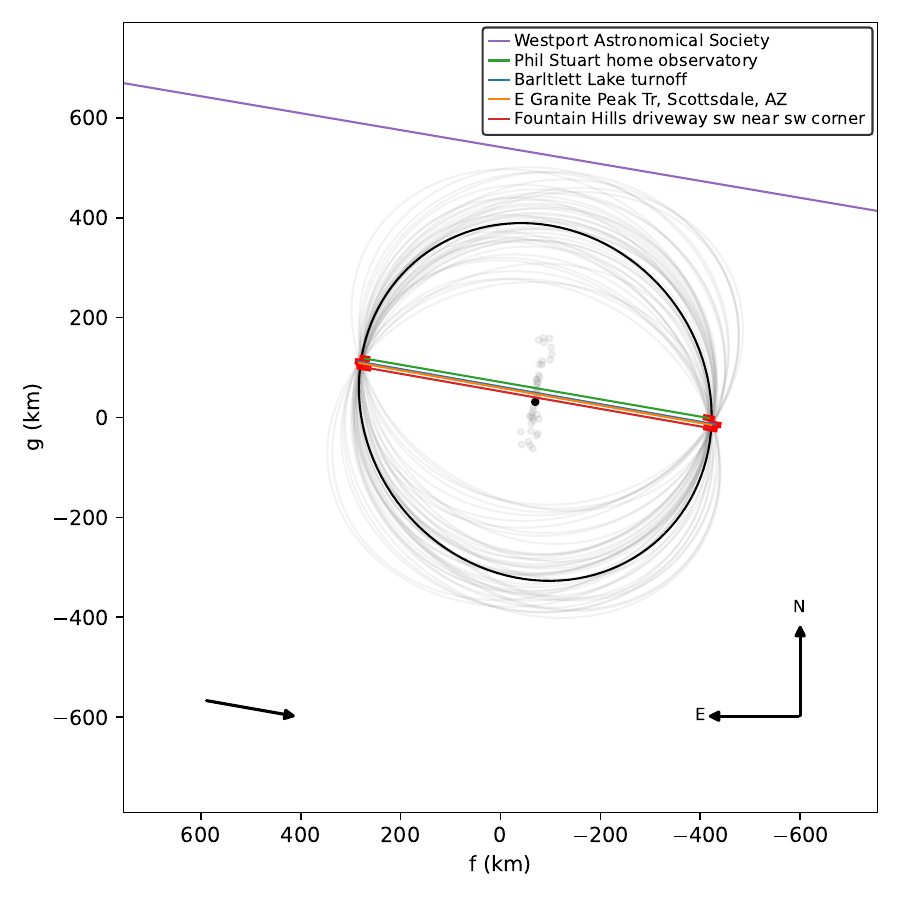}
  \caption{20 May 2021}\label{fig:ellipse_fits_d}
\end{subfigure}

\medskip

\begin{subfigure}[t]{0.24\textwidth}
  \includegraphics[width=\linewidth]{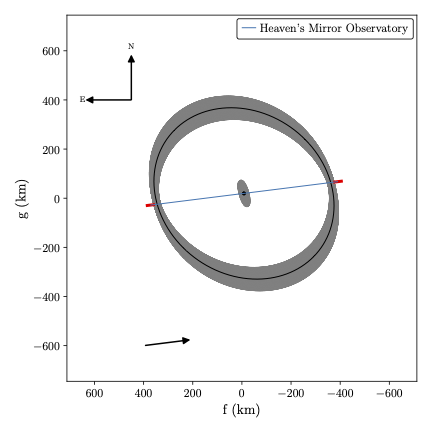}
  \caption{17 August 2021}\label{fig:ellipse_fits_e}
\end{subfigure}\hfill
\begin{subfigure}[t]{0.24\textwidth}
  \includegraphics[width=\linewidth]{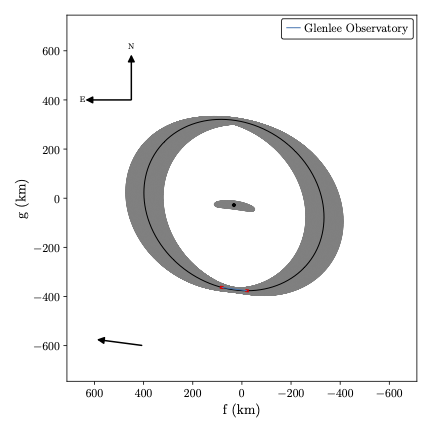}
  \caption{2 June 2022}\label{fig:ellipse_fits_f}
\end{subfigure}\hfill
\begin{subfigure}[t]{0.24\textwidth}
  \includegraphics[width=\linewidth]{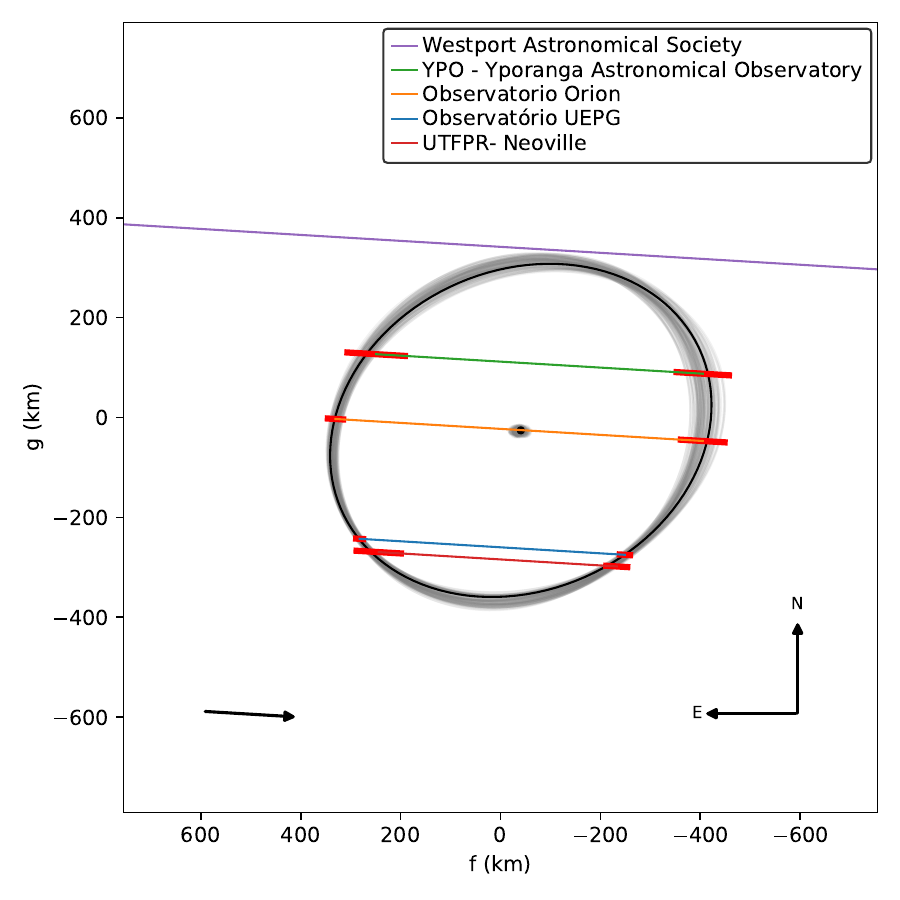}
  \caption{30 June 2022}\label{fig:ellipse_fits_g}
\end{subfigure}\hfill
\begin{subfigure}[t]{0.24\textwidth}
  \includegraphics[width=\linewidth]{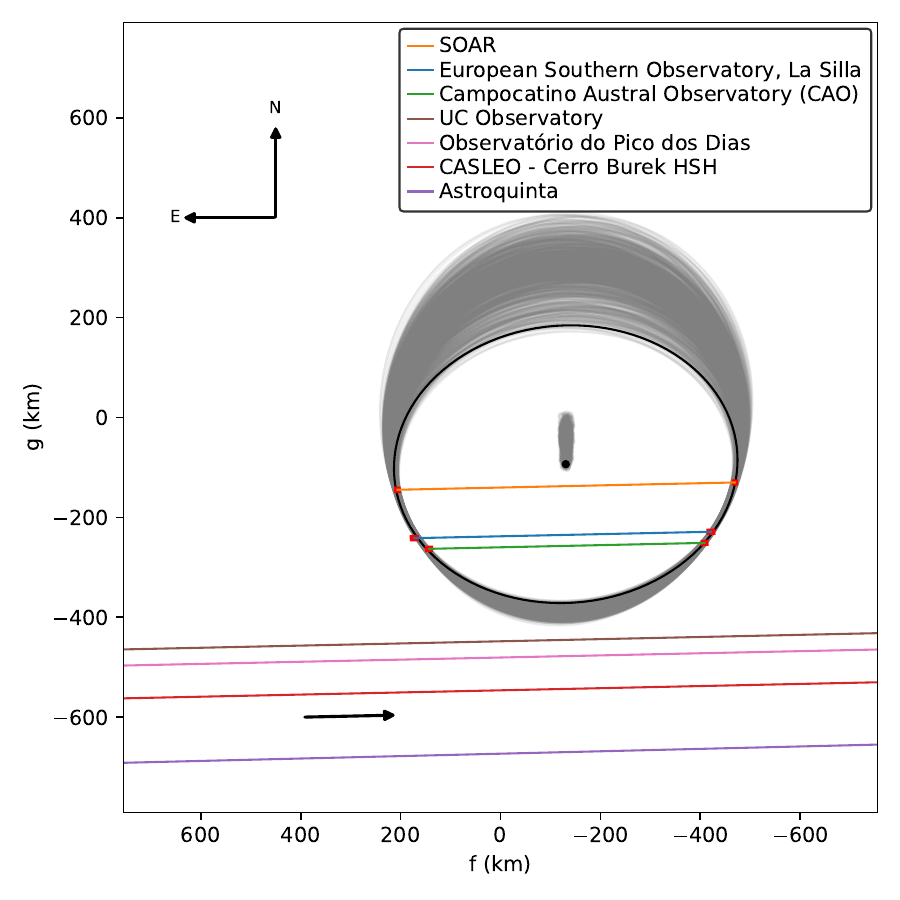}
  \caption{28 July 2023}\label{fig:ellipse_fits_h}
\end{subfigure}

\caption{
Limb-fit diagrams for the eight stellar occultations by (28978)~Ixion:  
(a)~17 August 2020, (b)~13 October 2020, (c)~28 April 2021, (d)~20 May 2021,  
(e)~17 August 2021, (f)~2 June 2022, (g)~30 June 2022, and (h)~28 July 2023.  
Each chord is shown in a different colour, corresponding to the observing sites listed in the legends.  
Positive detections define the limb of the body, while continuous full-length chords represent negative observations that constrain the maximum extent of the limb.  
The x- and y-axes correspond to the sky-plane coordinates $f$ and $g$ (in km), oriented along the fundamental plane of the occultation. For the single-chord events (e) and (f), the limb shape was fixed to the 
global solution; only the astrometric offsets $(f,g)$ were determined. The red segments at the ends of the chords represent the ingress and egress timing uncertainties.  
An elliptical fit was applied in each case, with the best-fit solution (1$\sigma$) shown.  
The grey-shaded regions around the ellipses indicate the 1$\sigma$ confidence domains obtained from the fitting procedure.  
The corresponding parameters are listed in Table~\ref{tab:ixion_limb_solutions_full}.  
The black arrow indicates the direction of the shadow motion, while the labelled N and S arrows denote celestial north and south, respectively.
}
\label{fig:ellipse_fits}
\end{figure*}

\begin{figure*}
\centering
\includegraphics[width=14cm]{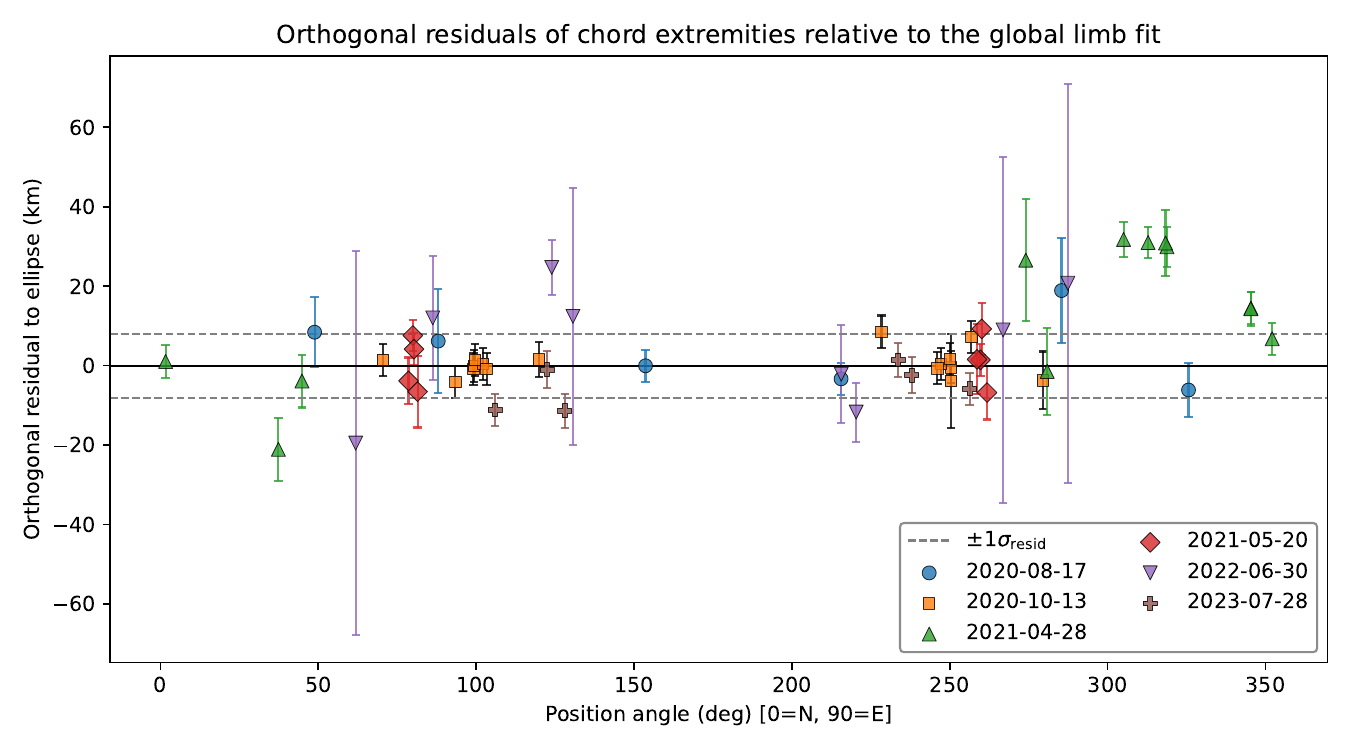}
\caption{
Orthogonal (normal) residuals of the chord extremities with respect to the global limb solution, plotted as a function of position angle in the ellipse-aligned frame. Residuals are measured along the local normal to the best-fitting ellipse, ensuring a geometrically consistent assessment of deviations over the entire limb. Each colour and marker corresponds to a different occultation epoch (same coding as in Fig.~\ref{fig:ixion_all_dates_fit}). No clear systematic trend with position angle is observed, indicating that the residuals are consistent with random scatter within the adopted observational and model uncertainties. The offsets observed for the 28 Apr 2021 and 30 Jun 2022 epochs are discussed in Sect.~\ref{sec:discussion} and may be associated with changes in the projected figure of the body at different rotational phases or aspect angles.
}
\label{fig:global_fit_residuals}
\end{figure*}
\end{document}